# Scalable Conformal $MoS_x$ Catalyst for Efficient Hydrogen Evolution at Industrial-Level Current Density in Alkaline Electrolyzers

Yong Zuo[1,2], Sebastiano Bellani[3,4,*], Meysoun Jabrane[2], Gabriele Saleh[2], Thi-Hong-Hanh Le[2], Michele Ferri[2,4], Davide Salusso[5], Zhanzhao Li[2], Valentina Mastronardi[3], Marilena I. Zappia[3], Manjunath Chatti[2], Mirko Prato[6], Silvia Dante[6], Francesco Bonaccorso[3], Yongsheng Han[1], Liberato Manna[2,4,7,*]

[1] School of Chemistry and Chemical Engineering, Chongqing University, Chongqing, 400044, China

[2] Nanochemistry Department, Istituto Italiano di Tecnologia, Via Morego 30, 16163 Genova, Italy

[3] BeDimensional S.p.A., Via Lungotorrente Secca, 30R, 16163 Genova, Italy

[4] Antares Electrolysis S.r.l., Piazza della Vittoria 14/19, 16121 Genova, Italy

[5] European Synchrotron Radiation Facility, CS 40220, Cedex 9 F-38043 Grenoble, France

[6] Materials Characterization Facility, Istituto Italiano di Tecnologia, Via Morego 30, 16163 Genova, Italy

[7] Lead contact

Correspondence: sebastiano.bellani@antares-electrolysis.com (S.B.); liberato.manna@iit.it (L.M.)

## SUMMARY

The development of simple and scalable fabrication strategies for cost-effective electrodes is crucial to advance water splitting in alkaline water electrolyzers (AWEs). Here, we present a coating-annealing method to conformally coat a $MoS_x$ catalyst layer onto a porous Ni foam (NF) substrate. By controlling the annealing process, the composition of the $MoS_x$ layer could be tuned from $MoS_2$ to $MoS_3$ and its catalytic performance for hydrogen evolution reaction (HER) in alkaline media was optimized. The $MoS_3$@NF synthesized by this method achieved industrially relevant HER current densities of 200 mA/cm$^2$ at a low overpotential of 246 mV, maintaining stable operation for over 240 h. The $MoS_3$@NF cathode, combined with a stainless steel anode, enabled an alkaline water electrolyzer (AWE) cell to operate steadily at 1.96 V and 1 A/cm$^2$ for 1000 h. This performance surpasses that of most of the previously reported water electrolyzers employing $MoS_x$-based cathodes. Our work demonstrates the potential of $MoS_3$ (with its abundant edge-sulfur atoms serving as active sites) as a high-performance cathode material for industrial AWEs.

## INTRODUCTION

Green hydrogen, produced through water electrolysis powered by electricity from renewable energy sources (such as solar, wind and others), has emerged as a promising alternative energy source to achieve climate neutrality[1,2]. According to energy supply and demand roadmaps, hydrogen could account for up to 10 % of the world's final energy consumption, contributing to a total reduction of 560 million metric tons of $CO_2$ by 2050[3–5]. This implies that, in the near future, green hydrogen may become the most economical and secure source of renewable energy[6]. In this context, alkaline water electrolyzers (AWEs) have been firmly established at the megawatt (MW)-scale for over a century, with single-stack capacities reaching several MWs[7]. However, commercial AWEs still face limitations in achieving cost-competitive hydrogen production due to limited current density and low volumetric performance, which constrain capital expenditure reduction and limit energy efficiency [8,9].

To boost the economic competitiveness of green hydrogen, it is crucial to design highly efficient and sustainable electrode catalysts for AWE deployment. In this context, huge efforts have been made to design high-performance catalysts based on noble metals, as they can substantially reduce operating costs compared to traditional catalysts, *e.g.,* Raney Ni, without compromising capital expenses at stack and system levels[10–13]. In particular, Ru-based electrocatalysts have recently garnered significant attention due to their relatively lower cost compared to Pt and their excellent catalytic activity for the hydrogen evolution reaction (HER) in alkaline media[14–17]. Despite the use of Ru in AWEs has been effective in reducing the levelized cost of hydrogen[8], its scarce availability poses a significant challenge for large-scale AWE deployment. This issue is further exacerbated by the high demand for noble metals in other clean energy technologies (*e.g.,* fuel cells). As a result, the development of efficient electrodes based on earth-abundant metals remains a critical research pursuit[18]. In this context, $MoS_2$ and its derivatives have emerged as promising cathode catalysts due to their theoretically Pt-like catalytic activity towards the HER[19,20].

$MoS_2$ was initially predicted by Nørskov and his co-workers to exhibit efficient HER activity, based on the Sabatier principle that an optimal HER catalyst should have a hydrogen adsorption free energy close to zero[21]. Since that prediction, extensive research has been devoted to $MoS_2$. It is widely recognized that the active sites for HER are located at the edges of $MoS_2$[21–23]. Therefore, maximizing edge exposure has been a key strategy to enhance HER activity of $MoS_2$ over the past decade[24–27]. However, most of the reported synthesis methods, such as bulk exfoliation[28,29] and chemical vapor deposition[30,31], are not directly applicable to large-scale industrial production. In addition, testing of $MoS_2$ or other $MoS_x$-based cathodes under industrially relevant conditions, including high current densities (up to 1 A/cm$^2$ or higher) in AWE, remains scarce. Hence, it has become critical to develop simple and scalable synthesis routes for the direct synthesis of $MoS_x$ materials with abundant Mo-S edges (*e.g.,* $MoS_3$)[32,33]. Furthermore, assessing their practical potential as cathodes in operational AWE devices is essential to bridge the gap between laboratory-scale research and industrial applications.

Here, we report a simple coating-annealing strategy to conformally deposit a $MoS_x$ catalyst layer onto a porous Ni foam (NF) substrate for alkaline HER applications. The coating protocol, annealing parameters and substrate characteristics were systematically investigated to achieve an optimal $MoS_3$ layer on NF. The resulting $MoS_3$@NF electrode demonstrates superior electrocatalytic performance, requiring a low overpotential of 246 mV to reach an alkaline HER current density of -200 mA/cm$^2$, outperforming the $MoS_2$@NF counterpart prepared at a higher annealing temperature (350 ℃ *vs.* 200 ℃). Moreover, the $MoS_3$@NF electrode exhibits remarkable stability, operating at -200 mA/cm$^2$ for over 240 h in a three-electrode cell configuration. Additionally, the cell can reach 1 A/cm$^2$ at a cell voltage of 1.96 V for at least 1000 h, surpassing the performance of most of the previously reported electrolyzers using a $MoS_x$-based cathode.

## RESULTS AND DISCUSSION

### Morphological Characterizations and Analyses

The preparation of the $MoS_x$ catalyst layer on a porous NF substrate involved a two-step process. First, the pre-cleaned substrate was uniformly coated with a $MoS_x$ precursor ($(NH_4)_2MoS_4$) solution. After drying, the coated substrate was subjected to an annealing treatment to promote the formation and adhesion of the $MoS_x$ layer on the NF surface (**Figure S1**), leading to what will be henceforth referred to as "$MoS_x$@substrate", or, in shorthand notation, $MoS_x$@NF, and as depicted in **Figure 1a,b**. A representative electrode prepared under mild annealing conditions (200 ℃ for 1 h in Ar) is referred to as $MoS_x$@NF_200C-1h. As shown in **Figure 1c,d**, the resulting electrode consists of a NF covered by a conformal $MoS_x$ catalyst that is approximately 2 μm thick (**Figure S2**). Scanning electron microscopy (SEM) elemental mapping of the $MoS_x$@NF_200C-1h sample indicated a uniform distribution of the Mo and S elements across the surface, attesting that the catalyst layer has homogeneous coverage and compositional uniformity (**Figure 1e**). Also, the in-situ growth of the $MoS_x$ layer is likely to create intimate interactions at the catalyst-NF interface, ensuring robust adhesion, as proven later in this work.

The preparation protocol for $MoS_x$@NF was systematically optimized to obtain a uniform $MoS_x$ coating on the substrate surface while enhancing the HER activity. Key parameters, including the concentration of the $MoS_x$ precursor, the coating method and the annealing process, were systematically investigated. The concentration of the $MoS_x$ precursor solution played a crucial role in controlling the surface coverage of the catalyst layer. As shown in **Figure S3**, a diluted $MoS_x$ precursor solution (5 mM) resulted in incomplete surface coverage, leaving significant areas of the bare NF exposed. In contrast, a 50 mM precursor solution was found to be optimal, leading to a uniform $MoS_x$ coating. The method used to deposit the $MoS_x$ precursor on the NF also had a significant impact on the coating quality. As shown in **Figure**

**S3**, dip coating was identified as the most effective approach, producing a conformal and continuous $MoS_x$ layer on the NF surface. In comparison, drop-casting resulted in the formation of $MoS_x$ flakes with pronounced cracks between them, which negatively affected the overall coating integrity.

The annealing process was found to be another critical factor. Higher annealing temperatures (*e.g.,* 500 ℃) or faster ramping rates (*e.g.*, 30 ℃/min) led to the formation of cracks separating agglomerates of $MoS_x$ flakes, while a continuous coating layer could be achieved at lower ramping rates (*e.g.*, 10 °C/min) (**Figure S4**). This cracked $MoS_x$ coating exhibited inferior HER performance compared to the $MoS_x$ layer uniformly coating the NF surface (**Figure S5**). To further explore the versatility of the process, other porous substrates were tested, including Cu foam, stainless steel felt, and Ti felt (**Figure S6**). Among these, NF coated with $MoS_x$ exhibited the best HER performance and good reproducibility (**Figure S7**). Given the simplicity and versatility of this coating-annealing process, we can state that the synthesis of $MoS_x$@NF electrodes should be readily scaled up to fabricate large-area electrodes suitable for industrial AWE systems.

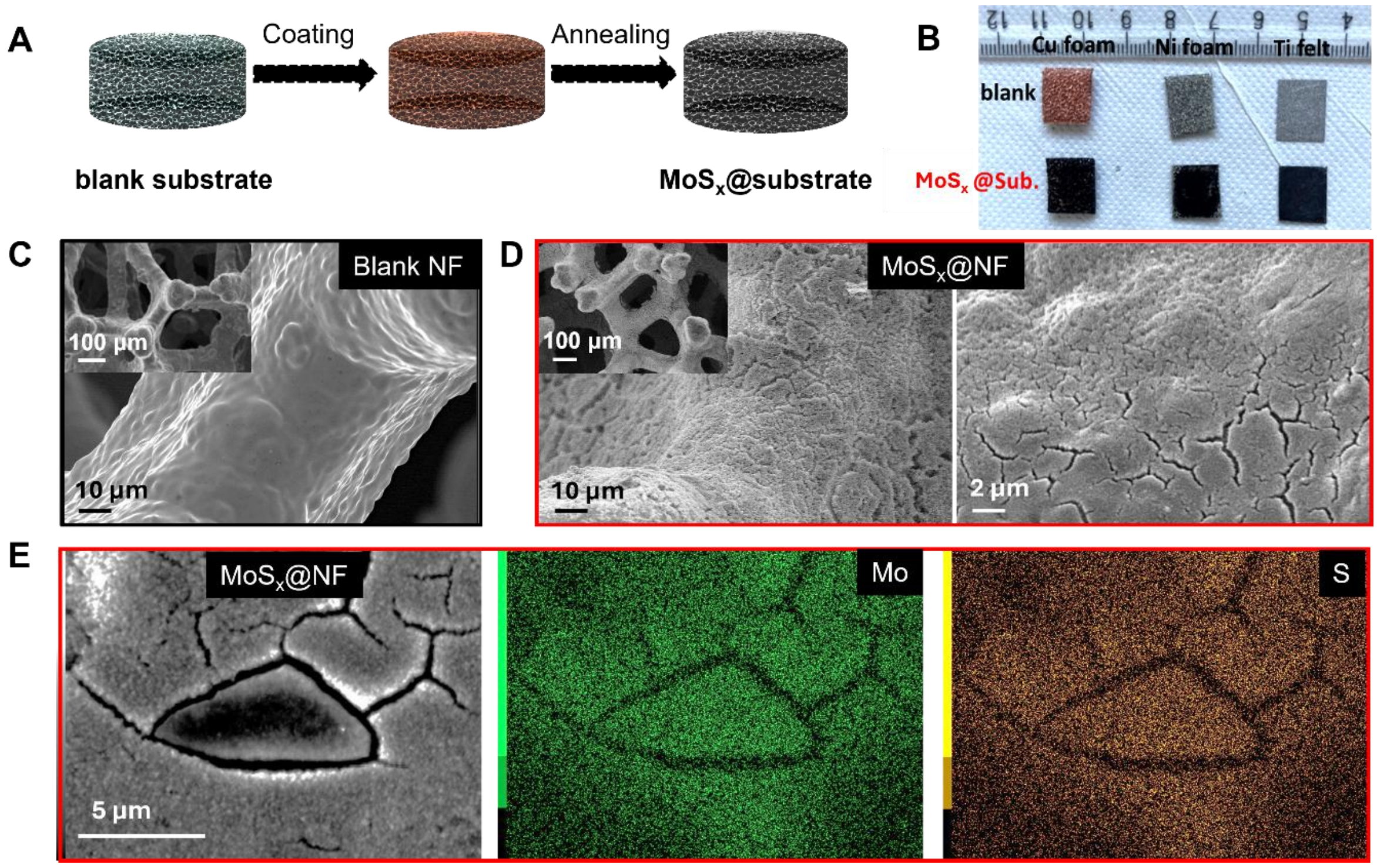

**Figure 1. Catalysts synthesis and morphological characterizations**

(A) Preparation scheme.

(B) photographs of electrodes prepared using different porous substrates.

(C) SEM images of blank NF.

(D) SEM images of $MoS_x$@NF at different magnifications.

(E) SEM-EDS elemental mappings of $MoS_x$@NF, showing the distribution of Mo and S elements.

## Spectral Characterizations and Analyses

The composition of $MoS_x$@NF produced *via* the optimized synthesis process was thoroughly characterized using Raman spectroscopy, X-ray diffraction (XRD) and X-ray photoelectron spectroscopy (XPS). Raman spectra, shown in **Figure 2a**, indicated that the composition of $MoS_x$@NF varied depending on the annealing temperature. Under mild annealing conditions (1h@200 °C, Ar), the $MoS_x$ layer was in the $MoS_3$ phase. It is notable that $MoS_3$ is a structurally complex molybdenum sulfide, which cannot be fully represented by the simple chemical formula "$MoS_3$". This material consists of sulfide clusters based on metal-metal bonded Mo(IV) triangles[34], and features three distinct types of sulfur atoms[35]. In some reports, it is referred to as $MoS_{3\,2/3}$[36] or $[Mo_3S_{13}]^{2-}$[24]. However, for the sake of clarity and simplicity, the term $MoS_3$ is used throughout this work. Harsher annealing conditions (for example 2h@350 °C, Ar) favoured the formation of the $MoS_2$ phase. When the annealing time was shortened to 1 h at 350 °C in Ar, a mixture of $MoS_3$ and $MoS_2$ phases was observed (**Figure 2a**). These findings were further supported by operando thermal Raman analysis, where the $MoS_x$ precursor solution was continuously monitored at increasing temperatures, from the initial dried stage up to 500 °C in $N_2$. The phase transition from $MoS_3$ to $MoS_2$ with increasing temperature was clearly observed (**Figure S8**).

The elemental composition of the $MoS_x$@NF electrodes was further examined using inductively coupled plasma optical emission spectroscopy (ICP-OES) and SEM-energy dispersive X-ray spectroscopy (EDS). The analyses revealed a decrease in the S/Mo ratio for samples prepared at higher annealing temperatures. This observation is consistent with the operando thermal Raman measurements and supports the gradual transformation of $MoS_3$ to $MoS_2$ at elevated temperatures. Detailed data can be found in **Table S1**. Based on XRD, both the $MoS_3$ and $MoS_2$ phases were amorphous, with no discernible crystalline diffraction peaks (**Figure S9**). This amorphous nature is typical for $MoS_x$ prepared using solution-based synthesis methods. XPS analysis was carried out to further investigate the chemical composition and oxidation state of the $MoS_x$ layer on the NF surface. As shown in **Figure 2b**, the S 2p region of $MoS_3$@NF exhibited two characteristic doublets corresponding to terminal $S_2^{2-}$ (S2$p_{3/2}$ and S 2$p_{1/2}$ components at 161.9 eV and 163.1 eV, respectively) and bridge $S_2^{2-}$ (and/or apical $S^{2-}$) ligands (components at 163.2 eV and 164.4 eV). These features are consistent with the known structure of $MoS_3$[33,37]. In contrast, the S 2p region of $MoS_2$@NF displayed a single doublet that can be assigned to the $S^{2-}$ of $MoS_2$ species[38–40]. For both $MoS_3$@NF and $MoS_2$@NF, the presence of $SO_4^{2-}$ and Mo-O species (**Figure S10**) was also detected, indicating partial oxidation of the $MoS_x$ layers. We further acquired information on the chemical composition of the electrode "bulk" using XPS depth profiling. To this aim, we combined a series of ion gun etching cycles to etch the surface layer of the $MoS_3$@NF electrode at various times (from 0 s to 1050 s), interleaved with XPS measurements. As demonstrated in **Figure S11**, after $Ar^+$ etching, the signals from $SO_4^{2-}$ species (ascribable to surface oxidation) disappeared and the intensity of the signals from $MoO_3$ species decreased significantly, while signals from $MoS_x$ species increased. These results demonstrate that oxidation products are mainly present at the $MoS_x$ surface, most likely due to air exposure.

Further insights into the sample's local structure were obtained by Mo K-edge X-ray absorption spectroscopy (XAS). The XAS spectra of commercial $MoO_3$ ($MoO_3$-comm) and commercial $MoS_2$ ($MoS_2$-comm) were analyzed as benchmark. As shown in **Figure 2c**, $MoO_3$-comm XANES spectra presented the characteristic 1s⟶4d transition (feature A) in the pre-edge region, which was absent in $MoS_3$@NF[41,42]. The spectra of $MoS_3$@NF and $MoS_2$-comm were clearly distinct from each other, yet consistent with previously reported reference data[42–46]. The two phases can be well distinguished by the presence of feature B in $MoS_2$-comm, which was absent in $MoS_3$@NF. In the Fourier-transformed extended X-ray absorption fine structure (FT-EXAFS) region (**Figure 2d**), $MoS_3$ displayed a Mo-S first coordination shell that is markedly different from the Mo-O coordination observed in $MoO_3$. Additionally, the $MoS_3$ spectrum presented a weak FT-EXAFS Mo-Mo second coordination shell, in contrast to the well-defined Mo–Mo contribution observed in $MoS_2$-comm. These features indicate a significant structural disorder in the $MoS_3$ phase. Notably, the $MoS_3$@NF electrode after 240 h stability test at -200 mA/cm$^2$ (hereafter denoted as $MoS_3$@NF-240h) retained nearly the same structural characteristics of $MoS_3$@NF before the stability test (**Figure S12c, Figure S13**), with XANES and FT-EXAFS profiles closely resembling those of the pristine sample, showing only a minor increase (<5%) in the $MoS_2$ fraction, as confirmed by linear combination fitting (LCF) analysis (**Figure S12b**).

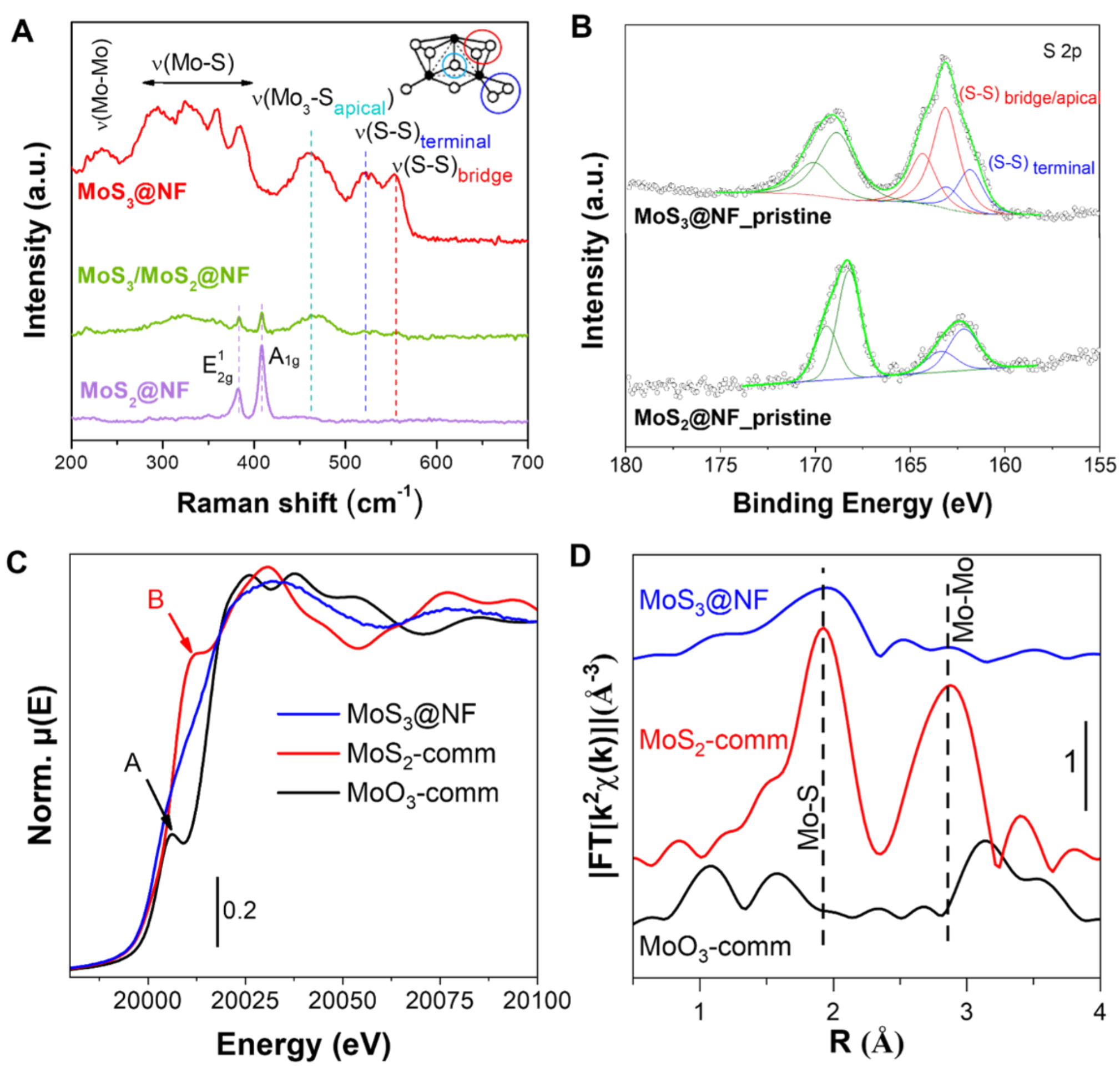

**Figure 2. Spectra characterizations**

(A) Raman spectra of $MoS_x$@NF electrodes synthesized under different annealing conditions. The inset illustrates the core structure of $MoS_3$, highlighting the presence of three distinct types of S atoms.

(B) XPS spectra of the S2p region for the synthesized $MoS_x$@NF electrodes, showing the characteristic binding energies associated with different sulfur species.

(C) Ex-situ Mo K-edge XANES of $MoS_3$@NF.

(D) (phase uncorrected) FT-EXAFS spectra of $MoS_3$@NF. Fourier Transform (FT) was performed in the 3-13 $Å^{-1}$ k-range. Experimental EXAFS and FT-EXAFS spectra imaginary components are reported in **Figure S13**.

## HER Performance Evaluation of $MoS_x$@NF Electrodes

The electrocatalytic performance of the $MoS_x$@NF electrodes for the alkaline HER was first analysed using a three-electrode cell configuration. To ensure fair HER performance comparison, a single homogeneous $(NH_4)_2MoS_4$@NF starting sample was cut into pieces. Such pieces were then annealed at different temperatures, yielding $MoS_x$@NF electrodes. As demonstrated in **Figure 3a**, $MoS_3$@NF required an overpotential of 132 mV (246 mV) to reach -10 mA/cm$^2$ (-200 mA/cm$^2$). In comparison, $MoS_2$@NF exhibited significantly higher overpotentials of 207 mV at -10 mA/cm$^2$ and 380 mV at -200 mA/cm$^2$. The intermediate $MoS_3$/$MoS_2$@NF electrode, which contains a mixture of $MoS_3$ and $MoS_2$ phases, consistently displayed a performance that was intermediate between those of $MoS_3$@NF and $MoS_2$@NF. This finding attests the superior HER activity of the $MoS_3$ phase compared to the $MoS_2$ phase. Although the $MoS_3$@NF electrode exhibits larger electrochemically active surface area (ECSA, **Figure 3b**), its enhanced HER performance can be primarily attributed to a higher intrinsic catalytic activity compared to $MoS_2$. This conclusion is supported by the current density normalized by ECSA (**Figure 3c, Figure S14**), which shows that $MoS_3$@NF maintains superior HER performance even after normalization. Further evidence of the enhanced HER activity of $MoS_3$ is provided by the Tafel slope analysis (inset figure of **Figure 3c**). The Tafel slope of $MoS_3$@NF was 74 mV/dec, significantly lower than that of $MoS_2$@NF (108 mV/dec). This indicates that the HER kinetics on $MoS_3$@NF is significantly faster than on $MoS_2$@NF. Additionally, EIS analysis revealed that the charge transfer resistance of $MoS_3$@NF was significantly lower than that of $MoS_2$@NF (**Figure 3d**), indicating faster electron transfer at the electrode-electrolyte interface.

The stability of the electrode during electrocatalysis is a critical performance metric for practical applications. As shown in **Figure 3e**, the $MoS_3$@NF electrode displayed a stable HER operation for over 240 h at a high current density of -200 mA/cm$^2$. In contrast, Pt/C-CPR exhibited a rapid decline in performance within the first few hours of operation, ultimately indicating a significantly lower stability. The rapid decay in HER activity of Pt/C-CPR can be attributed to the detachment of the powdered Pt/C catalyst from the carbon paper substrate during prolonged operation, a phenomenon widely reported in previous studies[8,47]. In contrast, SEM images of the $MoS_3$@NF electrode after the 240 h stability test revealed that, although some cracks had formed on the surface, the $MoS_3$ catalyst layer remained firmly attached to the NF substrate, further underscoring the effectiveness of the synthesis method developed in this study (**Figure 3f, Figure S15**). We performed a quantitative analysis of the elemental composition of the $MoS_3$@NF electrode before and after the stability test, using XPS (**Figure S16**), ICP and EDS. As summarized in **Table S2**, both S and Mo were partially lost from the $MoS_3$ catalyst after the stability test, but the loss of Mo was greater than that of S. The leaching of the catalyst may be ascribed to the oxidation of Mo and S, in agreement with previous studies indicating that OH- ions can attack Mo sites and form Mo-O species[48]. Such oxidation process led to the gradual dissolution of the $MoS_3$ catalyst and modified its tremella-like morphology at the edge (TEM, **Figure S17**). Notably, the $MoS_3$ catalyst after the stability test retained its amorphous nature, as seen by selected Area Electron Diffraction (SAED) in the TEM (**Figure S17**) and by XRD analysis (**Figure S18**). On the other hand, based on the Raman spectra, the $MoS_3$ structure was still present, although with attenuated ν(Mo–S), ν(Mo–Mo), $\nu(Mo_3{-}S_{apical})$ and $\nu(S{-}S)_{terminal}$ peak intensities and with the almost complete disappearance of the $\nu(S{-}S)_{bridge}$ peak (**Figure S19**). Overall, our $MoS_3$@NF electrode produced via a simple coating-annealing strategy demonstrated excellent catalytic activity and stability for the alkaline HER. Most importantly, its electrochemical performance outperforms many of the previously reported $MoS_x$-based electrocatalysts that were prepared using more complex or labor-intensive synthesis methods (**Figure 3g, Table S3**). Interestingly, such $MoS_3$@NF electrode demonstrated better performance than those of the $MoS_3$ catalysts deposited onto substrate post synthesis rather than in-situ grown.[33,49] We attribute this to the thick catalyst layer and intimate catalyst-substrate contact afforded by the conformal coating (**Figure S2**) that we achieve in this work.

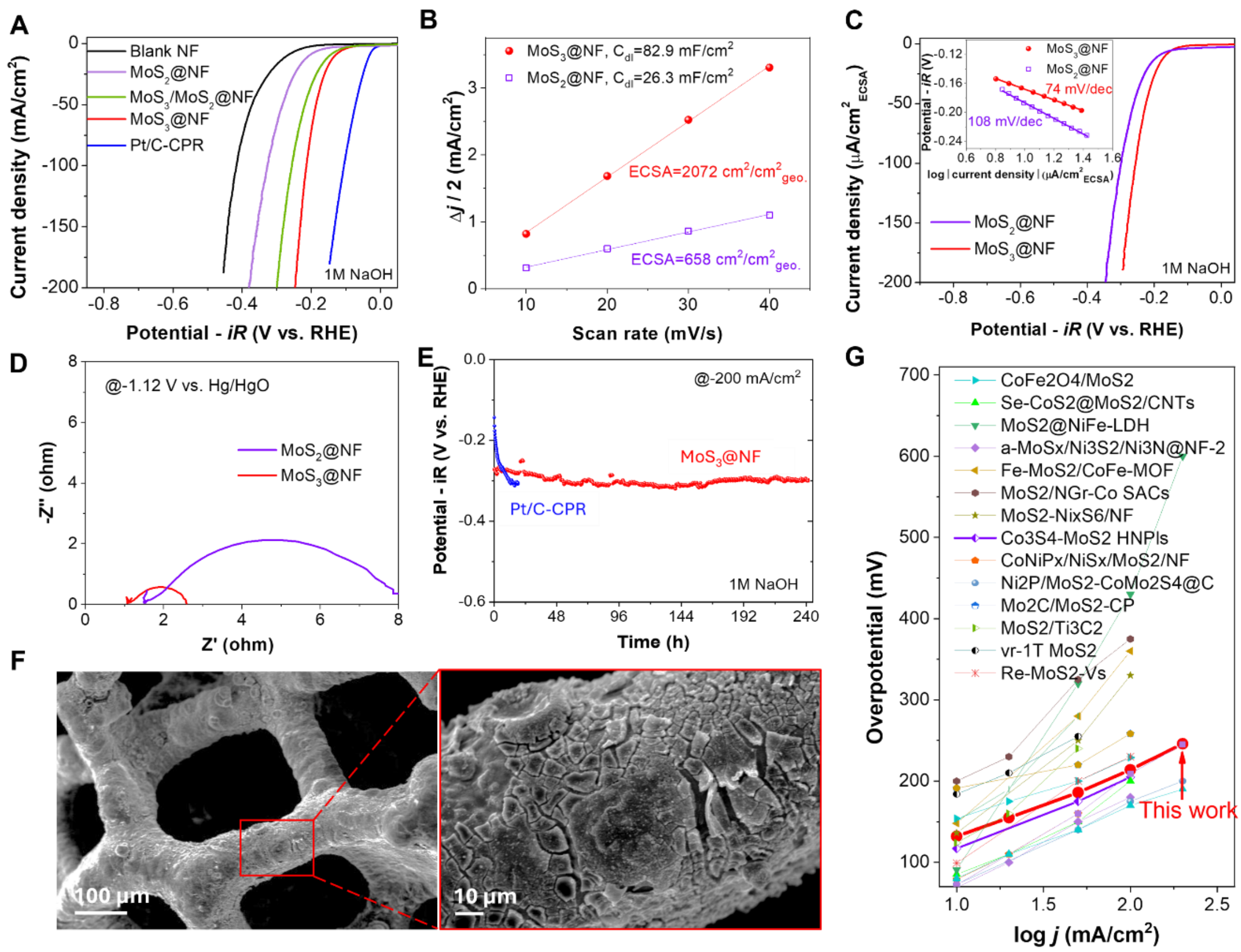

**Figure 3. Electrochemical characterization of the electrodes in three-electrode cell configuration**

(A) iR-corrected linear sweep voltammetry (LSV) curves of various electrodes, including blank NF, Pt/C-CPR benchmark (100 $\mu g_{Pt}/cm^2$), and different $MoS_x$@NF electrodes.

(B) Δj/2 *vs.* scan rate plot extracted from corresponding CV curves at -0.81 V *vs.* Hg/HgO, used for ECSA calculation.

(C) ECSA-normalized iR-corrected LSV curves and Tafel slopes of the electrodes.

(D) EIS plots of the investigated electrodes.

(E) Chronopotentiometric stability test for $MoS_3$@NF and Pt/C-CPR benchmark, showing long-term HER stability at -200 $mA/cm^2$.

(F) SEM images of $MoS_3$@NF electrode after 240 h CP test at −200 $mA/cm^2$.

(G) Comparison of the HER activity of the $MoS_3$@NF electrode with other metal sulfide-based electrocatalysts reported in the literature of the past two years for operation in 1 M KOH/NaOH, as summarized in **Table S3**.

The higher intrinsic HER activity of $MoS_3$ compared to $MoS_2$ can be attributed to the abundance of edge sulfur atoms in the $MoS_3$ structure. This phenomenon is supported by both theoretical calculations[48] and experimental evidence[33]. Recent studies have further revealed that the apical sulfur atoms play a crucial role in maintaining the integrity of the $MoS_3$ structure, whereas the bridging $S_2^{2-}$ species act as the active sites for HER[35]. This highlights the importance of investigating the dynamic evolution of these sulfur species (including terminal S, apical S, and bridge S) during the HER. To address this, we conducted operando electrochemical Raman measurement on the $MoS_3$@NS (the NF was replaced by thinner Ni sheet, NS) electrode in a custom-designed Raman cell under HER operating conditions. As shown in **Figure S20**, no significant change was observed in the characteristic Raman peaks corresponding to the ν(Mo-S) and ν(S-S) vibration modes when the electrode was kept at open circuit potential (OCP) or operated at -100 mV (*vs.* RHE). These results indicate that the structure of $MoS_3$ remains stable during HER operation.

## HER Mechanism from Atomistic Simulations

We carried out density functional theory (DFT) simulations to gain insights into the HER mechanism. To model the catalyst, we make the following considerations. The SEM results of **Figure 1d** and **Figures S1-S4** show that the Ni substrate is fully covered by a relatively thick (*ca.* 2 µm) layer of $MoS_x$, and the best HER performances are obtained for the $MoS_x$ with fewer cracks. Thus, we assume that the Ni substrate does not directly take part in the HER; rather, its primarily role is to act as a template for $MoS_x$ growth and as an electron-conducting scaffold. XAS analysis (Section 2.2) indicates that $MoS_x$ has no structural order, thus being either amorphous or a mix of many structures. Given this information, we modelled the catalyst as an ensemble of structures that in literature are generally associated to $MoS_x$ (**Figure 4a**), namely[34,48,50]: $Mo_3S_{13}$, $Mo_6S_{24}$ (a $Mo_3S_{13}$ dimer adopted as a model for $[Mo_3S_{11}]_\infty$ polymer), $Mo_3S_{10}$, $Mo_3S_9$, and a $MoS_3$ chain. We also considered $MoS_2$ as a reference, modelled as a nanoribbon, since the active sites of $MoS_2$ are known to be its edges[21–23].

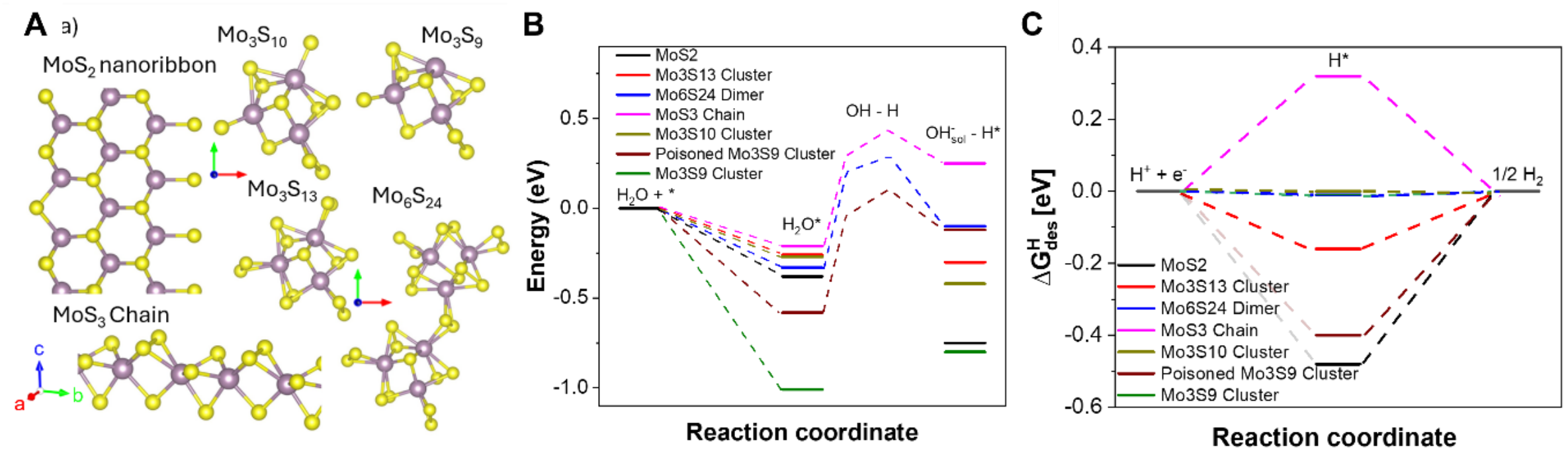

**Figure 4. Reaction energies on $MoS_x$**

(A) All the structures considered in the simulations.

(B) The energy diagram for the Volmer step.

(C) Free energy diagram for the hydrogen desorption.

For all the systems mentioned above, we investigated the energetics of the well-established reaction steps of alkaline HER[51], namely water dissociation (Volmer step), and $H_2$ formation from adsorbed hydrogen (Heyrovsky or Tafel step). In particular, we adopt the following three parameters to discuss the efficiency of electrocatalysts:

1. *Water dissociation*, known to be one of the main rate-determining steps of alkaline HER[52]. We calculated the energy of water dissociation ($\Delta E^{H2O}_{diss}$, see **Note S1.2**) and adopted it as a prompt for the dissociation barrier according to the Brønsted–Evans–Polanyi principle[53], which was shown to apply also to HER[54]. In particular, we calculated the energy of the reactions:

$$H_2O^* + * \rightarrow H^* + OH^* \quad (1),$$

and

$$OH^* + e^- \rightarrow OH^-_{sol} + * \quad (2),$$

where A* indicates a species A adsorbed on the catalyst, * represents an empty adsorption site, and $OH^-_{sol}$ is the $OH^-$ ion in solution (see Methods section and **Notes S1.2 - S1.3** for details, in particular for how we simulated reaction 2). We considered as $\Delta E^{H2O}_{diss}$ the most favourable energy value between two possible pathways: i) water dissociation into adsorbed H and OH (reaction 1), and ii) dissociation into H* and $OH^-_{sol}$ (reaction 1+2), according to ref. [54] (see next point).

2. *OH desorption*. For catalysts that bind OH strongly, ref. [54] showed that water dissociation occurs first through equation 1 (rather than directly forming solvated $OH^-$, as in the Volmer step), then OH desorbs from the surface. In these cases, the OH desorption becomes another important rate determining step. Thus, we consider the OH desorption energy as another important parameter for the efficiency of a catalyst for alkaline HER.

3. *$H_2$ desorption*. While the final $H_2$ formation can take place through two different mechanisms, Heyrovsky or Tafel[51] the hydrogen desorption Gibbs free energy ($\Delta G^H_{des}$), that is the free energy associated to the following reaction:

$$H^* \rightarrow \frac{1}{2} H_2 + * \quad (3),$$

represents the barrier for the final HER step. In fact, at the electrode potential at which the HER occurs, the free energy change associated to Heyrovsky and Tafel reaction steps are the same (see **Note S1.3**). Thus, $\Delta G^H_{des}$ is the third parameter that we adopt to discuss the efficiency of electrocatalysts.

In order to obtain a thorough energetic scenario of the three parameters above, we calculated the corresponding reaction energies on all adsorption sites for all systems considered (see **Note S1.2** and **Figure S21, Figure S22**). The results are summarized in **Table 1** and the energy profiles of the Volmer and Tafel steps are shown in **Figure 4b-c**. The binding sites are in general sulfur atoms, with the notable exception of $Mo_3S_9$ cluster, which exposes a Mo atom. This cluster was included as a model for Mo binding sites, which can form in $MoS_x$ under operando conditions following $S_2$ release, and may, in principle, be present at the certain edges of $MoS_2$ sheets[21,48,55,56]. Expectedly, these metal sites bind strongly H, $H_2O$, and OH, the latter being the most energetically favourable species, as our simulations revealed that swapping OH on Mo with H or $H_2O$ on another site is always energetically unfavourable. As a result, the catalyst regions which expose Mo sites readily dissociate water, but the resulting OH fragment poisons the site. Indeed, even taking into account the applied potential (reaction 2), 0.81 eV are required to desorb OH from the Mo binding site (**Table 1**). Thus, our simulations predict that, in aqueous environments, Mo sites that are not already bound to S atoms are passivated by OH. In fact, the presence of Mo-O bonds was detected in both XPS and Raman spectra[48]. We thus included in our study also a 'poisoned' $Mo_3S_9$ cluster, where the Mo site is passivated by OH (**Table 1**). Once all Mo sites are passivated, the remaining adsorption sites are all sulfur atoms. There, the OH desorption energy is always negative or close to zero, as expected for an electron-withdrawing group adsorbing onto an anionic site. Water will thus dissociate into an adsorbed H and a solvated $OH^-$ ion (see above and ref. [54]). For all $MoS_x$ systems except the $MoS_3$ chain (discussed below), water dissociation is significantly exothermic, which suggests a small kinetic barrier. Thus, water dissociation is not the rate determining step on $MoS_x$, nor is it for $MoS_2$, which can dissociate water even more efficiently than $MoS_x$, as evidenced by its more negative dissociation energy (**Table 1**).

As OH does not bind to the catalyst (see above), the only bottleneck for the HER on $MoS_x$ and $MoS_2$ is the H desorption, which can be significantly endothermic (**Table 1**), as discussed below. The $MoS_3$ chain structure of $MoS_x$ represents an exception to the energetic scenario described above, as there the bottleneck for the HER is the water dissociation rather that the H desorption. However, the amorphous nature of $MoS_x$ suggests that the $MoS_3$ chain does not represent a dominant portion of the $MoS_x$ sample, and that the most likely energetic scenario is that of the remaining $MoS_x$ structures. Interestingly, the H desorption represents the rate determining step for both $MoS_x$ and $MoS_2$, and in both alkaline and acidic[21] conditions. However, there is a key difference between these two systems. In $MoS_2$, H can bind strongly on the two different sites of one edge, while all the remaining sites cannot bind H ($\Delta G^H_{des}$ significantly exothermic, **Table 1**). This suggests that, upon water dissociation, H atoms can only migrate to sites with strong binding energy, thus creating a significant energy barrier for the HER completion. In $MoS_x$, instead, there are several sites that can bind H atoms, with a whole range of $\Delta G^H_{des}$. Some of them have $\Delta G^H_{des}$ close to zero, which makes the $H_2$ desorption fast. Therefore, as more water molecules dissociate, the binding sites with the lowest H adsorption become progressively occupied, leaving only those with $\Delta G^H_{des}$ close to zero available. Once this point is reached, HER can proceed at a fast rate as the $H_2$ desorption can occur without a significant energy barrier. We note that our simulations model low coverages scenario and, at high surface coverage, adsorption energies can change. In fact, Hinnemann *et al.*[21] showed that, as the H coverage on $MoS_2$ increases, $\Delta G^H_{des}$ decreases and becomes close to zero. However, we argue that the absolute value of calculated $\Delta G^H_{des}$ is significantly influenced by the adopted approximations (DFT functional, solvation and entropic corrections, etc.).

In summary, our systematic study of adsorption energies on various $MoS_x$ structures revealed that, for these systems, the bottleneck for the alkaline HER is not the Volmer step but rather the hydrogen desorption (Tafel/Heyrovsky), similarly to the acidic HER. Moreover, it emerged that $MoS_x$, unlike $MoS_2$, has a multitude of sites that can bind H atoms with a wide range of adsorption energies. This, together with the well-established greater number of active sites[34,48,50], confers to $MoS_x$ its superior electrocatalytic properties.

**Table 1.** Energies of water dissociation energy (see main text) and adsorption energies (the three lowest-energy values are reported) for various $MoS_x$ structures.

| System | $\Delta E^{H2O}_{diss}$ [a] | $\Delta E^{OH}_{des}$ (highest values) [b] | $\Delta G^{H}_{des}$ (highest values) [c] |
|---|---|---|---|
| $Mo_3S_{13}$ (cluster) | -0.30 | -0.27, -0.32, -0.37 | +0.26, +0.16, -0.87 |
| $Mo_3S_{10}$ | -0.42 | -0.57, -0.92, -1.41 | +0.40, 0.00, -0.19 |
| $Mo_3S_9$ | -0.80 | +0.81, -0.98, -1.03 | +0.71, +0.01, -0.22 |
| 'poisoned' $Mo_3S_9$ | -0.12 | -0.45, -0.58, -0.81 | +0.40. -0.05, -0.63 |
| $Mo_6S_{24}$ ($Mo_3S_{13}$ dimer) | -0.10 | -0.38, -0.44, -0.93 | +0.14, +0.12, +0.01 |
| $[MoS_3]_\infty$ | +0.25 | -0.27, -0.93, -0.97 | -0.32, -0.32, -0.42 |
| $[Mo_3S_8]_\infty$ ($MoS_2$ nanoribbon) | -0.75 | +0.04, +0.02, -0.18 | +0.65, +0.48, -0.58 |

[a] electronic energy change associated to water dissociation reaction, as defined by reactions 1 and 2 (see corresponding discussion in the main text).
[b] electronic energy change associated to OH desorption reaction, as defined by above-mentioned reaction 2 and reaction 1 (Methods section). The values corresponding to the three sites most energetically favourable for adsorption are reported.
[c] Gibbs free energy associated to the hydrogen desorption reaction, as defined by reaction 3 (see eq. 2 in Methods section for the vibrational and entropic contributions). The values corresponding to the three sites most energetically favorable for adsorption are reported.

## Performance Evaluation of AWE Cell based on $MoS_3$@NF Cathode

The optimized $MoS_3$@NF electrode was thereafter integrated as the cathode into an AWE single cell (Zirfon diaphragm as separator, operating at atmospheric-pressure), coupled with a SSM anode developed in our previous works (**Figure 5a**)[8,17,57]. For comparison, a commercial NF substrate was also tested as the cathode. Hereafter, the AWE configurations are denoted as cathode||anode. As demonstrated in **Figure 5b**, the $MoS_3$@NF||SSMs AWE reached current densities of 0.4 A/cm$^2$ and 1.0 A/cm$^2$ at cell voltages of 1.75 V and 1.90 V, respectively. These values correspond to energy efficiencies of 83.8 % and 77.2 % (based on the higher heating value of $H_2$) and voltage efficiencies of 67.6 % and 62.3 %, respectively (**Figure 5c**). In contrast, NF||SSMs AWE displayed poorer performance, requiring cell voltages of 1.97 V at 0.4 A/cm$^2$ and 2.27 V at 1 A/cm$^2$. The performance of the $MoS_3$@NF||SSMs AWE also surpassed that of other advanced electrolyzers, such as the previously reported Co-doped 1T-$MoS_2$||NiFe-LDH anion-exchange-membrane electrolyzer (0.1 A/cm$^2$@~2.0 V)[58], as well as other electrolyzers based on $MoS_x$ or non-noble metal-based cathodes (**Table S4**).

To evaluate the stability of our $MoS_3$@NF||SSMs AWE under intermittent operation, an accelerated stress test (AST) was applied, following protocols reported in previous studies[10,59]. As displayed in **Figure 5d**, the AWE single cell was subjected to consecutive alternating galvanostatic steps at 1 A/cm$^2$ and 0.05 A/cm$^2$, each lasting for 15 minutes. This protocol simulates the alternation between high- and low-load operation typically observed in systems powered by intermittent renewable energy sources. During the initial stages of AST, a gradual performance decay was observed, with a cell voltage increase ($\Delta V$) of 70 mV. The areal high-frequency resistance (HFR) of the cell also decreased by approximately 15 m$\Omega\cdot$cm$^2$, which would account for an additional 15 mV decrease in the cell voltage when operated at 1 A/cm$^2$.

Thus, the total kinetic performance decay, after excluding the contribution due to ohmic resistance decrease, was calculated to be 85 mV. The observed kinetic performance decay was linked to the partial detachment of $MoS_3$ layer from the NF substrate, as evidenced by SEM images of the electrodes after AST (**Figure S23a**). Noteworthy, the AWE disassembling operation may also break the $MoS_3$ layer. The decrease in areal HFR during the initial stage is likely due to the modifications of interfacial contact resistances, as well as to the modifications of $MoS_3$ layer. Interestingly, after the initial decay, a slight performance improvement was observed during the AST, which could be ascribed to the activation of SSM anode induced by AST cycling, as previously observed in similar systems[57]. Following the AST stabilization, a continuous stability test of 100 h was conducted at a fixed current density of 1 A/cm$^2$ (**Figure S23c**). We also carried out a continuous stability test for 1000 h without AST stabilization, during which the cell voltage also remained stable at approximately 1.96 V (**Figure 5e**). Both the 100 h+AST operation and continuous 1000 h test indicated excellent long-term stability of the $MoS_3$@NF||SSM AWE. Further SEM analysis of the $MoS_3$@NF electrode surface before and after the stability test revealed no significant morphological changes in the electrode surface (**Figure S23b** *vs.* **Figure S23a**), confirming that the $MoS_3$ coating remained structurally stable. This was further validated by Raman spectroscopy (**Figure S24**), which showed that, although the Raman peak intensity was attenuated after the test, the characteristic v(Mo–S) peaks of $MoS_3$ were still present.

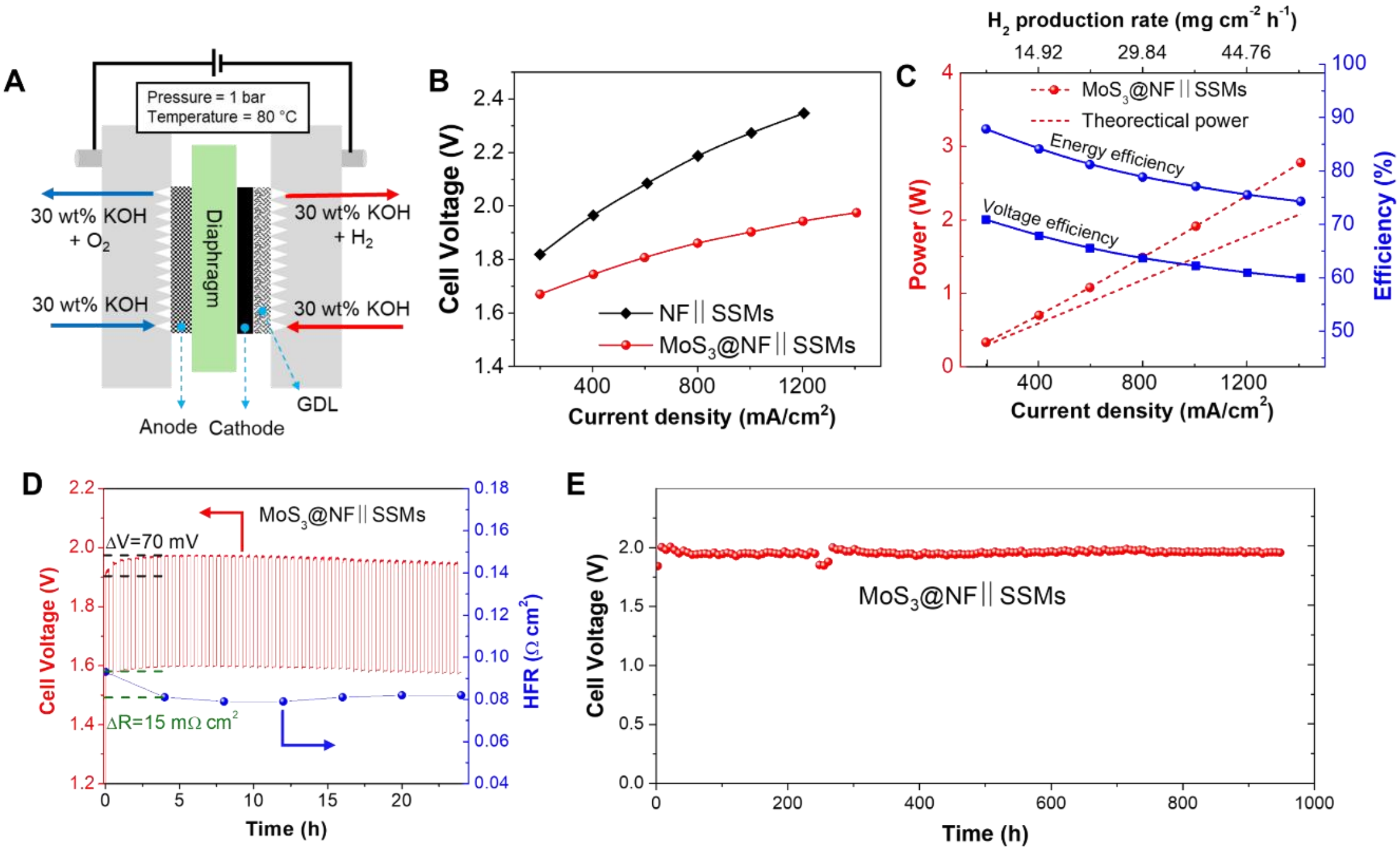

**Figure 5. Electrochemical characterization of AWE single cells based on the investigated electrodes**

(A) Schematic illustration of the AWE configuration used in this work.

(B) Polarization curves measured for the $MoS_3$@NF||SSMs and NF||SSMs AWE cells.

(C) Power consumption, $H_2$ production rate, energy efficiency (based on the hydrogen HHV) and voltage efficiency as a function of current density for the $MoS_3$@NF||SSMs cell. The theoretical power

consumption for water splitting is calculated based on the thermoneutral voltage at 80°C and 1 bar, as referenced in the literature[60].
(D) 24 h-AST for the $MoS_3$@NF||SSMs AWE. HFR: high-efficiency resistance.
(E) Continuous stability test for the $MoS_3$@NF||SSMs AWE cell at 1 A/cm$^2$ for 1000 h.
Note: All the AWE electrochemical data presented here are not iR-corrected.

## CONCLUSIONS

In this study, we developed a cost-effective and straightforward coating-annealing strategy to conformally deposit a $MoS_x$ catalyst layer on a Ni foam (NF) substrate for the hydrogen evolution reaction (HER) in alkaline media. By simply adjusting the annealing parameters, the composition of $MoS_x$ could be tuned from $MoS_2$ to the more HER-active $MoS_3$ phase, which exhibits a higher density of active edge-sulfur sites. The resulting $MoS_3$@NF electrode demonstrated outstanding HER performance in a three-electrode cell configuration, with low overpotentials, fast HER kinetics, and excellent stability. However, when the electrode was implemented as a cathode in an alkaline water electrolyzer (AWE) under industrially relevant conditions, partial detachment of the $MoS_3$ catalyst layer from the NF substrate was observed during the accelerated stress test (AST). This detachment led to a reduction in performance, emphasizing the importance of testing electrocatalysts under practical AWE operating conditions rather than relying solely on performance in a three-electrode setup. Despite the initial performance decay during the AST, the AST-stabilized $MoS_3$@NF electrode, when coupled with a 5-stacked stainless steel mesh anode, exhibited impressive long-term stability. The AWE system maintained a stable operating voltage of ~1.96 V at 1 A/cm$^2$ for 1000 h, outperforming most previously reported water electrolyzers utilizing $MoS_x$-based cathodes. This result highlights the potential of $MoS_3$-based cathodes for industrial AWE applications. Prospectively, further research should focus on enhancing adhesive stability to improve the mechanical robustness and long-term performance of the catalyst under dynamic AWE conditions. By addressing these two challenges, $MoS_3$@NF-based electrodes could serve as a viable alternative to noble-metal-based cathodes, thereby advancing the development of scalable, sustainable hydrogen production technologies.

## METHODS

### Materials

Ammonium tetrathiomolybdate [$(NH_4)_2MoS_4$, 99.97 %], Ni foam (NF, 1.6 mm thickness), sodium hydroxide (NaOH, 98 %), isopropanol (IPA, ⩾ 99.8 %), acetone (ACE, ⩾99.5 %), 20 wt% platinum on graphitized carbon (Pt/C) were purchased from Sigma Aldrich. Cu foam was purchased from Boegger Industech LTD. (China). Carbon paper (CPR, AvCarb MGL280) and Ti felt were purchased from FuelCell Store. Zirfon Perl UTP 220 used as diaphragm was purchased from Agfa. The as-received substrates were cleaned using HCl solution (1 M), IPA/ACE (1:1, vol/vol) and distilled water, and dried using a $N_2$-gun stream before use. All the chemicals were used as received.

### Synthesis of $MoS_x$@Substrate electrode

A 1.5cm×3cm pre-cleaned substrate (*e.g.,* NF, Cu foam or Ti felt) was immersed in a $(NH_4)_2MoS_4$ solution for 1 min. After drying, the substrate was cut into a required size and placed in a tubular furnace for annealing in Ar atmosphere at different temperatures and time durations. The optimal substrate type was found to be NF, while the optimal concentration of $(NH_4)_2MoS_4$ solution was determined to be 50 mM. The resulting electrode upon annealing in Ar at α °C for β h was denoted as $MoS_x$@NF_αC-βh.

The optimized electrode prepared using the abovementioned procedure, namely $MoS_x$@NF_200C-1h, is simply denoted by its actual composition: $MoS_3$@NF. Following this naming convention, the $MoS_x$@NF_350C-1h and $MoS_x$@NF_350C-2h are labelled as $MoS_3/MoS_2$@NF and $MoS_2$@NF, respectively.

### Preparation of Pt/C-CPR electrode

1 mg of commercial 20 wt% Pt/C (platinum on graphitized carbon, Sigma Aldrich) was added to a glass vial containing a mixture of water and IPA (0.1 mL/0.09 mL). Subsequently, 10 µl of Nafion (5 wt% Nafion 117 solution, Sigma Aldrich) was added. The mixture was sonicated for at least 30 min to form a homogeneous ink. A volume of 0.1 mL of this ink was then drop-cast onto the CPR, obtaining a Pt mass loading of 100 $\mu g_{Pt}/cm^2$. The resulting electrode, labelled Pt/C-CPR, was dried in air before the measurements.

### Characterization of electrode materials

X-ray diffraction (XRD) measurements were conducted on a PANalytical Empyrean diffractometer with Cu Kα radiation. X-ray photoelectron spectroscopy (XPS) measurements were carried out on a Kratos Axis UltraDLD spectrometer equipped with a monochromatic Al Kα source operating at 20 mA and 15 kV. High-resolution spectra were recorded at a pass energy of 10 eV, and all spectra were calibrated using the C 1s mainline at 284.8 eV. To further acquire information on the chemical composition of the electrode "bulk" (via XPS depth profiling), a series of ion gun etching cycles were combined with interleaved XPS measurements. The etching cycles used $Ar^+$ ions at 4 keV energy. Scanning electron microscopy (SEM) images were acquired on an FEI NanoLab 600 dual beam system with an acceleration voltage of 5-10 kV, while energy dispersive spectroscopy (EDS) analysis was conducted at 20 kV. Inductively coupled plasma optical emission spectroscopy (ICP-OES) analysis was performed using an iCAP 6500 Thermo spectrometer. Samples were prepared by dissolving either the entire catalyst electrode (of known size) or the collected catalyst powder from the substrate surface in 2.5 mL aqua regia ($HCl/HNO_3$ 3:1, v/v) and allowing it to digest overnight. The resulting solution was then diluted to 25 ml with Milli-Q water, and approximately 10 mL solution was collected after filtration using a 0.45 µm Nylon filter. The ICP measurements had a systematic error of approximately 5 %. Raman spectra were acquired using an inVia Renishaw Raman microscope with an Ar laser (633 nm) and a nominal power of 17 mW. The spectrometer was equipped with a back illuminated CCD detector and with a 1200 line/mm diffraction grating, providing a spectral resolution better than 1 $cm^{-1}$. All measurements were conducted in backscattering geometry at room temperature. The samples were focused by a 50× long working distance objective, resulting in a spot size of approximately 1 µm in diameter. For operando Raman characterizations, a specific cell purchased from Redox.me Corp was used to facilitate the cell mounting and testing. To ensure compatibility, the porous NF substrate was replaced by a 0.1 mm-thick Ni sheet (NS), resulting in the sample denoted as $MoS_3$@NS.

### Electrochemical characterization of electrodes

Cyclic voltammetry (CV) curves were recorded to evaluate the electrode activity towards the HER. Measurements were performed at a scan rate of 5 mV/s in a 1 M NaOH solution. The presence or absence of iR correction is indicated on the title of the axis; for instance, "Potential - iR" denotes that iR correction has been applied. To determine the series resistance ($R_s$) of the electrodes, electrochemical impedance spectroscopy (EIS) was conducted at -1.12 V *vs.* Hg/HgO over a frequency range of 0.1 Hz to 100 kHz. The durability of the working electrodes was evaluated through chronopotentiometric (CP) measurements at fixed working currents. All CP plots are presented with iR correction applied. When required, recorded potentials were converted to the reversible hydrogen electrode (RHE) scale using the Nernst equation, *i.e.*:

$$E_{RHE} = E_{obs} + E^{o}_{Hg/HgO} + 0.0591 \times pH = E_{obs} + 0.89\ V$$

The value of ($E^{o}_{Hg/HgO}$ + 0.0591 × pH) was determined to be 0.89 V based on the calibration protocol described in the report by Xu *et al.*[61]

### Quantification of the electrochemical active surface area (ECSA)

The ECSA of $MoS_3$@NF and $MoS_2$@NF was determined by performing CV measurements in a non-Faradaic potential window (-0.78 V to -0.84 V *vs.* Hg/HgO) at various potential scan rates: 10, 20, 30, and 40 mV/s. The double-layer capacitance ($C_{dl}$) of the electrodes was calculated as the slope of the linear fit of the Δj/2 *vs.* scan rate (v) plot, where Δj is the difference in current density between the cathodic and anodic scans at a potential of -0.81 V *vs.* Hg/HgO. The ECSA was then calculated using the following equation:

$$ECSA = C_{dl}/C_s,$$

where $C_s$ represents the specific electrochemical double-layer capacitance of a material with a smooth atomic surface. According to McCrory *et al.*, the $C_s$ values for different catalysts varied between 0.022 and 0.130 mF/$cm^2$ in alkaline solutions[62], indicating that $C_s$ varies significantly depending on the chemical properties of the material under investigation. Despite this variability, and consistent with common practice in the literature[63–65], we assumed $C_s$ to be equal to 0.04 mF/$cm^2$ for our calculations. This assumption provides a standard reference for comparing the ECSA of $MoS_3$@NF and $MoS_2$@NF with other catalysts reported in the field.

## X-ray absorption spectroscopy measurements

Mo K-edge X-ray Absorption Spectra (XAS) were measured at the BM23 beamline of the European Synchrotron Radiation Facility (ESRF)[66] during beamtime IH-CH-1901. A monochromatic beam was obtained with a Si(111) Double Crystal Monochromator (DCM) while a Rh-coated double reflection mirror set at 2.6 mrad was used for harmonic rejection. Three ion chambers were used to measure the incoming beam ($I_0$), beam transmitted by the sample ($I_1$) and beam transmitted by reference metal foil ($I_2$). $I_0$ was filled with 0.6 mbar of Ar (top up to 2.2 bar in He) while $I_1$ and $I_2$ were filled with 2.2 bar of Ar. Mo metal foil was placed between $I_1$ and $I_2$ and its spectrum was measured together with each XAS scan for further energy calibration and alignment. Commercial $MoS_2$ and $MoO_3$ were measured on mass optimized pellets of Ø ≈ 13 mm in transmission mode. Samples spectra were instead measured in fluorescence mode with a Silicon Drift Vortex detector on the sample as prepared at RT in air. All scans were performed in the 19.6-21.2 keV energy range with 0.3 eV/point energy resolution. Transmission measurements were performed with 20 ms/point integration time and three successive scans were averaged to obtain the reported final spectrum. Fluorescence measurements were instead recorded with a 50 ms/point integration time and 20 successive measurements were averaged. Due to the lower Mo content, 240 successive scans were averaged for the $MoS_3$-240h sample. Spectra were energy calibrated and aligned, background subtracted and edge jump normalized with the Athena software from the Demeter package[67]. Linear Combination Fit (LCF) analysis was performed using Athena using $MoS_2$-commercial and $MoS_3$ spectra as pure components. The weight of the single component ($w_i$) was constrained to be $0 < w_i < 1$ while the sum of the component weights was forced to 1.

## Atomistic simulations

All simulations were run with the Vienna ab initio software package (VASP)[68–70], see **Note S1.1** for details. The results discussed in the main text were obtained adopting the Perdew–Burke–Ernzerhof (PBE)[71] exchange-correlation functional augmented with the D2 dispersion correction[72]. Hydrogen desorption energies on selected $MoS_x$ structures were further evaluated using the r²SCAN+rVV10 functional[73,74], which offers improved treatment of dispersion and intermediate-range correlation effects. The comparison is reported in **Figure S22**. Although the absolute $\Delta G^{H}_{des}$ values can differ between the two approaches, the overall trends and site activity rankings are preserved. Notably, the r²SCAN+rVV10 results confirm the main findings on $\Delta G^{H}_{des}$ being close to thermoneutrality for $Mo_3S_{13}$ (and its dimer) and relatively far from it in $MoS_2$.

The energy change associated to reaction 2 was calculated exploiting the thermodynamical considerations reported in ref. [17] and recalled in **Note S1.3**, which demonstrate that, at the electrode potential adopted for HER, the free energy of that reaction is equivalent to that of reaction 1:

$OH^* + ½ H_2 \rightarrow * + H_2O$ (1).

The Gibbs free energy for hydrogen adsorption ($\Delta G_H^{ads}$) of reaction 3 in the main text was calculated by adding to the adsorption internal energy calculated by DFT (sometimes referred to as 'electronic energy' or 'total energy') a correction term from ref. [21] to account for entropic and vibrational contributions:

$$\Delta G_H^{des} = \Delta E_H^{des} - 0.29 \text{ eV} \qquad (2)$$

Other details on the methodology and approximations to model adsorption and desorption processes can be found in **Note S1**.

## Test in alkaline water electrolyzer (AWE)

The AWE single cells were fabricated using commercially available zero-gap single electrolysis cell hardware (Dioxide Materials), featuring corrosion-resistant nickel-based anode and cathode flow field end plates, O-ring seals, and polytetrafluoroethylene (PTFE) gaskets. A Zirfon Perl UTP 220 diaphragm was utilized as the separator. The anode comprised 5-stacked stainless steel meshes (SSMs)[8,17,57]. The exposed active area of electrodes was 1 $cm^2$. During assembly, the cell components were compressed to achieve zero-gap AWE configurations, with the thickness of the PTFE spacers set to 85 % of the electrode thickness. The AWE single cells were connected to a custom-built test station. Both anodic and cathodic compartments were supplied with a 30 wt% KOH solution via a peristaltic pump (Masterflex L/S Series). The electrolyte flow rate was maintained at 30 mL/(min×$cm^2$). The temperature of the electrolyte was controlled at 80 °C using a proportional-integral-derivative (PID) controller coupled with an Omega SA3 self-adhesive thermocouple, which was attached to one of the cell end plates. The AWE single cells, operating at atmospheric pressure (1 bar), were powered by a potentiostat/galvanostat (VMP3, Biologic) equipped with an external high-current booster channel rated at 20 A. Galvanostatic polarization curves were recorded using a multistep chronopotentiometric protocol, in which the cell voltage was measured and averaged over 5 min for each current step to generate points on the polarization curve. To ensure reproducibility, the AWE cells were preconditioned by performing 20 CV cycles between cell voltages of 1 V and 2 V at a scan rate of 5 mV/s. The stability of the AWE cells was evaluated through an accelerated stress test (AST), which consisted of alternating steps set at current densities of 1000 mA/$cm^2$ and 50 mA/$cm^2$, with each galvanostatic step lasting for 15 min. The total duration of test was 24 h.

The energy efficiency of the AWEs, based on the higher heating value (HHV) of the hydrogen (141.7 kJ/$g_{H2}$), was calculated assuming a Faradaic efficiency for the HER equal to 1, as:

$$\text{Energy efficiency}_{HHV} = (M_{H2} \times HHV)/\text{energy}_{input} = E_{th}^0/E_{cell},$$

where $M_{H2}$ is the hydrogen gas mass produced by the AWE cell, $E_{th}^0$ (V) is the thermoneutral voltage for water electrolysis at standard temperature and pressure (1.48 V), $E_{cell}$ is the cell voltage. $\text{Energy}_{input}$ is the electric energy consumed to produce hydrogen, which is calculated by multiplying the operating power of AWE cell by time. It is important to note that $\text{energy}_{input}$ excludes certain energy contributions from the electrolyzer, such as energy consumption from the peristaltic pumps and auxiliary components. Therefore, the energy efficiency values presented should be considered approximate and intended for laboratory use or for comparison with literature results[75].

### 1000 h test in alkaline water electrolyzer

A 1000-hour stability assessment of the AWE system was conducted using a zero-gap cell fixture (exposed active area: 1 $cm^2$) supplied by Antares Electrolysis S.r.l. The fixture incorporates corrosion-resistant Ni anodic plates with integrated flow fields, ethylene–propylene–diene monomer (EPDM) O-ring seals, and metallic fittings. In addition, the use of springs for bolt tightening ensures that the compression of the cell components remains constant throughout the entire test. The membrane–electrode assembly (MEA) consisted of the $MoS_3$-based cathode, a stainless steel-based anode, and a Zirfon® Perl UTP 220 diaphragm, compressed to achieve the zero-gap configuration. Polytetrafluoroethylene spacers were positioned in each electrode compartment to control the compression of the cathode/diaphragm/anode stack. Spacer thickness was optimized to achieve a gas diffusion layer (GDL) compression of 15 % ± 5 %. An aqueous 30 wt. % KOH solution was employed as the electrolyte and continuously circulated through both compartments at 30 mL/min using a recirculation system. The electrolyte and cell were maintained at 80 °C *via* hotplates coupled to precision thermal controllers. Water losses due to evaporation and electrolysis were compensated by periodic water refilling. Electrochemical data were collected with a Maccor 4200 tester.

## RESOURCE AVALABILITY

### Lead contact

Further information and requests should be directed to and will be fulfilled by the Lead Contact, Liberato Manna (liberato.manna@iit.it).

### Materials availability

This study did not generate new unique reagents.

### Data and code availability

Supplemental data are available in the supplemental information. Requests for additional data or code will be fulfilled by the lead contact upon reasonable request.

## ACKNOWLEDGMENTS

This work has been funded by the European Union's Horizon 2020 “Proof of Concept” program under Grant Agreement No. 899412 (HyCat); the Chongqing Talents program under Grant Agreement CSTB2024YCJH-KYXM0089; the starting grant of Chongqing University under Grant Agreement No. 0220005203006. The authors acknowledge the CINECA award under the ISCRA initiative, for the availability of high-performance computing resources and support.

The authors thank Filippo Drago (Nanochemistry department – IIT) for the support in ICP analysis. The authors acknowledge the ESRF for beamtime provision (IH-CH-1901) at BM23 XAS beamline.

## AUTHOR CONTRIBUTIONS

Y.Z. prepared the electrodes and performed the physical, chemical and electrochemical characterizations, analyses, organized figures and wrote the paper. S.B., V.M. and M.I.Z carried out the tests and analyses of AELs. S.B. co-wrote the paper. M.J. and G.S. carried out the theoretical calculation of DFT, and wrote the related part. D.S. contributed to the XAS measurements. T.-H.-H. L., M.F., M.C., F.B. and Y.H. significantly contributed to the results discussion and manuscript drafting. Z.L. performed the TEM characterization and analyses. M.P. carried out the XPS characterization. S.D. contributed to the Raman measurement and analysis. L.M. and S.B. directed the research and conceived the project. The manuscript was corrected and improved by all authors.

## DECLARATION OF INTERESTS

V.M., M.I.Z, F.B. are researchers of BeDimensional S.p.A., a company commercializing two-dimensional materials. S.B., M.F., LM. are CTO, CSO and scientific advisor of Antares Electrolysis S..rl., a company designing and manufacturing electrolyzer stacks. The other authors declare no competing interests.

## SUPPLEMENTAL INFORMATION

Supplemental information can be found online at http://doi.org/10.XXXXXXXX.

## REFERENCES

1. Guerra, O.J., Eichman, J., Kurtz, J., and Hodge, B.M. (2019). Cost Competitiveness of Electrolytic Hydrogen. Joule *3*, 2425–2443. https://doi.org/10.1016/J.JOULE.2019.07.006.

2. Xiang, L., Li, N., Zhao, L., Wang, K., Pang, B., Liu, Z., and Guo, J. (2024). Boosting alkaline hydrogen evolution via in-plane heterostructure construction with Ultra-Exposed heterointerfaces. Chem. Eng. J. *499*, 155833. https://doi.org/10.1016/J.CEJ.2024.155833.

3. Communication COM/2020/301: A hydrogen strategy for a climate-neutral Europe | Knowledge for policy https://knowledge4policy.ec.europa.eu/publication/communication-com2020301-hydrogen-strategy-climate-neutral-europe_en.

4. Global Hydrogen Review 2022 – Analysis - IEA https://www.iea.org/reports/global-hydrogen-review-2022.

5. Terlouw, T., Rosa, L., Bauer, C., and McKenna, R. (2024). Future hydrogen economies imply environmental trade-offs and a supply-demand mismatch. Nat. Commun. *15*, 7043. https://doi.org/10.1038/s41467-024-51251-7.

6. Velazquez Abad, A., and Dodds, P.E. (2020). Green hydrogen characterisation initiatives: Definitions, standards, guarantees of origin, and challenges. Energy Policy *138*, 111300. https://doi.org/10.1016/j.enpol.2020.111300.

7. Ehlers, J.C., Feidenhans'l, A.A., Therkildsen, K.T., and Larrazábal, G.O. (2023). Affordable Green Hydrogen from Alkaline Water Electrolysis: Key Research Needs from an Industrial Perspective. ACS Energy Lett. *8*, 1502–1509. https://doi.org/10.1021/ACSENERGYLETT.2C02897/ASSET/IMAGES/LARGE/NZ2C02897_0005.JPEG.

8. Zuo, Y., Bellani, S., Ferri, M., Saleh, G., Shinde, D. V., Zappia, M.I., Brescia, R., Prato, M., De Trizio, L., Infante, I., et al. (2023). High-performance alkaline water electrolyzers based on Ru-perturbed Cu nanoplatelets cathode. Nat. Commun. *14*, 4680. https://doi.org/10.1038/s41467-023-40319-5.

9. Bartels, J.R., Pate, M.B., and Olson, N.K. (2010). An economic survey of hydrogen production from conventional and alternative energy sources. Int. J. Hydrogen Energy *35*, 8371–8384. https://doi.org/10.1016/J.IJHYDENE.2010.04.035.

10. Zappia, M.I., Bellani, S., Zuo, Y., Ferri, M., Drago, F., Manna, L., and Bonaccorso, F. (2022). High-current density alkaline electrolyzers: The role of Nafion binder content in the catalyst coatings and techno-economic analysis. Front. Chem. *10*, 1045212. https://doi.org/10.3389/FCHEM.2022.1045212/BIBTEX.

11. Yu, W., Zhang, Y., Qin, Y., Zhang, D., Liu, K., Bagliuk, G.A., Lai, J., and Wang, L. (2023). High-Density Frustrated Lewis Pair for High-Performance Hydrogen Evolution. Adv. Energy Mater. *13*, 2203136. https://doi.org/10.1002/AENM.202203136.

12. Li, H., Han, Y., Zhao, H., Qi, W., Zhang, D., Yu, Y., Cai, W., Li, S., Lai, J., Huang, B., et al. (2020). Fast site-to-site electron transfer of high-entropy alloy nanocatalyst driving redox electrocatalysis. Nat. Commun. *11*, 5437. https://doi.org/10.1038/S41467-020-19277-9;TECHMETA.

13. Gu, Y., Nie, N., Liu, J., Yang, Y., Zhao, L., Lv, Z., Zhang, Q., and Lai, J. (2023). Enriching H2O through boron nitride as a support to promote hydrogen evolution from non-filtered seawater. EcoEnergy *1*, 405–413. https://doi.org/10.1002/ECE2.9.

14. Liu, Y., Liu, S., Wang, Y., Zhang, Q., Gu, L., Zhao, S., Xu, D., Li, Y., Bao, J., and Dai, Z. (2018). Ru Modulation Effects in the Synthesis of Unique Rod-like Ni@Ni2P-Ru Heterostructures and Their Remarkable Electrocatalytic Hydrogen Evolution Performance. J. Am. Chem. Soc. *140*, 2731–2734. https://doi.org/10.1021/JACS.7B12615.

15. Li, W., Zhao, Y., Liu, Y., Sun, M., Waterhouse, G.I.N., Huang, B., Zhang, K., Zhang, T., and Lu, S. (2021). Exploiting Ru-Induced Lattice Strain in CoRu Nanoalloys for Robust Bifunctional Hydrogen Production. Angew. Chem. Int. Ed. *60*, 3290–3298. https://doi.org/10.1002/ANIE.202013985.

16. Chen, C.H., Wu, D., Li, Z., Zhang, R., Kuai, C.G., Zhao, X.R., Dong, C.K., Qiao, S.Z., Liu, H., and Du, X.W. (2019). Ruthenium-Based Single-Atom Alloy with High Electrocatalytic Activity for Hydrogen Evolution. Adv. Energy Mater. *9*, 1803913. https://doi.org/10.1002/AENM.201803913.

17. Zuo, Y., Bellani, S., Saleh, G., Ferri, M., Shinde, D. V., Zappia, M.I., Buha, J., Brescia, R., Prato, M., Pascazio, R., et al. (2023). Ru-Cu Nanoheterostructures for Efficient Hydrogen Evolution Reaction in Alkaline Water Electrolyzers. J. Am. Chem. Soc. *145*, 21419–21431. https://doi.org/10.1021/JACS.3C06726.

18. Wu, D., Du, H., Liu, Z., Bagliu, G.A., Lai, J., and Wang, L. (2024). Construction of high-loading WO3-x sub-nanometer clusters via orderly-anchored top–down strategy boost acidic hydrogen evolution. EcoEnergy *2*, 724–735. https://doi.org/10.1002/ECE2.63.

19. Niu, S., Cai, J., and Wang, G. (2021). Two-dimensional MOS2 for hydrogen evolution reaction catalysis: The electronic structure regulation. Nano Res. *14*, 1985–2002. https://doi.org/10.1007/S12274-020-3249-Z/METRICS.

20. Cao, Y. (2021). Roadmap and Direction toward High-Performance MoS2Hydrogen Evolution Catalysts. ACS Nano *15*, 11014–11039. https://doi.org/10.1021/ACSNANO.1C01879.

21. Hinnemann, B., Moses, P.G., Bonde, J., Jørgensen, K.P., Nielsen, J.H., Horch, S., Chorkendorff, I., and Nørskov, J.K. (2005). Biomimetic hydrogen evolution: MoS2 nanoparticles as catalyst for hydrogen evolution. J. Am. Chem. Soc. *127*, 5308–5309. https://doi.org/10.1021/JA0504690.

22. Benck, J.D., Hellstern, T.R., Kibsgaard, J., Chakthranont, P., and Jaramillo, T.F. (2014). Catalyzing the hydrogen evolution reaction (HER) with molybdenum sulfide nanomaterials. ACS Catal. *4*, 3957–3971. https://doi.org/10.1021/CS500923C.

23. Dong, Y., Wang, T., Jie, P., Li, M., Wu, T., and Yang, W. (2024). Graphdiyne oxide-sandwiched MoS2 heterostructure with sufficient hetero-interphase and highly expanded interlayer for efficient hydrogen evolution. Chem. Eng. J. *484*, 149457. https://doi.org/10.1016/J.CEJ.2024.149457.

24. Kibsgaard, J., Jaramillo, T.F., and Besenbacher, F. (2014). Building an appropriate active-site motif into a hydrogen-evolution catalyst with thiomolybdate [Mo3S13]2− clusters. Nat. Chem. *6*, 248–253. https://doi.org/10.1038/nchem.1853.

25. Xie, J., Zhang, H., Li, S., Wang, R., Sun, X., Zhou, M., Zhou, J., Lou, X.W., and Xie, Y. (2013). Defect-Rich MoS2 Ultrathin Nanosheets with Additional Active Edge Sites for Enhanced Electrocatalytic Hydrogen Evolution. Adv. Mater. *25*, 5807–5813. https://doi.org/10.1002/ADMA.201302685.

26. Kong, D., Wang, H., Cha, J.J., Pasta, M., Koski, K.J., Yao, J., and Cui, Y. (2013). Synthesis of MoS2 and MoSe2 films with vertically aligned layers. Nano Lett. *13*, 1341–1347. https://doi.org/10.1021/NL400258T.

27. Chatti, M., Gengenbach, T., King, R., Spiccia, L., and Simonov, A.N. (2017). Vertically Aligned Interlayer Expanded MoS2 Nanosheets on a Carbon Support for Hydrogen Evolution Electrocatalysis. Chem. Mater. *29*, 3092–3099. https://doi.org/10.1021/ACS.CHEMMATER.7B00114.

28. Kiran, P.S., Kumar, K.V., Pandit, N., Indupuri, S., Kumar, R., Wagh, V.V., Islam, A., and Keshri, A.K. (2024). Scaling up Simultaneous Exfoliation and 2H to 1T Phase Transformation of MoS2. Adv. Funct. Mater. *34*, 2316266. https://doi.org/10.1002/ADFM.202316266.

29. Zhang, Z., Liu, W., Zhao, W., Xue, H., Chen, Z., Wang, D., and Ji, J. (2024). Exceptionally high-rate production of few-layer MoS2 by a scalable and cost-effective gas-driven shear exfoliation method and its application for supercapacitors. Chem. Eng. J. *479*, 147551. https://doi.org/10.1016/J.CEJ.2023.147551.

30. Liu, H.F., Wong, S.L., and Chi, D.Z. (2015). CVD Growth of MoS2-based Two-dimensional Materials. Chem. Vap. Depos. *21*, 241–259. https://doi.org/10.1002/CVDE.201500060.

31. Zheng, J., Yan, X., Lu, Z., Qiu, H., Xu, G., Zhou, X., Wang, P., Pan, X., Liu, K., Jiao, L., et al. (2017). High-Mobility Multilayered MoS2 Flakes with Low Contact Resistance Grown by Chemical Vapor Deposition. Adv. Mater. *29*, 1604540. https://doi.org/10.1002/ADMA.201604540.

32. Merki, D., Fierro, S., Vrubel, H., and Hu, X. (2011). Amorphous molybdenum sulfide films as catalysts for electrochemical hydrogen production in water. Chem. Sci. *2*, 1262–1267. https://doi.org/10.1039/C1SC00117E.

33. Vrubel, H., Merki, D., and Hu, X. (2012). Hydrogen evolution catalyzed by MoS3 and MoS2 particles. Energy Environ. Sci. *5*, 6136–6144. https://doi.org/10.1039/C2EE02835B.

34. Hibble, S.J., and Wood, G.B. (2004). Modeling the Structure of Amorphous MoS3: A Neutron Diffraction and Reverse Monte Carlo Study. J. Am. Chem. Soc. *126*, 959–965. https://doi.org/10.1021/JA037666O.

35. Lee, C.H., Lee, S., Lee, Y.K., Jung, Y.C., Ko, Y. Il, Lee, D.C., and Joh, H.I. (2018). Understanding the origin of formation and active sites for thiomolybdate [Mo3S13]2-Clusters as hydrogen evolution catalyst through the selective control of sulfur atoms. ACS Catal. *8*, 5221–5227. https://doi.org/10.1021/ACSCATAL.8B01034.

36. Daeneke, T., Dahr, N., Atkin, P., Clark, R.M., Harrison, C.J., Brkljača, R., Pillai, N., Zhang, B.Y., Zavabeti, A., Ippolito, S.J., et al. (2017). Surface Water Dependent Properties of Sulfur-Rich Molybdenum Sulfides: Electrolyteless Gas Phase Water Splitting. ACS Nano *11*, 6782–6794. https://doi.org/10.1021/ACSNANO.7B01632.

37. Weber, T., Muijsers, J.C., and Niemantsverdriet, J.W. (1995). Structure of amorphous MoS3. J. Phys. Chem. *99*, 9194–9200. https://doi.org/10.1021/J100022A037.

38. Madhusudan, P., Shi, R., Chandrashekar, B.N., Xiang, S., Smitha, A.S., Wang, W., Zhang, H., Zhang, X., Amini, A., and Cheng, C. (2020). Highly Efficient Visible-Light-Driven Photocatalytic Hydrogen Production Using Robust Noble-Metal-Free Zn0.5Cd0.5S@Graphene Composites Decorated with MoS2 Nanosheets. Adv. Mater. Interfaces *7*, 2000010. https://doi.org/10.1002/ADMI.202000010.

39. Kong, L., Gao, C., Liu, Z., Pan, L., Yin, P., and Lin, J. (2024). Cerium-doped 1 T phase enriched MoS2 flower-like nanoflakes for boosting hydrogen evolution reaction. Chem. Eng. J. *479*, 147725. https://doi.org/10.1016/J.CEJ.2023.147725.

40. Zhang, S., Dong, H., An, C., Li, Z., Xu, D., Xu, K., Wu, Z., Shen, J., Chen, X., and Zhang, S. (2021). One-pot synthesis of array-like sulfur-doped carbon nitride with covalently crosslinked ultrathin MoS2 cocatalyst for drastically enhanced photocatalytic hydrogen evolution. J. Mater. Sci. Technol. *75*, 59–67. https://doi.org/10.1016/J.JMST.2020.10.030.

41. Leliveld, R.G., Van Dillen, A.J., Geus, J.W., and Koningsberger, D.C. (1997). A Mo–K Edge XAFS Study of the Metal Sulfide-Support Interaction in (Co)Mo Supported Alumina and Titania Catalysts. J. Catal. *165*, 184–196. https://doi.org/10.1006/JCAT.1997.1480.

42. Gaur, A., Stehle, M., Serrer, M.A., Stummann, M.Z., La Fontaine, C., Briois, V., Grunwaldt, J.D., and Høj, M. (2021). Using Transient XAS to Detect Minute Levels of Reversible S-O Exchange at the Active Sites of MoS2-Based Hydrotreating Catalysts: Effect of Metal Loading, Promotion, Temperature, and Oxygenate Reactant. ACS Catal. *12*, 633–647. https://doi.org/10.1021/ACSCATAL.1C04767.

43. Garcia-Esparza, A.T., Park, S., Abroshan, H., Paredes Mellone, O.A., Vinson, J., Abraham, B., Kim, T.R., Nordlund, D., Gallo, A., Alonso-Mori, R., et al. (2022). Local Structure of Sulfur Vacancies on the Basal Plane of Monolayer MoS2. ACS Nano *16*, 6725–6733. https://doi.org/10.1021/ACSNANO.2C01388.

44. Letourneau, S., Young, M.J., Bedford, N.M., Ren, Y., Yanguas-Gil, A., Mane, A.U., Elam, J.W., and Graugnard, E. (2018). Structural Evolution of Molybdenum Disulfide Prepared by Atomic Layer Deposition for Realization of Large Scale Films in Microelectronic

Applications. ACS Appl. Nano Mater. *1*, 4028–4037. https://doi.org/10.1021/ACSANM.8B00798.

45. Tang, M.L., Grauer, D.C., Lassalle-Kaiser, B., Yachandra, V.K., Amirav, L., Long, J.R., Yano, J., and Alivisatos, A.P. (2011). Structural and Electronic Study of an Amorphous MoS3 Hydrogen-Generation Catalyst on a Quantum-Controlled Photosensitizer. Angew. Chem. Int. Ed. *50*, 10203–10207. https://doi.org/10.1002/ANGE.201104412.

46. Lassalle-Kaiser, B., Merki, D., Vrubel, H., Gul, S., Yachandra, V.K., Hu, X., and Yano, J. (2014). Evidence from in Situ X-ray Absorption Spectroscopy for the Involvement of Terminal Disulfide in the Reduction of Protons by an Amorphous Molybdenum Sulfide Electrocatalyst. J. Am. Chem. Soc. *137*, 314–321. https://doi.org/10.1021/JA510328M.

47. Shinde, D. V., Dang, Z., Petralanda, U., Palei, M., Wang, M., Prato, M., Cavalli, A., De Trizio, L., and Manna, L. (2018). In Situ Dynamic Nanostructuring of the Cu-Ti Catalyst-Support System Promotes Hydrogen Evolution under Alkaline Conditions. ACS Appl. Mater. Interfaces *10*, 29583–29592. https://doi.org/10.1021/ACSAMI.8B09493.

48. Tran, P.D., Tran, T. V., Orio, M., Torelli, S., Truong, Q.D., Nayuki, K., Sasaki, Y., Chiam, S.Y., Yi, R., Honma, I., et al. (2016). Coordination polymer structure and revisited hydrogen evolution catalytic mechanism for amorphous molybdenum sulfide. Nat. Mater. *15*, 640–646. https://doi.org/10.1038/nmat4588.

49. Sekar, A., Muthukumar, K., Rajendran, S., and Li, J. (2024). Synthesis of defect-engineered molybdenum sulfides on reduced graphene oxide for enhanced hydrogen evolution reaction kinetics. Int. J. Hydrogen Energy *51*, 1387–1396. https://doi.org/10.1016/J.IJHYDENE.2023.11.107.

50. Ting, L.R.L., Deng, Y., Ma, L., Zhang, Y.J., Peterson, A.A., and Yeo, B.S. (2016). Catalytic Activities of Sulfur Atoms in Amorphous Molybdenum Sulfide for the Electrochemical Hydrogen Evolution Reaction. ACS Catal. *6*, 861–867. https://doi.org/10.1021/ACSCATAL.5B02369.

51. Zhu, Y., Li, L., Cheng, H., and Ma, J. (2024). Alkaline Hydrogen Evolution Reaction Electrocatalysts for Anion Exchange Membrane Water Electrolyzers: Progress and Perspective. JACS Au *4*, 4639–4654. https://doi.org/10.1021/JACSAU.4C00898.

52. Fei, H., Liu, R., Zhang, Y., Wang, H., Wang, M., Wang, S., Ni, M., Wu, Z., Wang, J., Fei, H., et al. (2023). Extending $MoS_2$-based materials into the catalysis of non-acidic hydrogen evolution: challenges, progress, and perspectives. Mater. Futur. *2*, 022103-. https://doi.org/10.1088/2752-5724/ACC51D.

53. Evans, M.G., and Polanyi, M. (1937). On the introduction of thermodynamic variables into reaction kinetics. Trans. Faraday Soc. *33*, 448–452. https://doi.org/10.1039/TF9373300448.

54. McCrum, I.T., and Koper, M.T.M. (2020). The role of adsorbed hydroxide in hydrogen evolution reaction kinetics on modified platinum. Nat. Energy *5*, 891–899. https://doi.org/10.1038/S41560-020-00710-8.

55. Escalera-López, D., Iffelsberger, C., Zlatar, M., Novčić, K., Maselj, N., Van Pham, C., Jovanovič, P., Hodnik, N., Thiele, S., Pumera, M., et al. (2024). Allotrope-dependent activity-stability relationships of molybdenum sulfide hydrogen evolution electrocatalysts. Nat. Commun. *15*, 3601. https://doi.org/10.1038/S41467-024-47524-W.

56. Niu, C., Song, H., Chang, Y., Hou, W., Li, Y., Zhao, Y., Xiao, Y., and Han, G. (2022). Amorphous MoSx electro-synthesized in alkaline electrolyte for superior hydrogen evolution. J. Alloys Compd. *900*, 163509. https://doi.org/10.1016/J.JALLCOM.2021.163509.

57. Zuo, Y., Mastronardi, V., Gamberini, A., Zappia, M.I., Le, T.H.H., Prato, M., Dante, S., Bellani, S., and Manna, L. (2024). Stainless Steel Activation for Efficient Alkaline Oxygen Evolution in Advanced Electrolyzers. Adv. Mater. *36*, 2312071. https://doi.org/10.1002/ADMA.202312071.

58. Liu, H.J., Zhang, S., Chai, Y.M., and Dong, B. (2023). Ligand Modulation of Active Sites to Promote Cobalt-Doped 1T-MoS2 Electrocatalytic Hydrogen Evolution in Alkaline Media. Angew. Chem. Int. Ed. *62*, e202313845. https://doi.org/10.1002/ANIE.202313845.

59. Mastronardi, V., Gamberini, A., Zappia, M.I., Zuo, Y., Abruzzese, M., Bagheri, A., Beydaghi, H., Gabatel, L., Thorat, S., Ferri, M., et al. (2024). Alkaline electrochemical deposition of Pt for alkaline electrolyzers. Int. J. Hydrogen Energy *80*, 261–269. https://doi.org/10.1016/J.IJHYDENE.2024.07.086.

60. Lamy, C., and Millet, P. (2020). A critical review on the definitions used to calculate the energy efficiency coefficients of water electrolysis cells working under near ambient temperature conditions. J. Power Sources *447*, 227350. https://doi.org/10.1016/J.JPOWSOUR.2019.227350.

61. Niu, S., Li, S., Du, Y., Han, X., and Xu, P. (2020). How to Reliably Report the Overpotential of an Electrocatalyst. ACS Energy Lett. *5*, 1083–1087. https://doi.org/10.1021/ACSENERGYLETT.0C00321.

62. McCrory, C.C.L., Jung, S., Peters, J.C., and Jaramillo, T.F. (2013). Benchmarking heterogeneous electrocatalysts for the oxygen evolution reaction. J. Am. Chem. Soc. *135*, 16977–16987. https://doi.org/10.1021/JA407115P.

63. Kibsgaard, J., and Jaramillo, T.F. (2014). Molybdenum Phosphosulfide: An Active, Acid-Stable, Earth-Abundant Catalyst for the Hydrogen Evolution Reaction. Angew. Chem. Int. Ed. *53*, 14433–14437. https://doi.org/10.1002/ANIE.201408222.

64. Hu, J., Zhang, C., Jiang, L., Lin, H., An, Y., Zhou, D., Leung, M.K.H., and Yang, S. (2017). Nanohybridization of MoS2 with Layered Double Hydroxides Efficiently Synergizes the Hydrogen Evolution in Alkaline Media. Joule *1*, 383–393. https://doi.org/10.1016/J.JOULE.2017.07.011.

65. Fei, H., Dong, J., Arellano-Jiménez, M.J., Ye, G., Dong Kim, N., Samuel, E.L.G., Peng, Z., Zhu, Z., Qin, F., Bao, J., et al. (2015). Atomic cobalt on nitrogen-doped graphene for hydrogen generation. Nat. Commun. *6*, 8668. https://doi.org/10.1038/ncomms9668.

66. Mathon, O., Beteva, A., Borrel, J., Bugnazet, D., Gatla, S., Hino, R., Kantor, I., Mairs, T., Munoz, M., Pasternak, S., et al. (2015). The time-resolved and extreme conditions XAS (Texas) facility at the European Synchrotron Radiation Facility: The general-purpose EXAFS bending-magnet beamline BM23. J. Synchrotron Radiat. *22*, 1548–1554. https://doi.org/10.1107/S1600577515017786.

67. Ravel, B., and Newville, M. (2005). ATHENA, ARTEMIS, HEPHAESTUS: data analysis for X-ray absorption spectroscopy using IFEFFIT. J. Synchrotron Radiat. *12*, 537–541. https://doi.org/10.1107/S0909049505012719.

68. Kresse, G., and Joubert, D. (1999). From ultrasoft pseudopotentials to the projector augmented-wave method. Phys. Rev. B *59*, 1758. https://doi.org/10.1103/PhysRevB.59.1758.

69. Kresse, G., and Furthmüller, J. (1996). Efficiency of ab-initio total energy calculations for metals and semiconductors using a plane-wave basis set. Comput. Mater. Sci. *6*, 15–50. https://doi.org/10.1016/0927-0256(96)00008-0.

70. Kresse, G., and Furthmüller, J. (1996). Efficient iterative schemes for *ab initio* total-energy calculations using a plane-wave basis set. Phys. Rev. B *54*, 11169. https://doi.org/10.1103/PhysRevB.54.11169.

71. Perdew, J.P., Burke, K., and Ernzerhof, M. (1996). Generalized Gradient Approximation Made Simple. Phys. Rev. Lett. *77*, 3865. https://doi.org/10.1103/PhysRevLett.77.3865.

72. Grimme, S. (2006). Semiempirical GGA-type density functional constructed with a long-range dispersion correction. J. Comput. Chem. *27*, 1787–1799. https://doi.org/10.1002/JCC.20495.

73. Furness, J.W., Kaplan, A.D., Ning, J., Perdew, J.P., and Sun, J. (2020). Accurate and Numerically Efficient r2SCAN Meta-Generalized Gradient Approximation. J. Phys. Chem. Lett. *11*, 8208–8215. https://doi.org/10.1021/ACS.JPCLETT.0C02405.

74. Ning, J., Kothakonda, M., Furness, J.W., Kaplan, A.D., Ehlert, S., Brandenburg, J.G., Perdew, J.P., and Sun, J. (2022). Workhorse minimally empirical dispersion-corrected density functional with tests for weakly bound systems: <math xmlns="http://www.w3.org/1998/Math/MathML"><mrow><msup><mrow><mi

mathvariant="normal">r</mi></mrow><mn>2</mn></msup><mi>SCAN</mi><mo>+ </mo><m…. Phys. Rev. B *106*, 075422. https://doi.org/10.1103/PhysRevB.106.075422.

75. Hanesch, M. (2009). Raman spectroscopy of iron oxides and (oxy)hydroxides at low laser power and possible applications in environmental magnetic studies. Geophys. J. Int. *177*, 941–948. https://doi.org/10.1111/J.1365-246X.2009.04122.X.

# Supplemental Information

## Scalable Conformal $MoS_x$ Catalyst for Efficient Hydrogen Evolution at Industrial-Level Current Density in Alkaline Electrolyzers

*Yong Zuo, Sebastiano Bellani*, Meysoun Jabrane, Gabriele Saleh, Thi-Hong-Hanh Le, Michele Ferri, Davide Salusso, Zhanzhao Li, Valentina Mastronardi, Marilena I. Zappia, Manjunath Chatti, Mirko Prato, Silvia Dante, Francesco Bonaccorso, Yongsheng Han, and Liberato Manna**

[1] School of Chemistry and Chemical Engineering, Chongqing University, Chongqing, 400044, China

[2] Nanochemistry Department, Istituto Italiano di Tecnologia, Via Morego 30, 16163 Genova, Italy

[3] BeDimensional S.p.A., Via Lungotorrente Secca, 30R, 16163 Genova, Italy

[4] Antares Electrolysis S.r.l., Piazza della Vittoria 14/19, 16121 Genova, Italy

[5] European Synchrotron Radiation Facility, CS 40220, Cedex 9 F-38043 Grenoble, France

[6] Materials Characterization Facility, Istituto Italiano di Tecnologia, Via Morego 30, 16163 Genova, Italy

[7] Lead contact

Correspondence: sebastiano.bellani@antares-electrolysis.com (S.B.); liberato.manna@iit.it (L.M.)

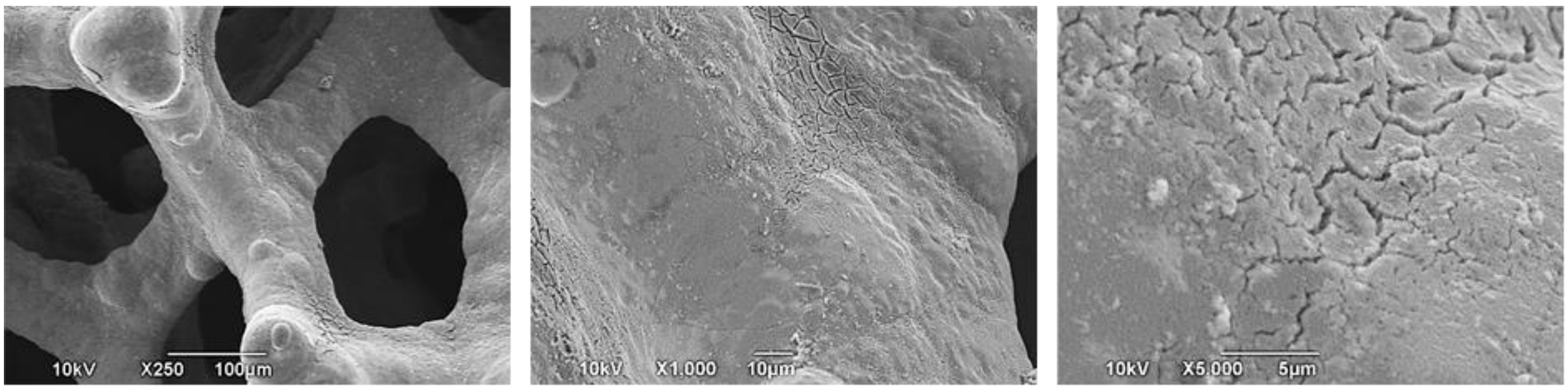

**Figure S1.** SEM images of NF after dip-coating in a 50 mM $(NH_4)_2MoS_4$ solution for 1 min, followed by natural air drying.

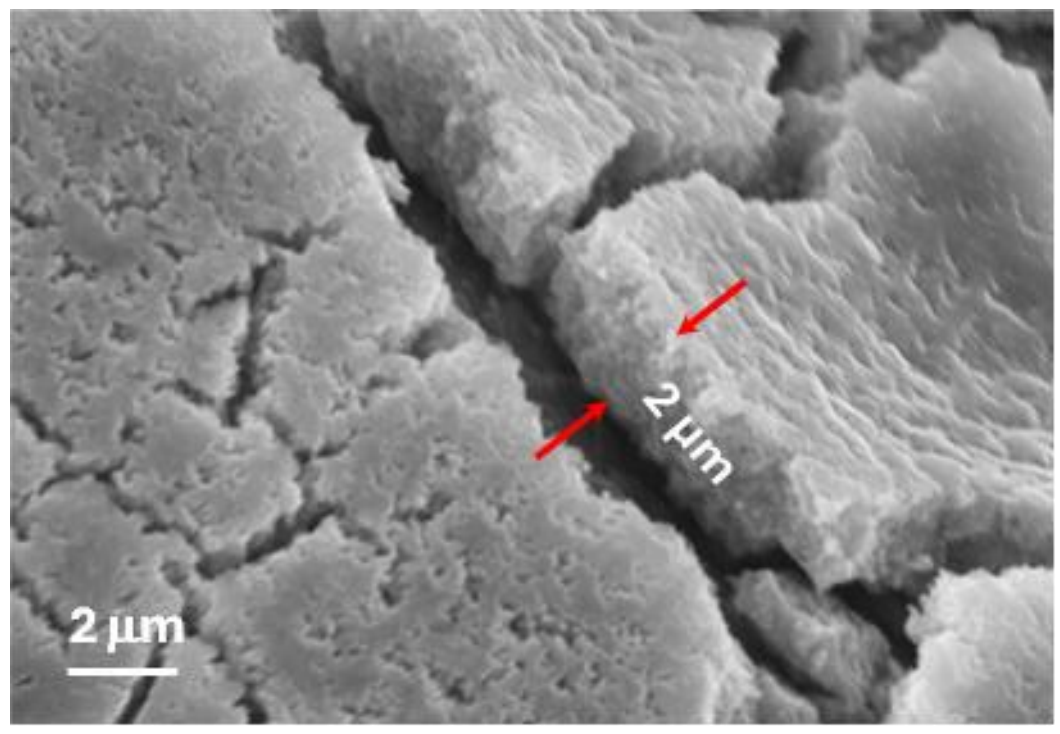

**Figure S2.** Cross-section SEM image of the $MoS_x$@NF_200C-1h electrode, showing the thickness of the $MoS_x$ layer on the NF surface.

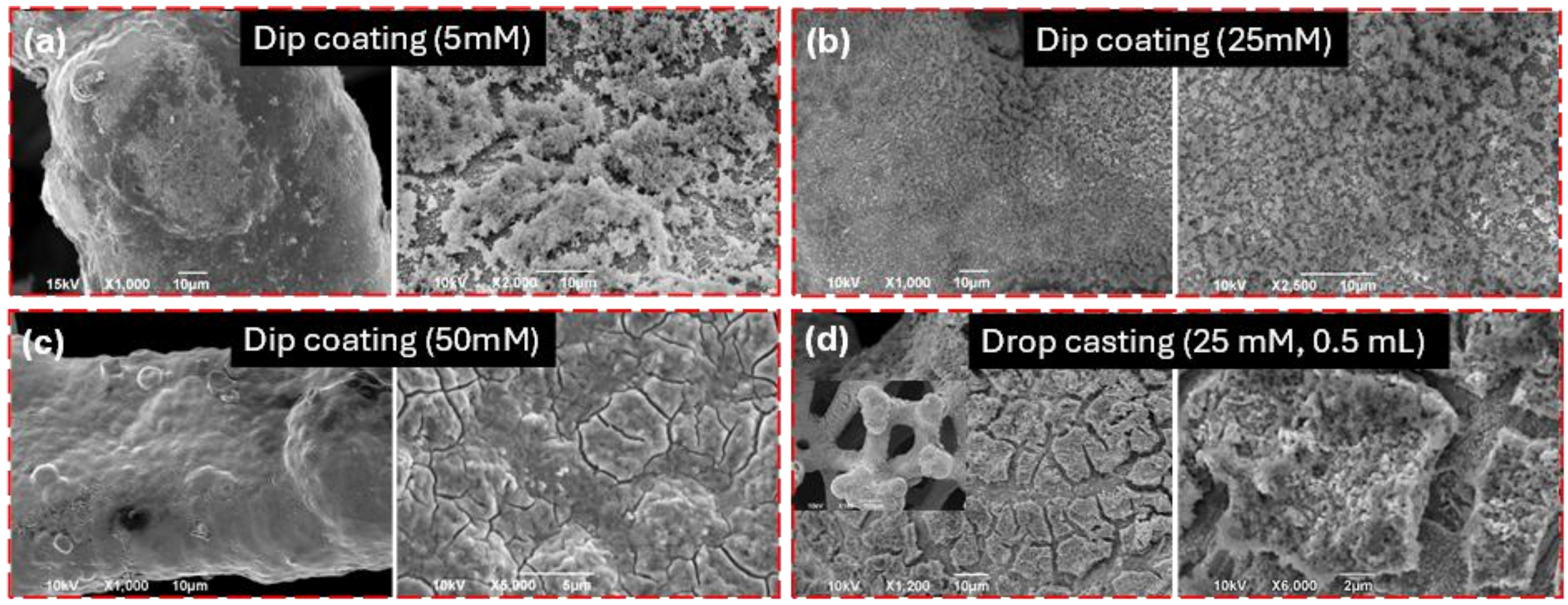

**Figure S3.** SEM images of the $MoS_x$@NF electrodes produced using different wetting strategy: (a) dip coating in 5 mM $(NH_4)_2MoS_4$ solution for 1 min; (b) dip coating in 25 mM $(NH_4)_2MoS_4$ solution for 1 min; (c) dip coating in 50 mM $(NH_4)_2MoS_4$ solution for 1 min; (d) drop casting using 0.5 mL 25 mM $(NH_4)_2MoS_4$ solution.

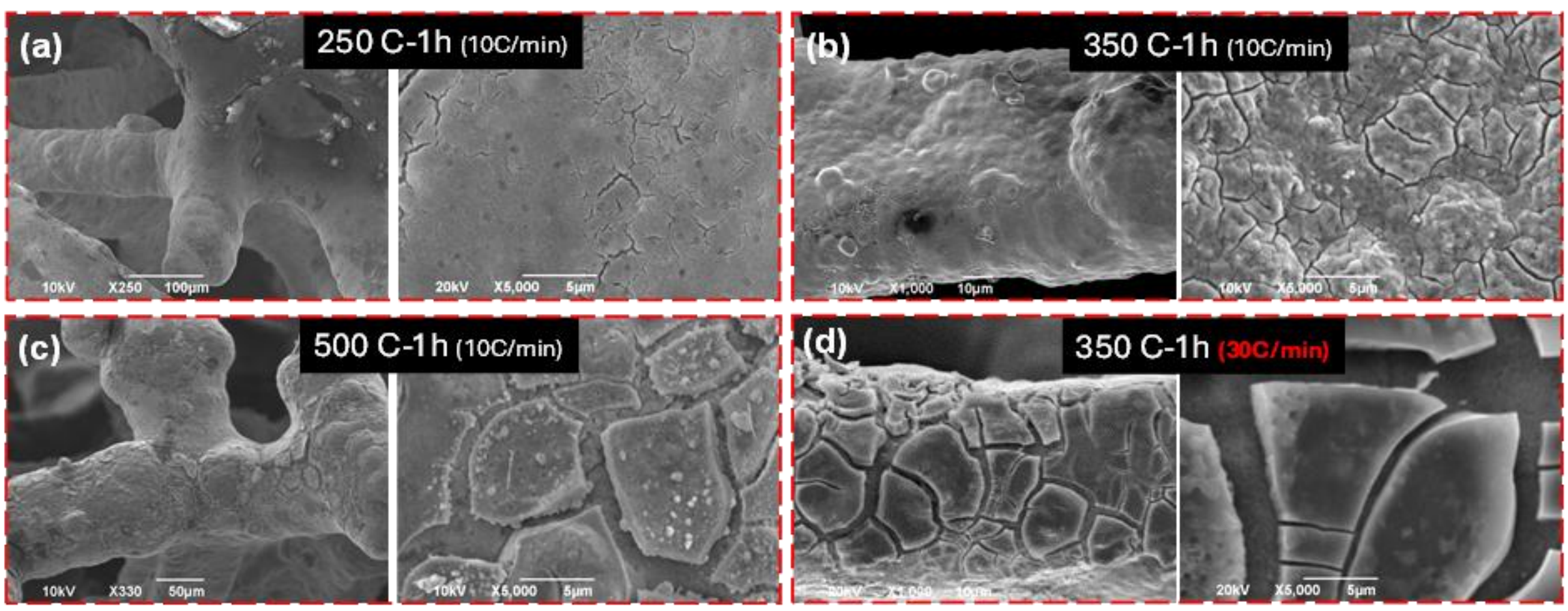

**Figure S4.** SEM images of the $MoS_x$@NF electrodes produced with different annealing treatments. The coating strategy involved dip-coating NF in a 50 mM $(NH_4)_2MoS_4$ solution for 1 min.

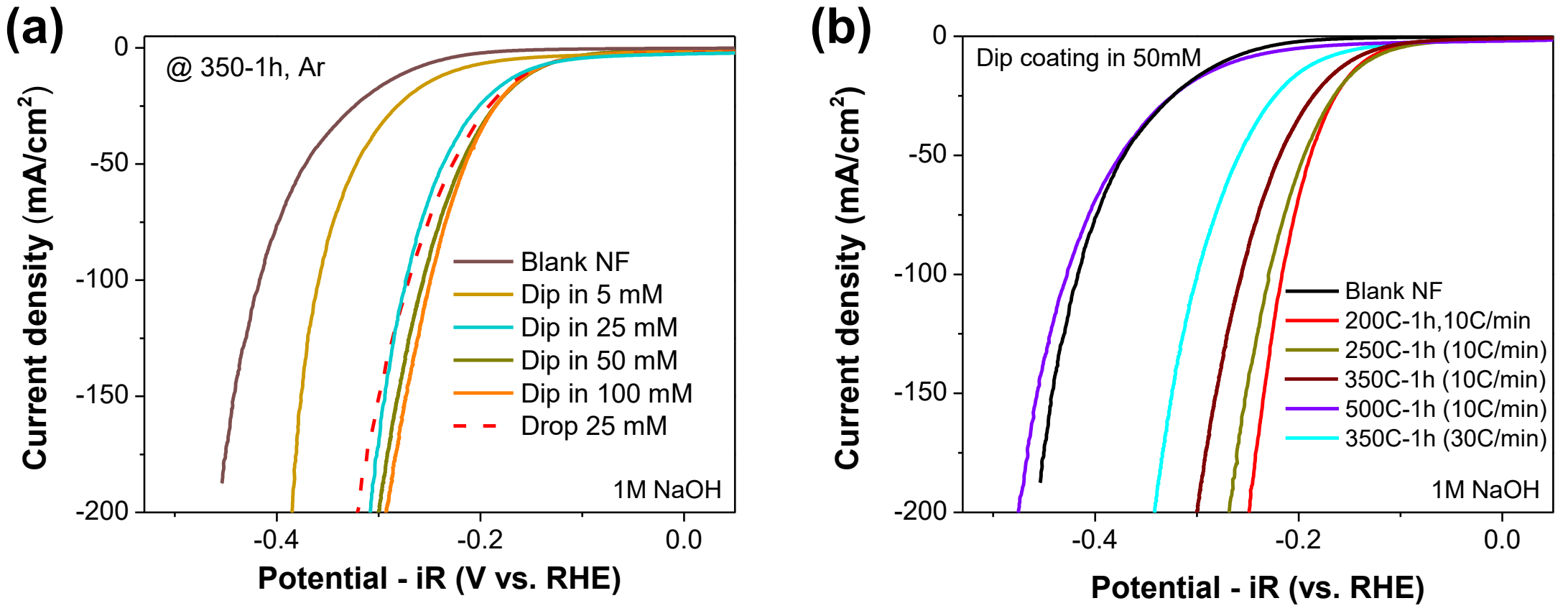

**Figure S5.** Polarization curves with iR-correction measured for the $MoS_x$@NF electrodes synthesized using: (a) different coating strategies, with annealing parameters set at 350 $^oC$ for 1h in Ar; (b) different annealing strategies, using a dip-coating strategy where NF was immersed in a 50 mM $(NH_4)_2MoS_4$ solution for 1 min.

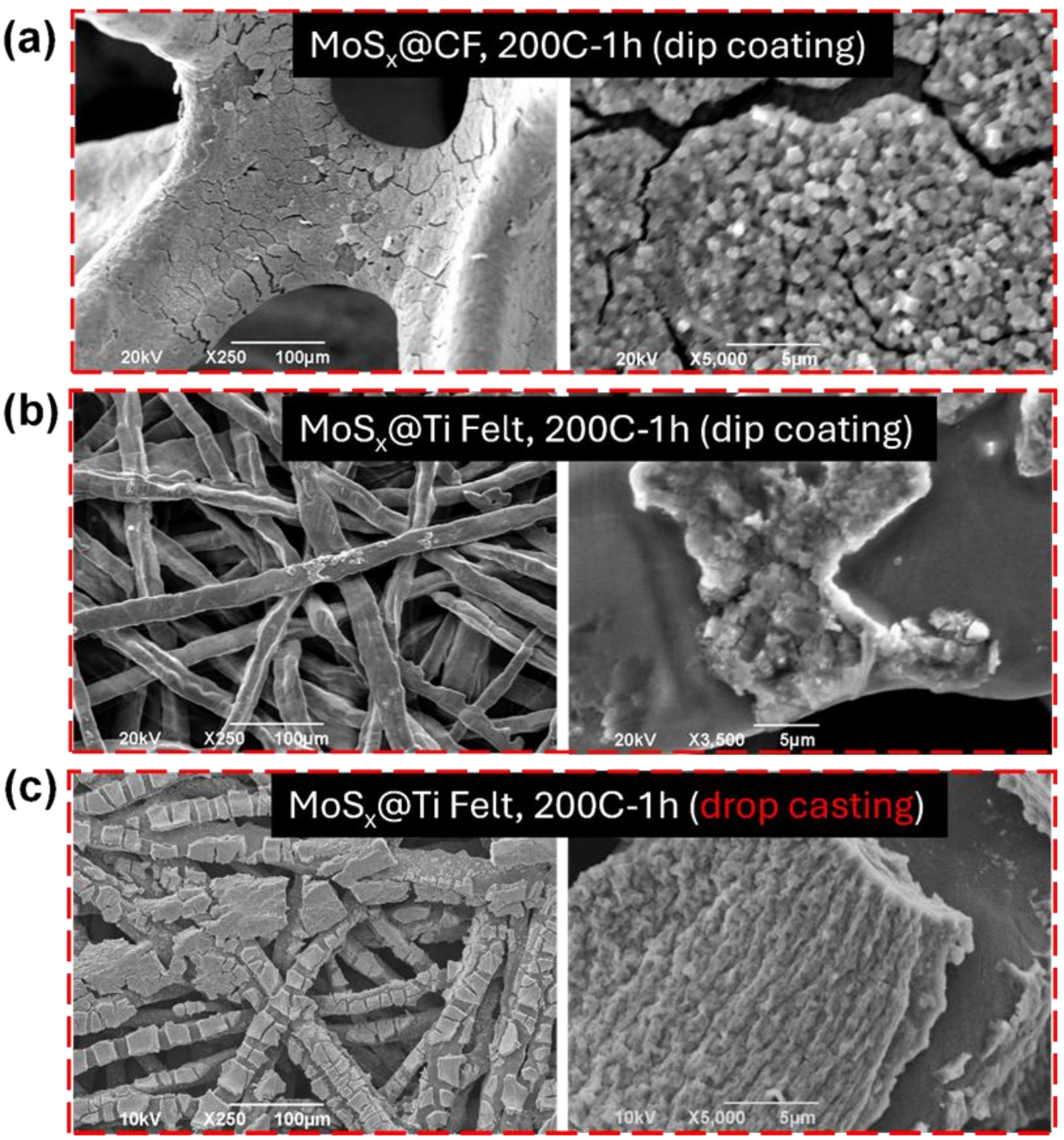

**Figure S6.** $MoS_x$ coated on different substrate under annealing conditions of 200 °C for 1 h in Ar: (a) Cu foam; (b-c) Ti felt.

Interestingly, the pre-cleaned Ti felt exhibited an insufficient $MoS_x$ coating when using the dip coating strategy. To address this, a drop-casting method was employed, where 0.5 mL of 50 mM $(NH_4)_2MoS_4$ solution was drop-cast onto the Ti felt. The sample was then air-dried naturally before being annealed in an Ar atmosphere at 200 °C for 1 h.

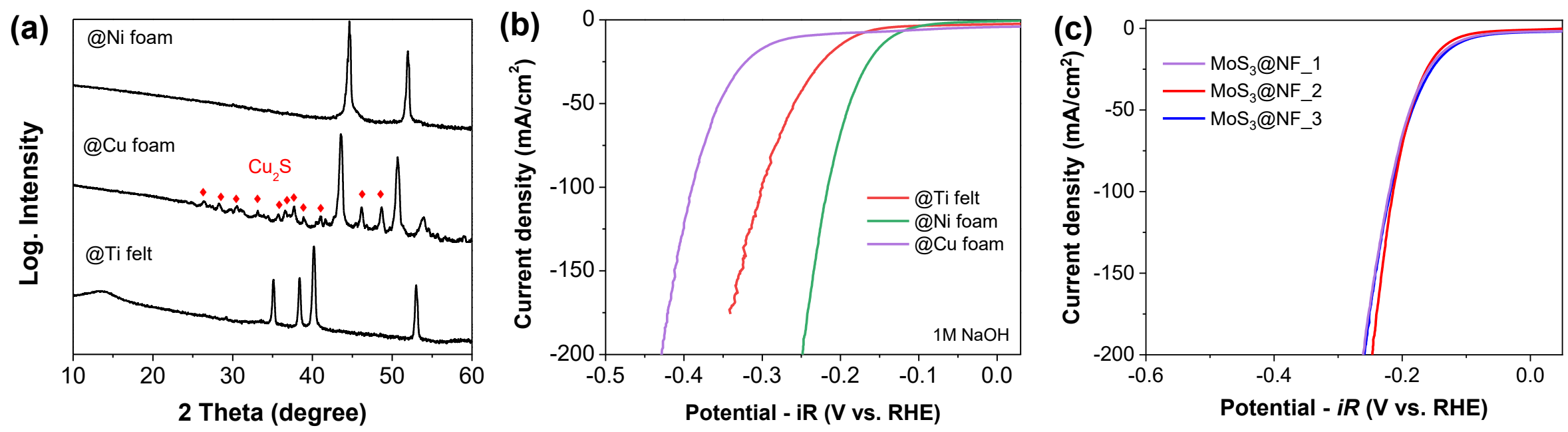

**Figure S7.** $MoS_x$ coated on different substrate: (a) XRD patterns; (b) Polarization curves. (c) LSV curves with iR-correction measured on $MoS_3$@NF electrodes from three synthesis batches. Annealing parameters: 200 °C, 1h, Ar.

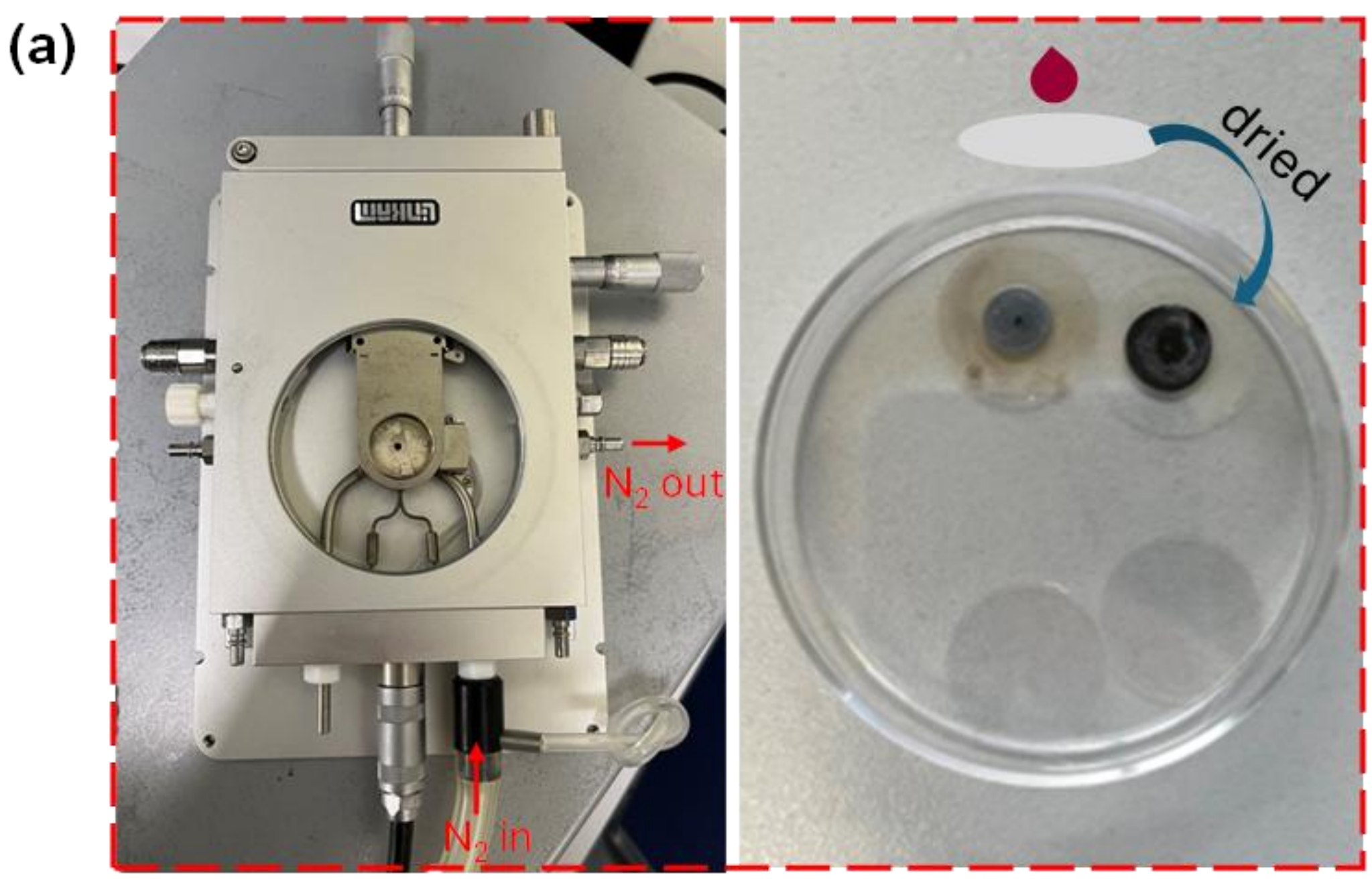

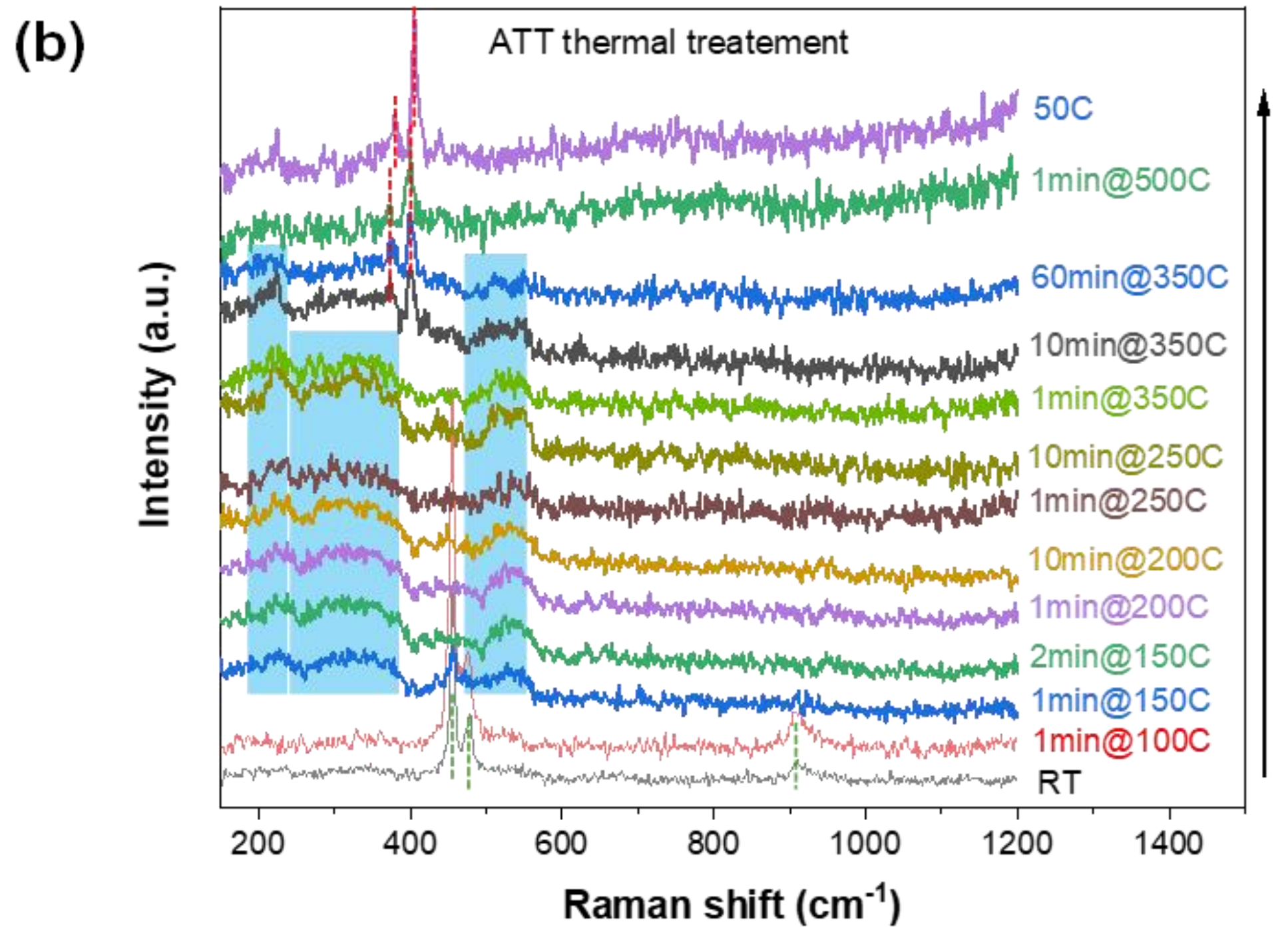

**Figure S8.** (a) Setup used for the thermal-operando Raman measurement. A drop of 50 mM $(NH_4)_2MoS_4$ solution was naturally dried on the surface of quartz glass and used as the sample. (b) Raman spectra acquired during the annealing in $N_2$ under elevating temperatures. The blue-shaded regions indicate the characteristic Raman peaks of $MoS_3$ species. The green dashed line indicates the Raman peaks of $(NH_4)_2MoS_4$ species. The red dashed lines indicate the Raman peaks of $MoS_2$ species. ATT: ammonium tetrathiomolybdate.

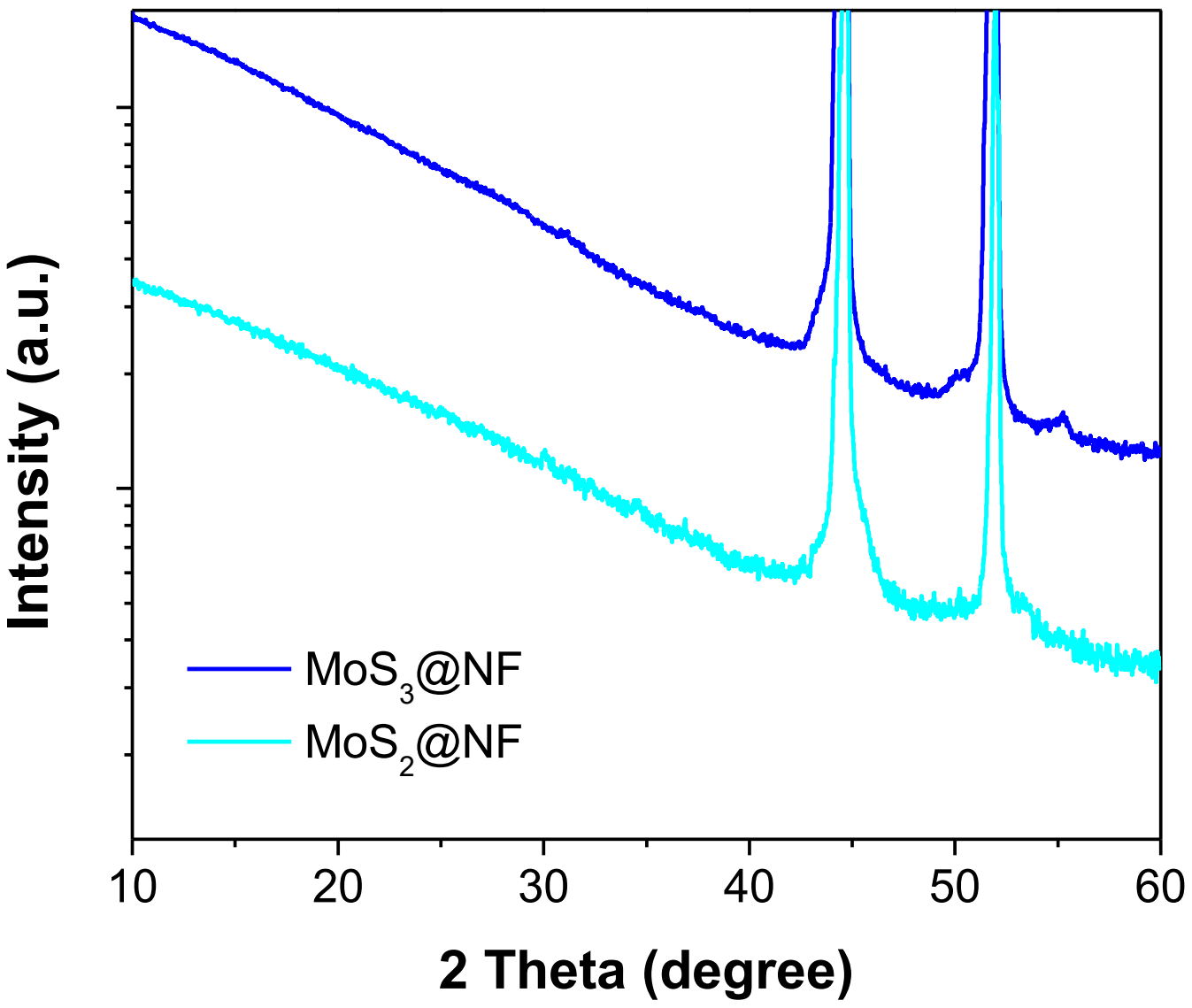

**Figure S9.** XRD patterns of the as-prepared $MoS_3$@NF and $MoS_2$@NF electrodes. In both cases, only the XRD peaks corresponding to the NF substrate were observed, indicating the amorphous nature of the $MoS_3$ and $MoS_2$ layers.

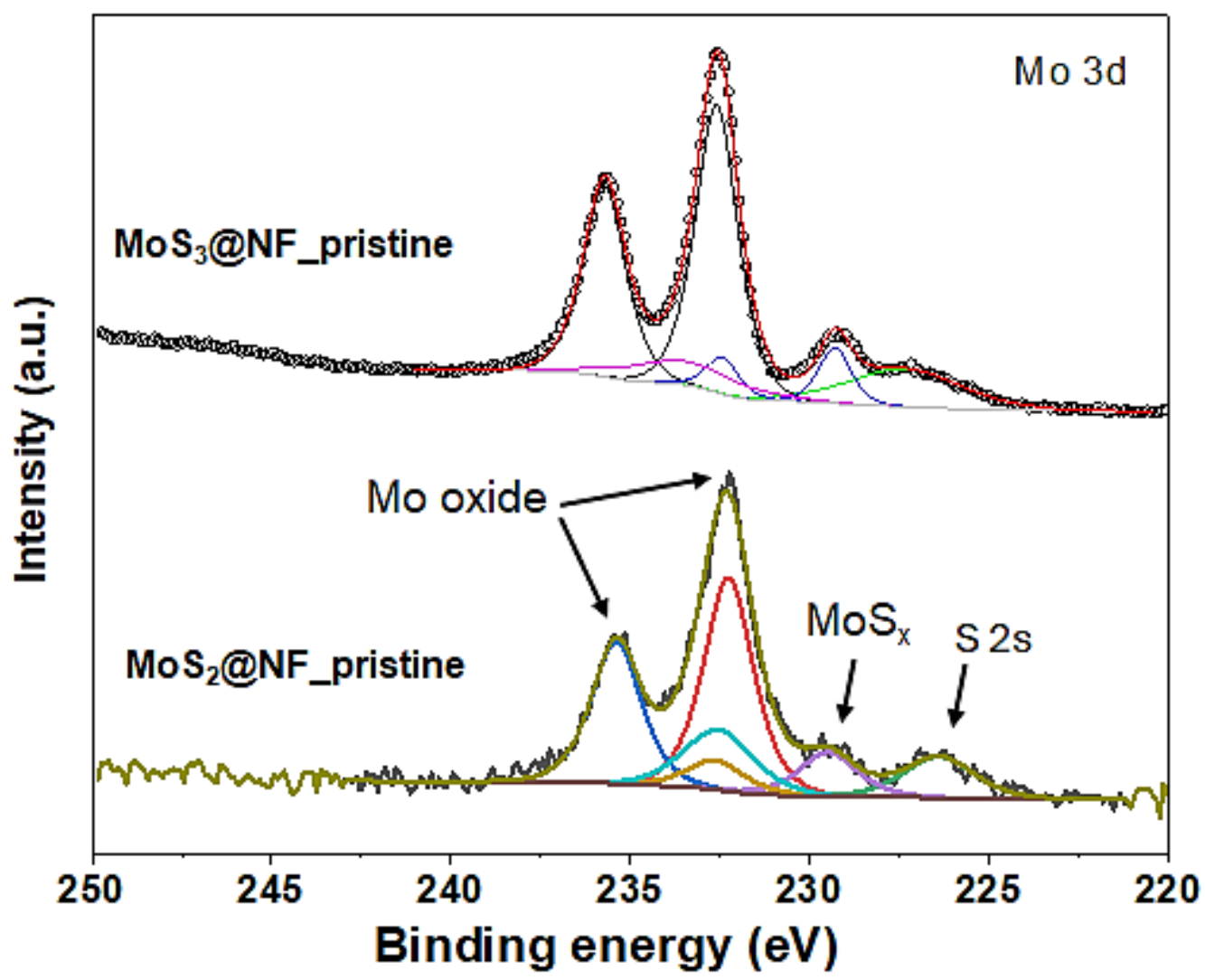

**Figure S10.** XPS spectra of Mo 3d region recorded on the $MoS_3$@NF_pristine and $MoS_2$@NF_pristine.

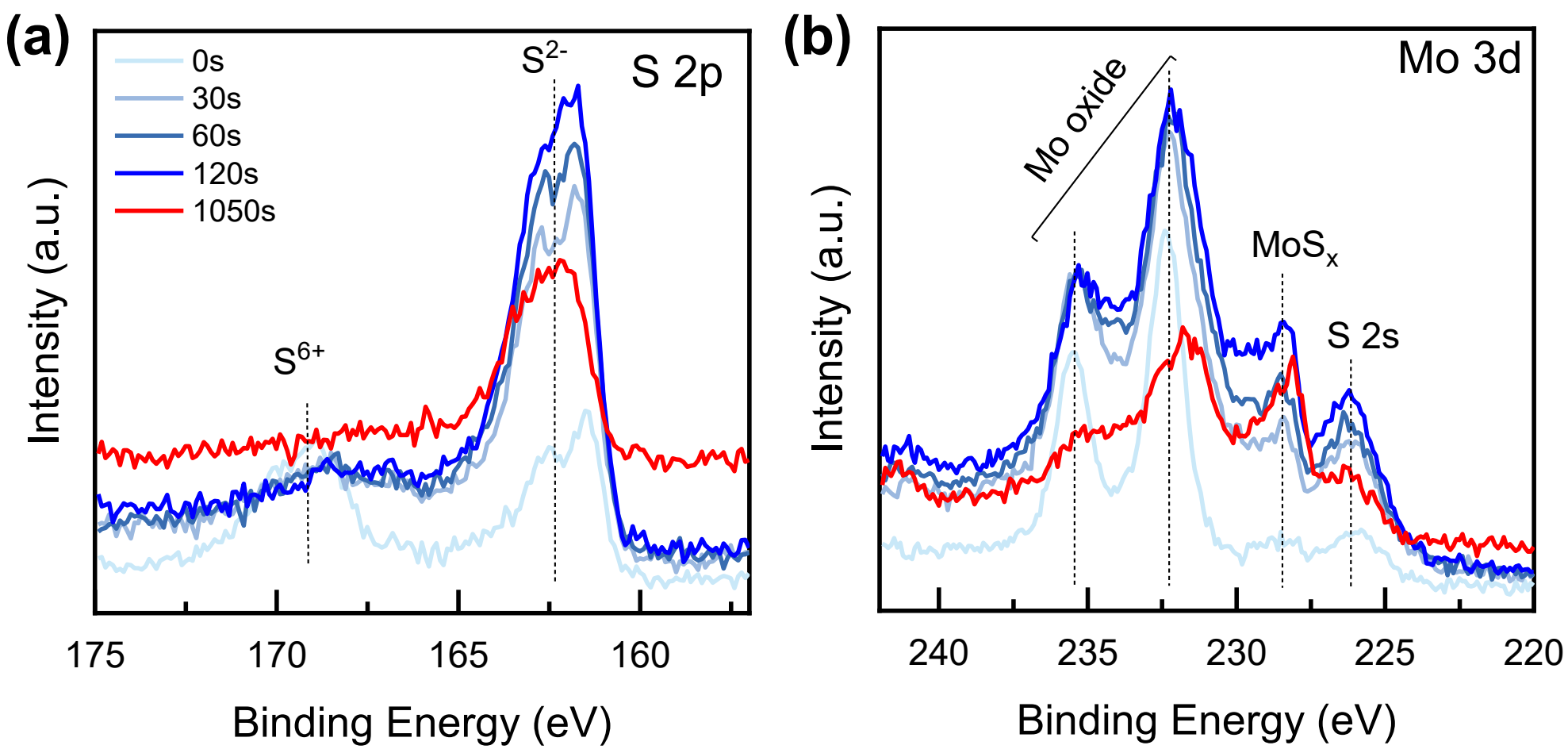

**Figure S11.** Ar etched XPS spectra of Mo 3d region recorded on the $MoS_3$@NF_pristine electrode.

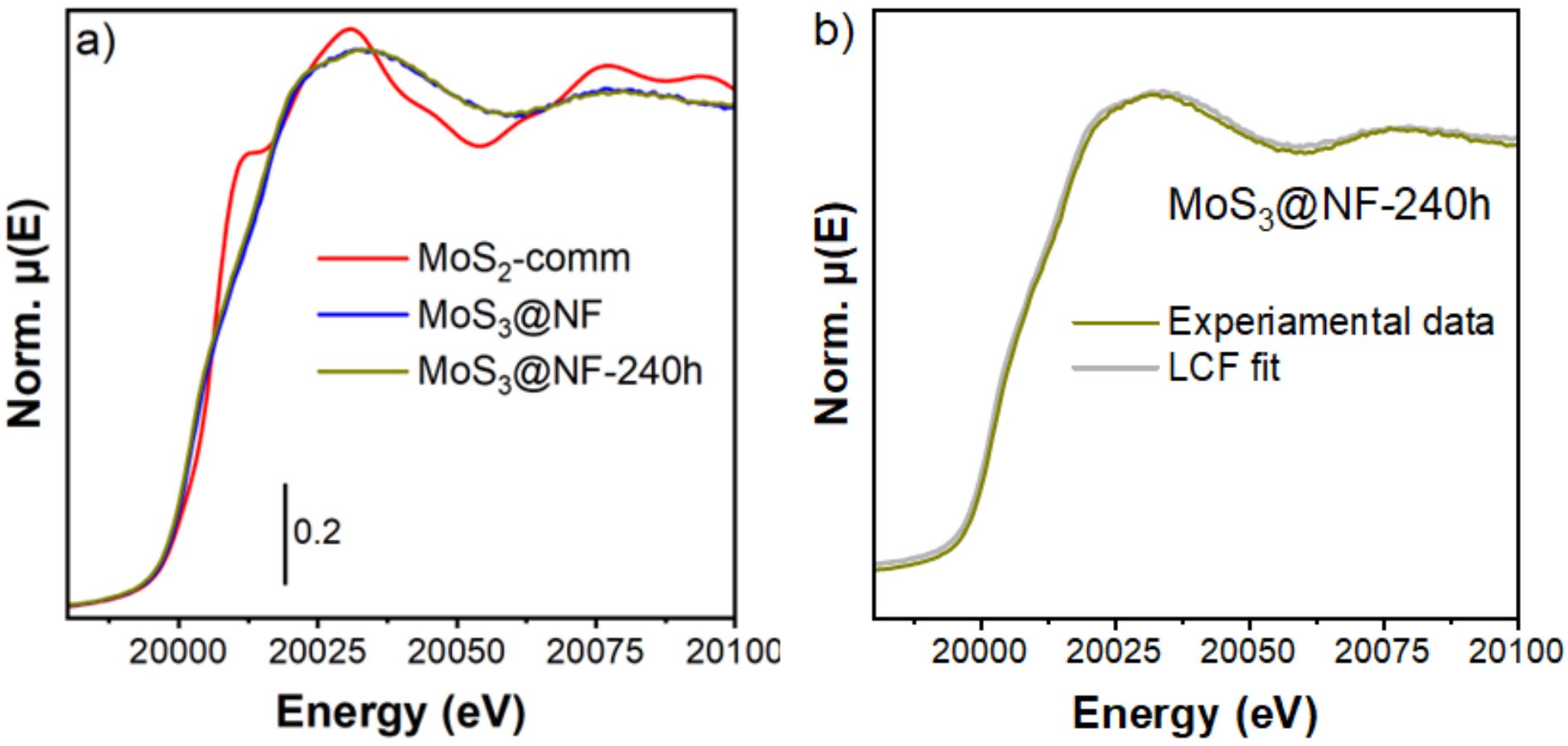

**Figure S12**. Ex-situ Mo K-edge of $MoS_3$@NF and $MoS_3$@NF-240h: a) XANES; (b) Stacked comparison of Mo K-edge XANES experimental spectra of $MoS_3$@NF-240h (colored line) with the LCF best fit curve (gray line).

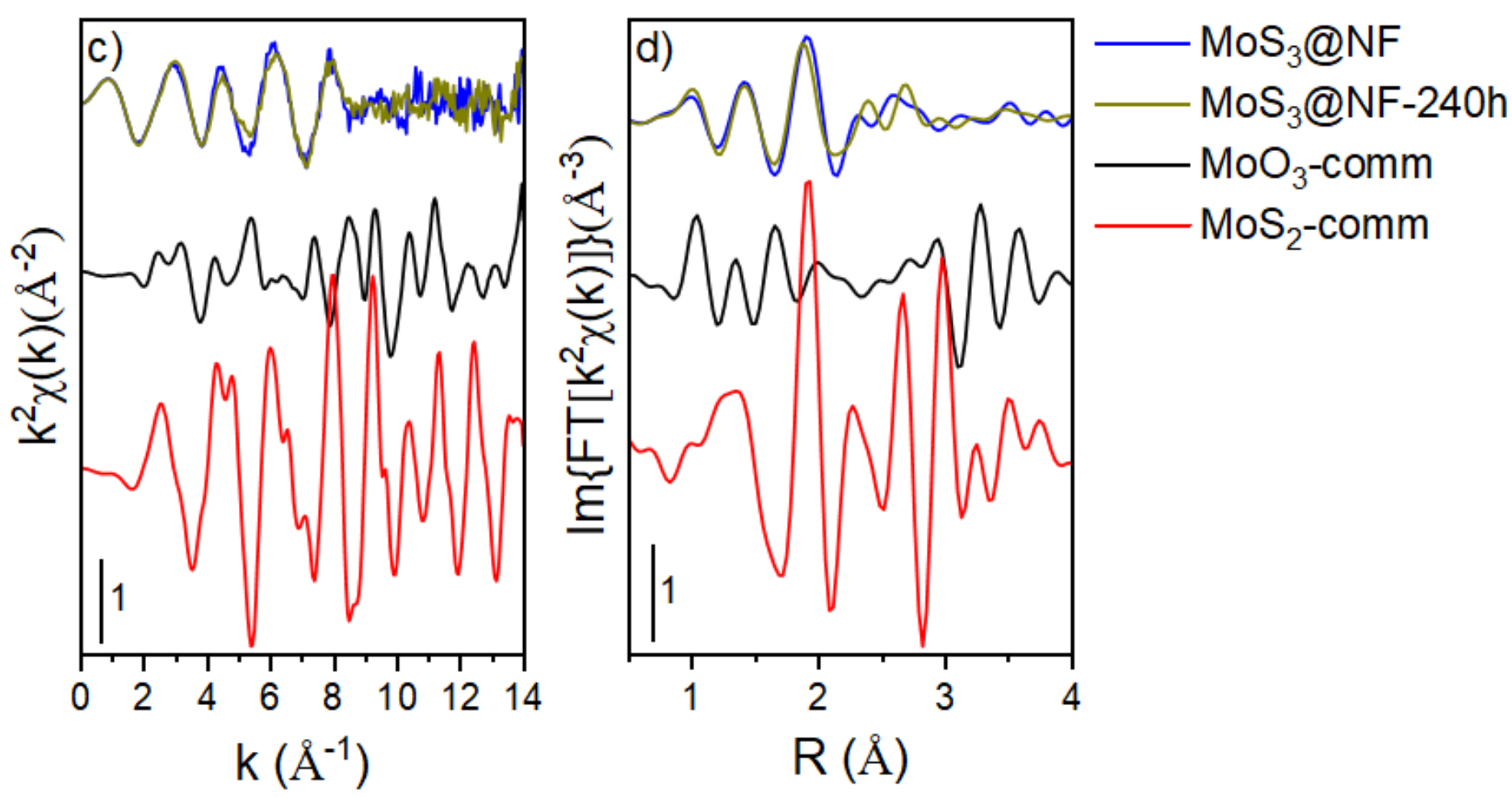

**Figure S13**. Ex-situ Mo K-edge: (a) extended X-ray absorption fine structure EXAFS and (b) (phase uncorrected) Fourier-transformed EXAFS (FT-EXAFS) spectra imaginary component.

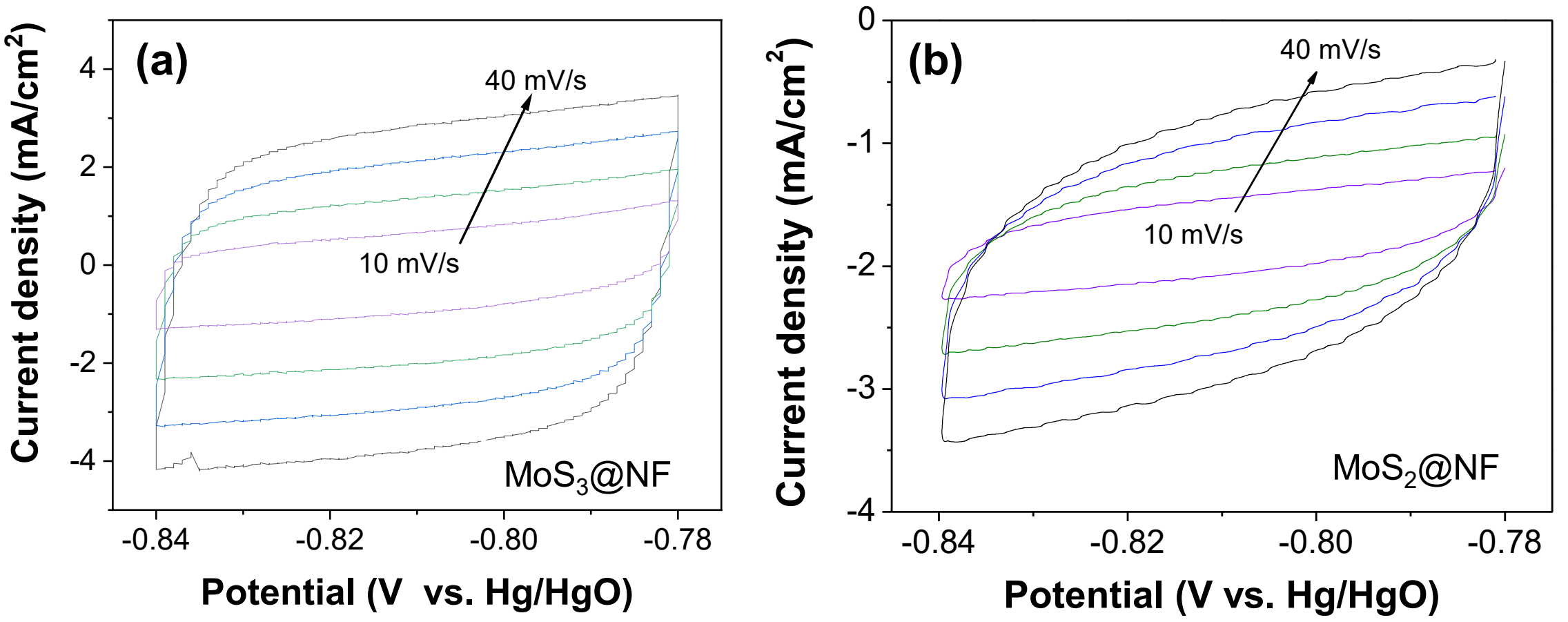

**Figure S14.** CV curves measured for (a) $MoS_3$@NF and (b) $MoS_2$@NF at potential scan rates of 10, 20, 30, and 40 mV/s.

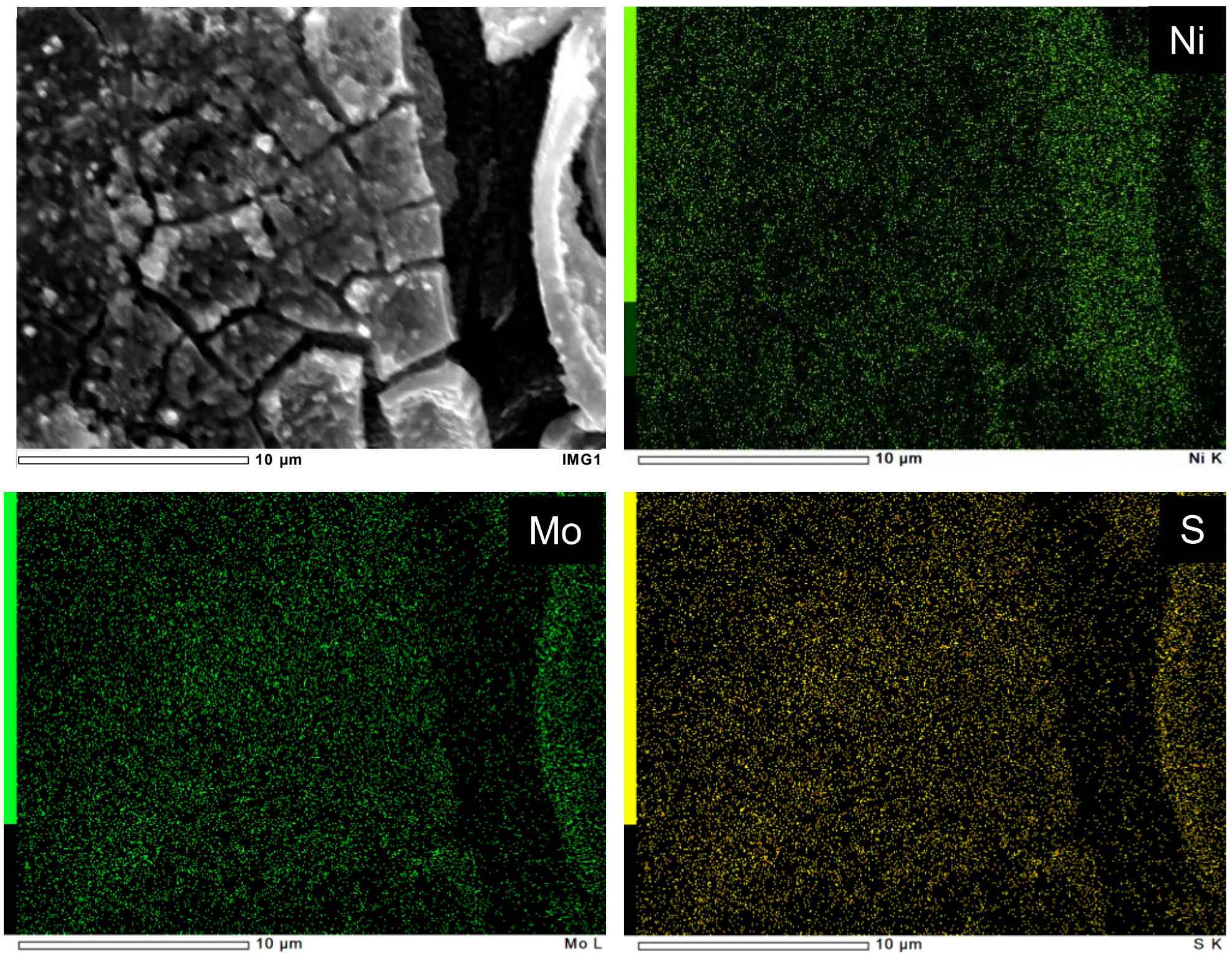

**Figure S15.** SEM-EDS characterization of the $MoS_3$@NF electrode after HER operation in 1 M NaOH at -200 mA/cm$^2$ for 240 h.

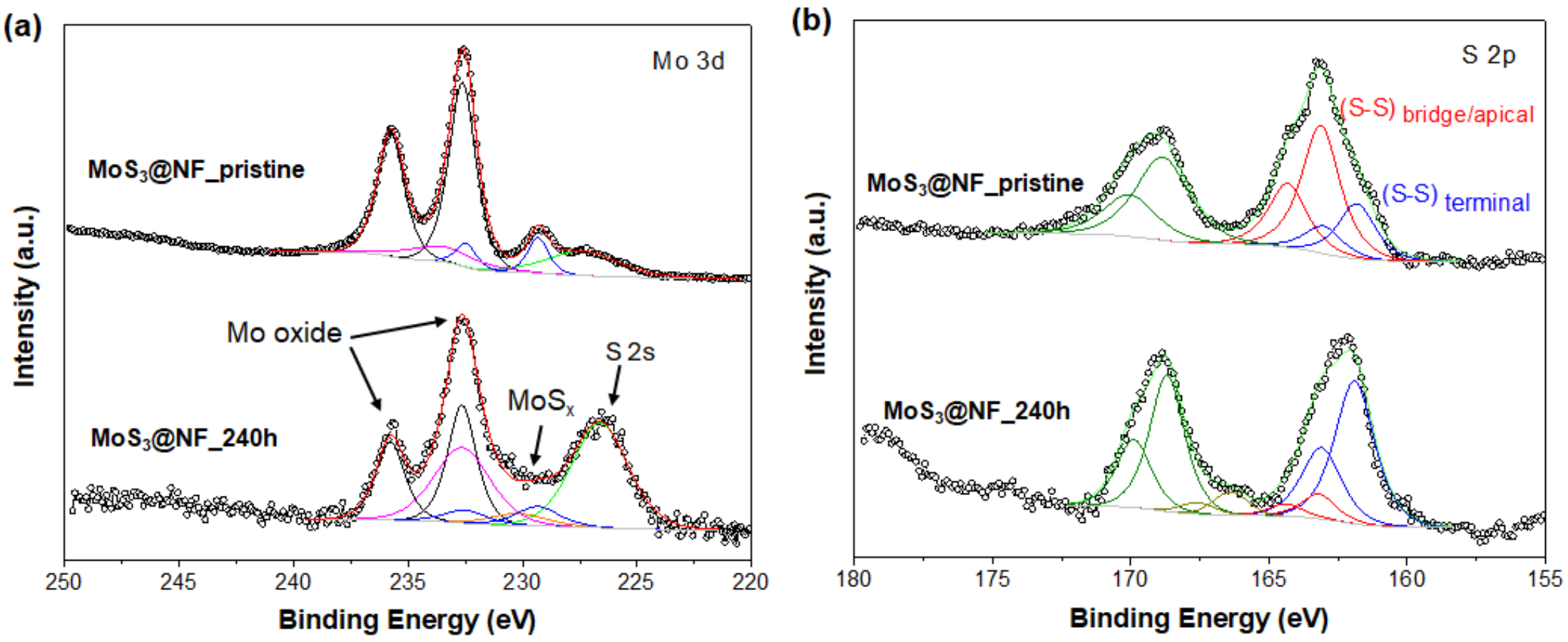

**Figure S16.** XPS spectra of: (a) Mo 3d and (b) S2p region recorded on the $MoS_3$@NF electrode before and after the HER stability test at -200 mA/cm$^2$ for 240 h.

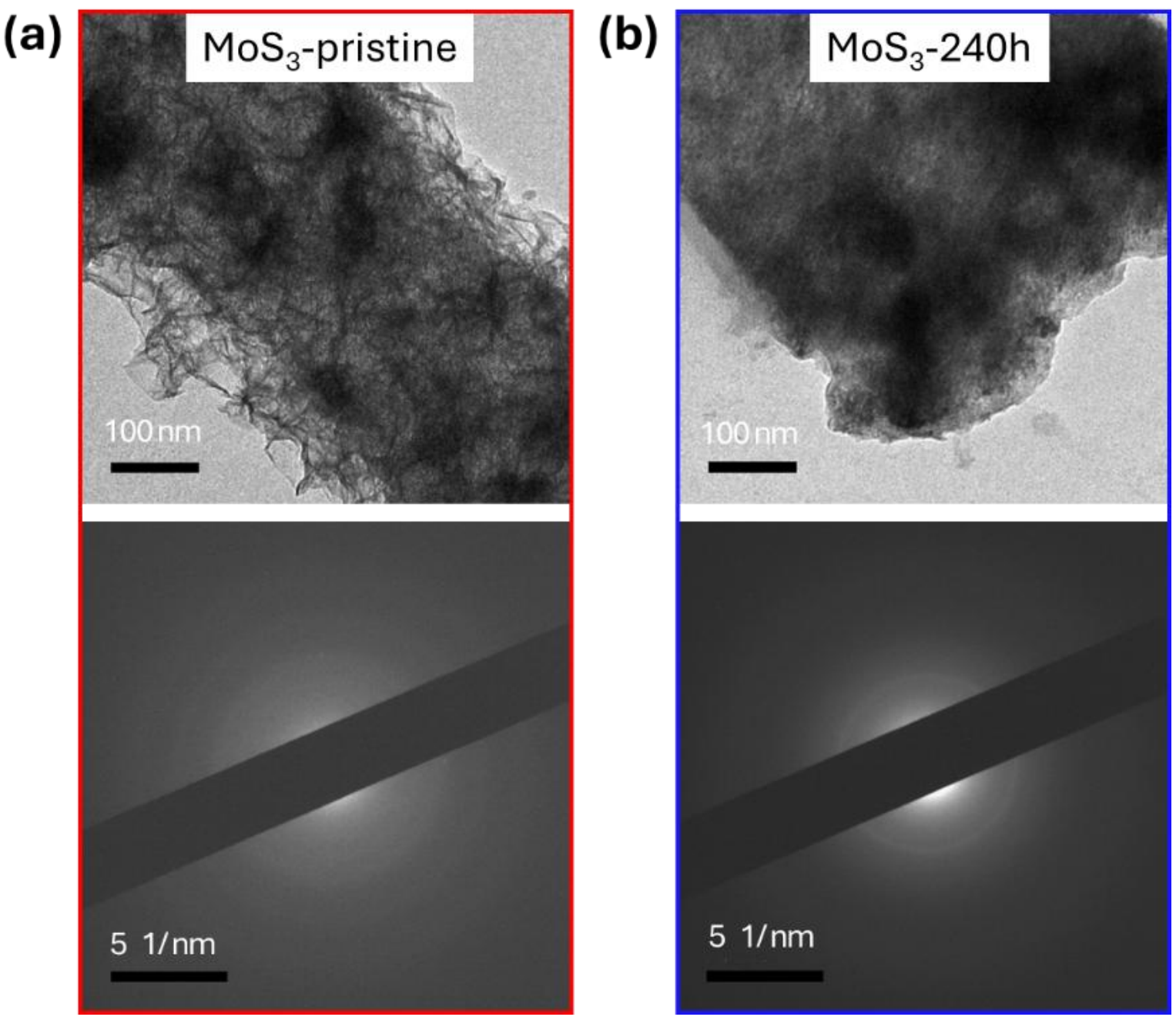

**Figure S17.** Transmission electron microscopy (TEM) image and corresponding selected area electron diffraction (SAED) pattern of the $MoS_3$ catalyst detached from NF substrate (a) before and (b) after the HER stability test at -200 mA/cm$^2$ for 240 h.

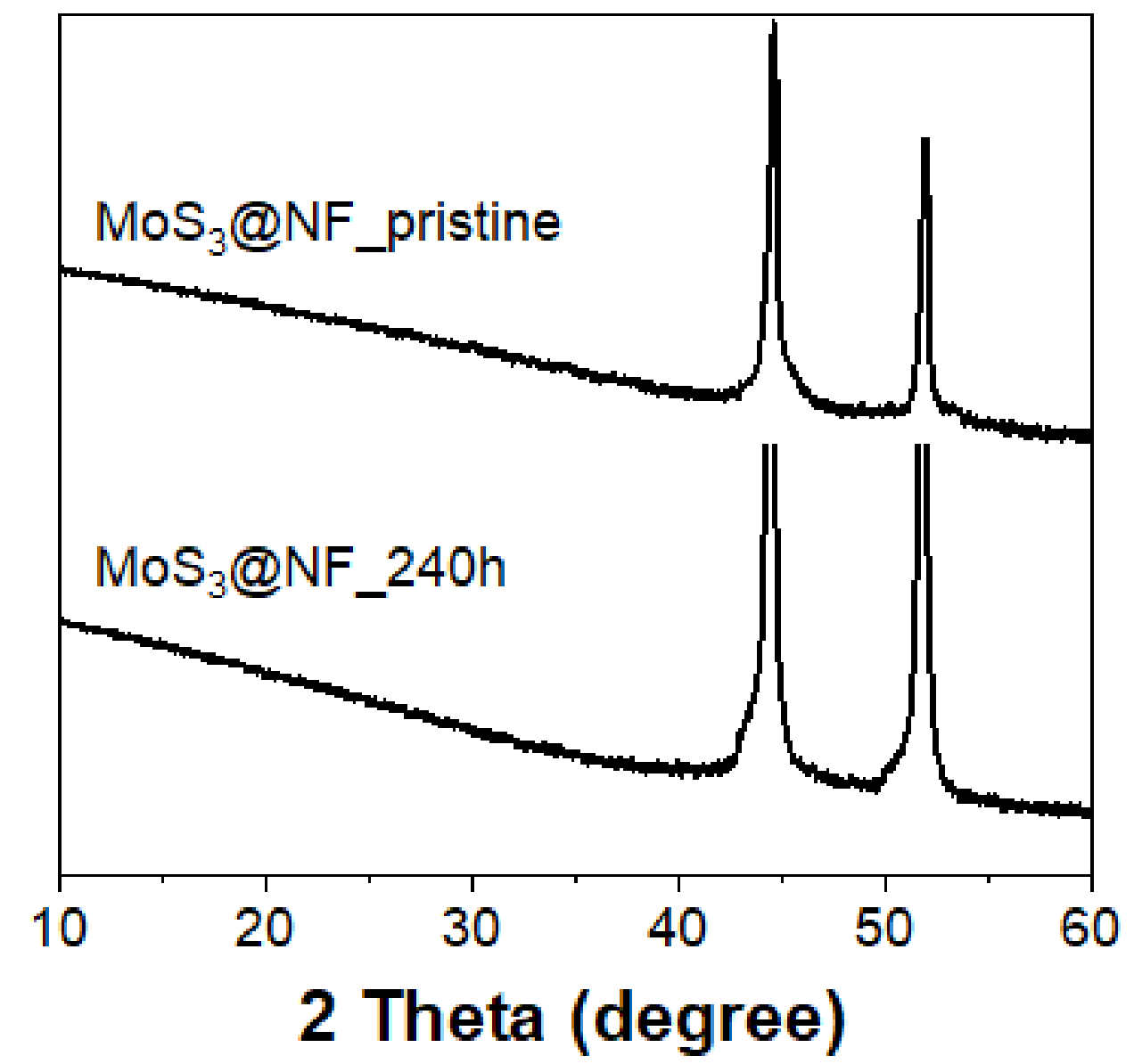

**Figure S18.** XRD pattern of the $MoS_3$@NF electrode: (a) before and (b) after the HER stability test at -200 mA/cm$^2$ for 240 h.

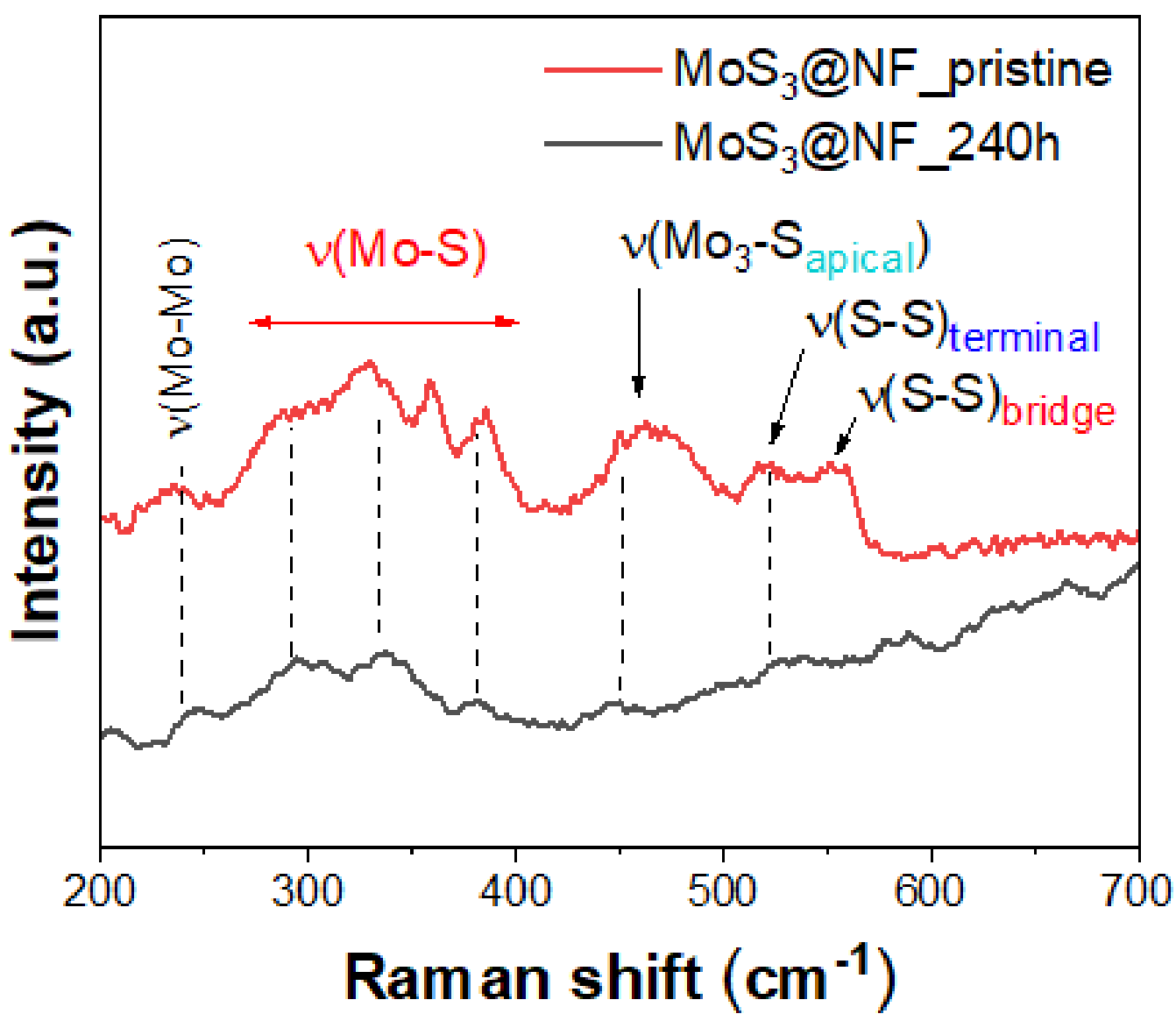

**Figure S19.** Raman spectra of the $MoS_3$@NF electrode before and after the HER stability test at -200 $mA/cm^2$ for 240 h.

As shown in **Figure S19**, $MoS_3$@NF electrode after the 240 h stability test ($MoS_3$@NF_240h) still displayed the characteristic ν(Mo–S) and ν(Mo–Mo) peaks of $MoS_3$, confirming the retention of the $MoS_3$ phase. However, if compared to pristine $MoS_3$@NF electrode, we noticed that: 1) 4 times more accumulations were needed during Raman characterization to acquire high-quality spectra for $MoS_3$@NF_240h electrode, indicating the loss of $MoS_3$ catalyst; 2) ν($Mo_3$–$S_{apical}$), ν(S–S)$_{terminal}$ and ν(S–S)$_{bridge}$ peaks are hardly to see, especially for the ν(S–S)$_{bridge}$ peak. This observation is quite consistent with Artero *et al.* work, in which the signal assigned to bridging disulfide significantly decreased and the signal attributed to terminal disulfide ligands completely disappeared during HER process[1].

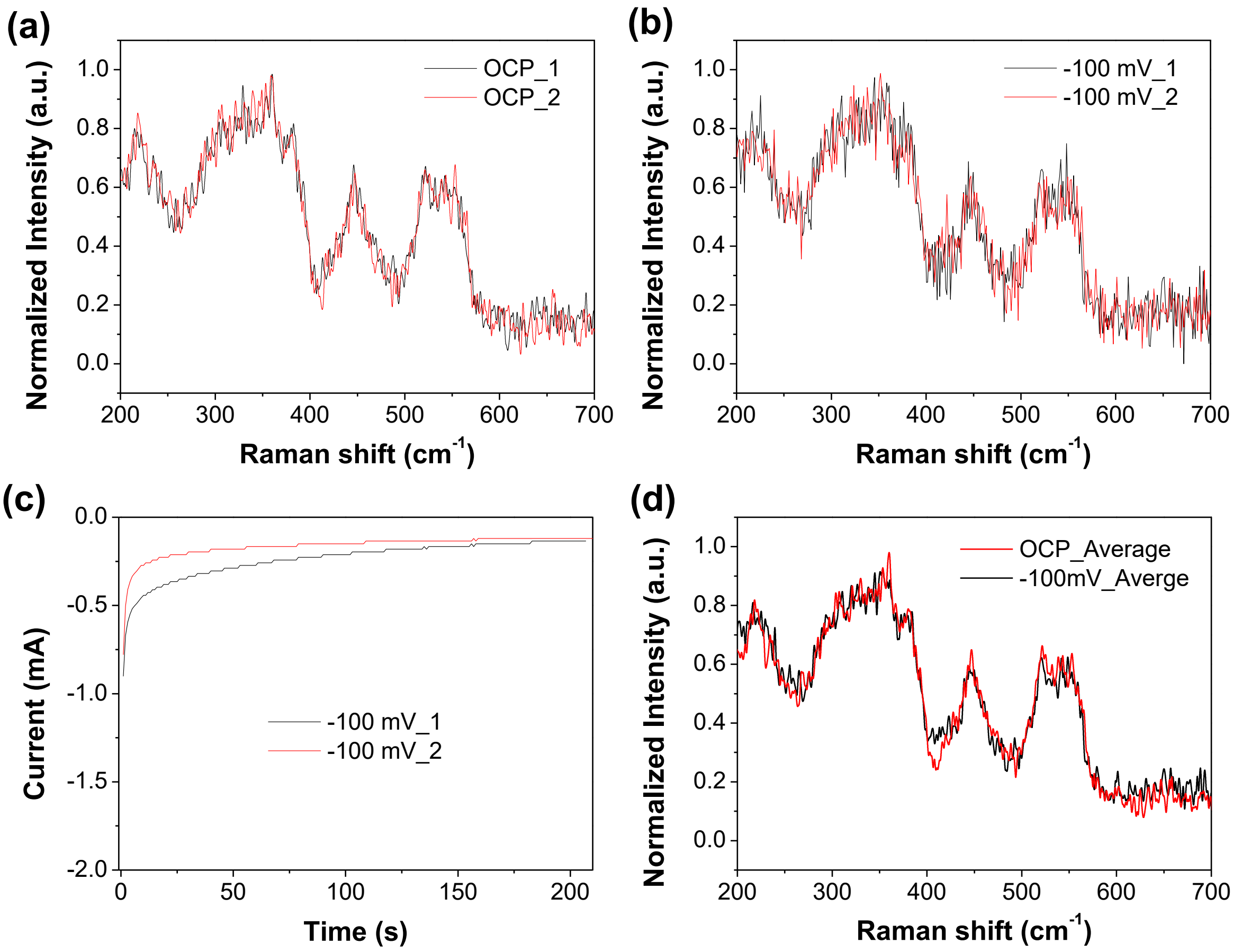

**Figure S20.** Operando Raman spectra acquired on $MoS_3$@Ni sheet at (a) OCP and (b) -100 mV vs. RHE (without iR compensation). (c) Chronoamperometric plots corresponding to the Raman data shown in (b), illustrating the recorded current over time. (d) Averaged Raman plots derived from the data in (a) and (b), enabling a direct comparison of the spectral features at OCP and -100 mV vs. RHE.

For each potential condition, two Raman spectra were recorded. As shown in **Figure S20**, shifting the potential from open circuit potential (OCP) to -100 mV vs. RHE (without iR correction) resulted in an increase in spectral noise. This change is likely attributed to the formation of tiny gas bubbles during the HER process, as evidenced by the corresponding HER current shown in **Figure S20c**. Further increasing the HER potential led to a significant increase in noise in the Raman spectra, likely due to the intensified evolution of hydrogen bubbles on the electrode surface.

**Note S1. Details of atomistic simulations**.

**Note S1.1. DFT settings.**

In this work, all the DFT calculations were performed using the Vienna Ab initio Simulation Package (VASP)[2–4]. To describe the interaction between valence electrons and ionic cores, we used the projector augmented-wave (PAW) method, with a plane-wave cutoff energy of 500 eV[2,5]. For the exchange–correlation functional, we adopted the Perdew–Burke–Ernzerhof (PBE) form of the generalized gradient approximation (GGA)[6], combined with Grimme's D2 dispersion correction (PBE+D2)[7], which provides a computationally efficient and fairly accurate description of van der Waals (dispersion) interactions. Selected PBE+D2 adsorption energies were compared with r2SCAN+rVV10[8,9], which offers improved accuracy for binding energies and weak interactions, greater numerical stability than SCAN [10], and a more realistic description of water structure and hydrogen bonding[11]. The *K*-point meshes were sampled as Γ-centered 1 × 1 × 1 , 1 × 2 × 1 and 1 × 3 × 1 for the $MoS_x$ clusters, the MoS3 chain and $MoS_2$ nanoribbon, respectively. To avoid fictious interactions among the periodic images of the systems, the vacuum spacing was added according to the system type. The isolated clusters (the dimer) were modelled with an average vacuum of 16.5Å (14 Å) along the z-direction and 9 Å (10 Å in average) along the x and y directions. The chain was modelled with 9 Å and 12 Å of vacuum along the x and z directions, respectively, while the nanoribbon was modelled with 11 Å and 16 Å along x and z, respectively (the periodic direction being parallel to the y axis). These vacuum spacings were chosen after a series of test calculations. During the geometry optimizations, all the atoms in the system were allowed to relax, while the cell shape and volume were kept fixed. Exceptions to this were the $MoS_2$ nanoribbon and the $MoS_3$ chain, for which the periodic cell edge (y) was manually varied until the minimum energy is reached, to simulate a 1D system at equilibrium geometry.

**Note S1.2. Structural models.**

*$MoS_x$ systems*. Since the exact geometry of $MoS_3$ is not well established in the literature, and our structural characterization pointed towards amorphous $MoS_x$ structure, we modelled several structures based on common Mo–S clusters that have been reported[12–15]. We began with the well-known $Mo_3S_{13}$ cluster consisting of three Mo atoms forming a triangle linked by an apical S atom. Each Mo–Mo edge is bridged by two sulfur atoms, and each Mo atom is further coordinated by two terminal sulfur atoms. We also considered a $Mo_3S_{13}$ dimer, formed by linking two clusters through 2 terminal S atoms, as well as $Mo_3S_{10}$ and $Mo_3S_9$ clusters obtained by removing S atoms from $Mo_3S_{13}$ according to the reported geometries[13,15]. For each $MoS_x$ system, only the most stable geometry is considered in the remainder of this work. In addition to the clusters, we included the $MoS_3$ chain[12], which consists of a linear backbone of molybdenum atoms connected by alternating long and short Mo–Mo distances. Each Mo–Mo pair is bridged by three S atoms: three unbonded S atoms or two form a disulfide (S–S) bond, and the third remains unbonded to other S atoms, respectively (**Figure 4a** of the main text). Finally, we modelled the $MoS_2$ as a nanoribbon, *i.e.,* a section of monolayer cut from the bulk lattice. The choice to use a nanoribbon rather than a fully 2D periodic $MoS_2$ was dictated by necessity to include the $MoS_2$ edges, which are known to be the active sites in photocatalysis[16]. The ribbon is two hexagons wide, forming two parallel stripes in which S atoms terminate the edges. Two distinct types of sulfur atoms are exposed along these edges, providing potential active sites.

*Adsorption of H, OH, and $H_2O$ and $H_2O$ dissociation*. Computationally, the reactions (1)-(3) of the main text were formulated to represent the processes under study, and the corresponding reaction energies were obtained directly from the DFT-calculated energies following these equations (eqs. E2-E4, respectively):

$$\Delta E^{H2O}_{ads} = E_{MoSx\text{-}H2O} - E_{MoSx} - E_{H2O} \quad (E1)$$

$$\Delta E^{H2O}_{diss} = E_{MoSx\text{-}OH} + E_{MoSx\text{-}H} - E_{MoSx\text{-}H2O} - E_{MoSx} \quad (E2)$$

$$\Delta E^{OH}_{des} = -\Delta E^{OH}_{ads} = E_{MoSx} + E_{H2O} - E_{MoSx\text{-}OH} - \tfrac{1}{2} E_{H2} \quad (E3)$$

$$\Delta E^{H}_{des} = -\Delta E^{H}_{ads} = E_{MoSx\text{-}H} - \tfrac{1}{2} E_{H2} - E_{MoSx} \quad (E4)$$

Where $E_{MoSx\text{-}OH}$, $E_{MoSx\text{-}H}$ and $E_{MoSx\text{-}H2O}$ are energies of the $MoS_x$ systems with adsorbed OH, H and $H_2O$, respectively, $E_{MoSx}$ is the energy of the bare $MoS_x$ systems, which can be a cluster, a chain, or a nanoribbon (see above). $E_{H2O}$ and $E_{H2}$ are the energies of gas-phase $H_2O$ and $H_2$, respectively. For each $MoS_x$ system, a systematic scan of adsorption sites and geometries was performed. An example for $Mo_6S_{24}$ (*i.e.* a $Mo_3S_{13}$ dimer, see above) is reported in **Figure S21** below. Note that, by employing eqs. E1-E4, we are implicitly making two assumptions: i) that the coverage of H and OH in the real catalyst is always low enough to make the adsorbate-adsorbate interactions negligible, and ii) that, once water dissociates, H and OH quickly reach their most stable positions, moving away from each other. Indeed, our test simulations on $Mo_3S_{13}$ and $Mo_6S_{24}$ showed that, when H and OH lie in nearby adsorption sites, the energy of the system is invariably higher than when H and OH are far away from each other (we model this large distance by adsorbing them on separate $MoS_x$ clusters). Note, however, that this assumption (ii) has hardly any relevance for the overall HER scenario, since in almost all systems our simulations found that $H_2O$ dissociates into adsorbed H and $OH^-$ in solution (see the discussion on computational results in the main text).

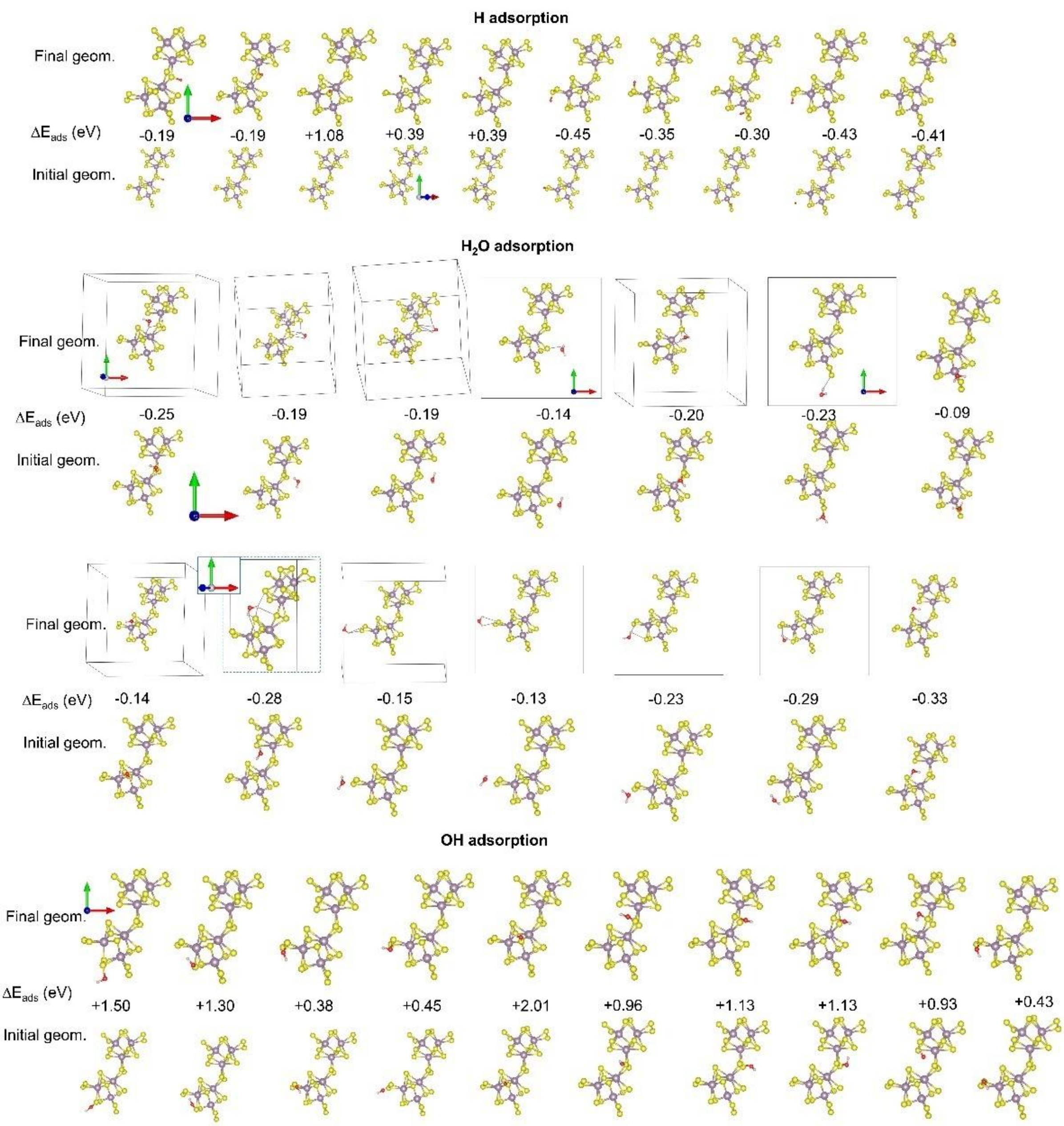

**Figure S21.** Adsorption configurations considered for H, OH, and $H_2O$ on $Mo_6S_{24}$. Starting and DFT-optimized geometries, along with the adsorption energies, are reported. In selected cases, hydrogen bonds are shown as thin grey lines. Note that, for hydrogen, Table 1 of the main text reports adsorption Gibbs free energies so the values are different (see Methods). Note also that for hydrogen adsorption geometries only, H atoms are shown in red to make it easier for the reader to spot them.

Following common practice in the literature on $MoS_x$ catalysts[1,16,17], the Gibbs free energy of hydrogen desorption ($\Delta G^{H}_{des}$) is obtained from the H desorption energy ($\Delta E^{H}_{des}$) calculated by equation (E4), by correcting it for zero-point vibrational energy and entropy change between the adsorbed and the gas-phase $H_2$. Nørskov and co-workers estimate this correction for $MoS_2$ to be[18]:

$$\Delta G^{H}_{des} = \Delta E^{H}_{des} - 0.29eV \qquad (E5)$$

Where $\Delta E^{H}_{des}$ is the energy of H desorption calculated by equation (E4).

**Note S1.3. Thermodynamic considerations.**

Following ref. [19] , we exploit the fact that reaction R1 below, at the electric potential adopted for the hydrogen evolution reaction (HER potential), is at equilibrium, *i.e.* its Gibbs free energy change (ΔG) is zero.

$2H_2O + 2e- \rightleftharpoons H_2 + 2OH^-$ (R1).

It follows that:

$G(e^-) = ½ G(H_2) + G(OH^-) - G(H_2O)$ (E6),

where the terms in equation E6 are, from left to right, the free energies of: electron, $H_2$ molecule, solvated $OH^-$, and a water molecule. Based on this, it can be demonstrated that, at the HER potential:

A. *The free energy changes associated to reactions R2 and R3 below are equivalent*. This demonstration can be found in Note S3.1 of ref. [19]. It has been exploited in the main text to evaluate the desorption of OH from $MoS_x$ in realistic conditions (R2) while keeping the computational cost manageable (R3).

   $OH^* + e^- \rightarrow OH^- + *$ (R2)

   $OH^* + ½ H_2 \rightarrow * + H_2O$ (R3)

B. The free energy changes of (alkaline) Heyrovsky reaction step (R4) is equivalent to that of the simple hydrogen desorption reaction (Tafel step, R5).

   $H^* + H_2O + e- \rightarrow H_2 + OH^- + *$ (R4),

   $H^* \rightarrow ½ H_2 + *$ (R5).

   Indeed:

   $$\Delta G^{Heyr.} = G(H_2) + G(OH^-) + G(*) – G(H^*) – G(H_2O) – G(e^-) =$$
   $$G(H_2) + G(OH^-) + G(*) – G(H^*) – G(H_2O) – [1/2\, G(H_2) + G(OH^-) – G(H_2O)]$$
   $$=$$
   $$1/2\, G(H_2) + G(*) – G(H^*) = \Delta G^{Tafel}$$

**Note S1.4. Additional simulation results**

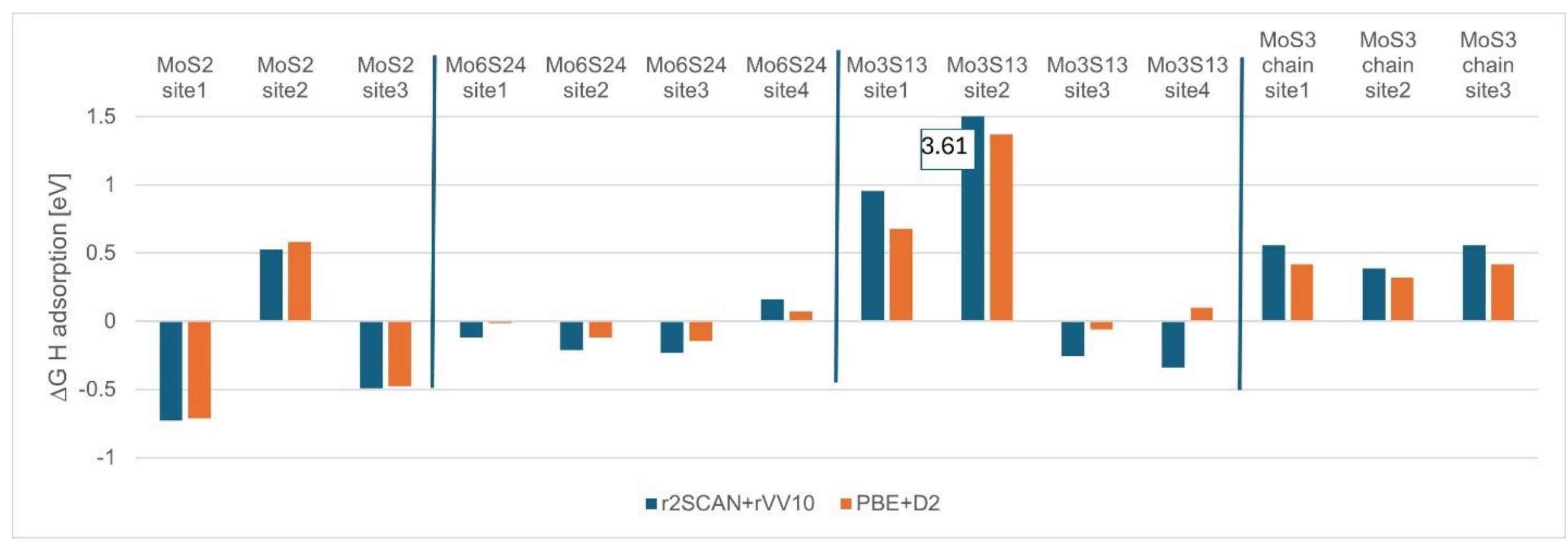

**Figure S22.** Comparison of the free energy of hydrogen adsorption of PBE+D2 vs. r2SCAN+rVV10 methods.

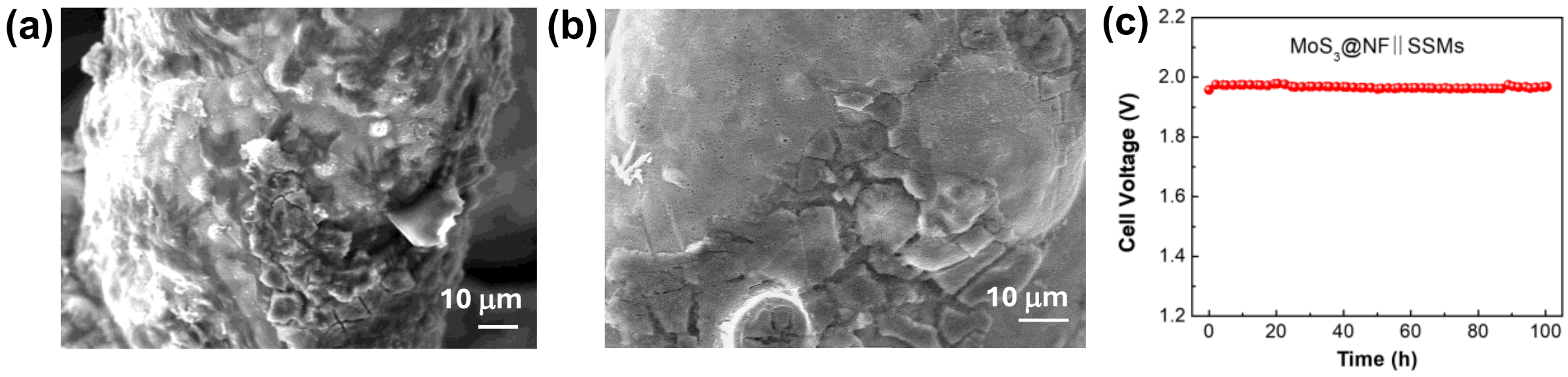

**Figure S23.** SEM image of the $MoS_3$@NF cathode surface after (a) the AST and(b) the additional stability test at 1 A/cm$^2$ for continuous 100 h in $MoS_3$@NF||SSM AWE single cell. (c) continuous stability test for the AST-stabilized $MoS_3$@NF||SSMs AWE cell at 1 A/cm$^2$ for 100 h. AST: accelerated stress test.

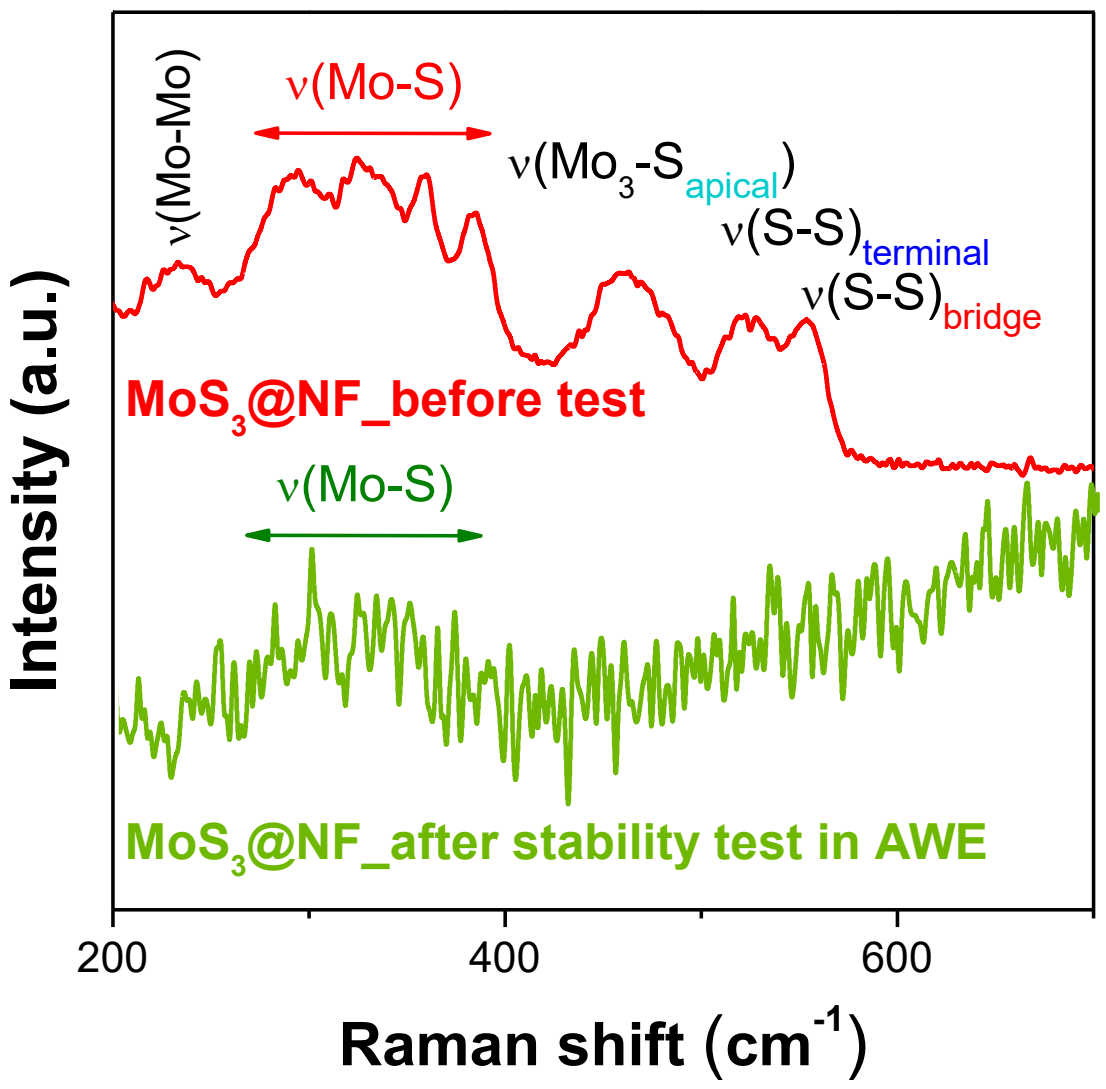

**Figure S24.** Raman spectra of the $MoS_3$@NF electrode before and after the continuous stability test at 1 A/cm$^2$ for 100 h in AWE single cell.

Although the Raman intensity was significantly attenuated, the spectra acquired from the $MoS_3$@NF electrode surface after the stability test still displayed the characteristic ν(Mo–S) peaks of $MoS_3$, confirming the partial retention of the $MoS_3$ phase. The reduction in Raman signal intensity may be attributed to two possible factors: 1) loss of surface $MoS_3$ species due to partial detachment or degradation during the long-term test; 2) contamination or surface coverage of $MoS_3$ by carbonate ions formed from the reaction between $CO_2$ in air and the alkaline electrolyte during the prolonged operation.

**Table S1.** ICP and SEM-EDS data measured for the $MoS_x$@NF electrode after the annealing for 1 h at various temperatures (250 °C, 350 °C, 500 °C).

| Sample | ICP S/Mo, atm% | *EDS[a] S/Mo, atm%* |
|---|---|---|
| ATM@NF_RT | 3.91 | *2.79* |
| $MoS_x$@NF_250 | 3.52 | *1.90* |
| $MoS_x$@NF_350 | 3.08 | *1.45 & 1.61* |
| $MoS_x$@NF_500 | 2.59 | *0.80* |

[a.] Less reliable due to the close dispersive energy of Mo and S

As discussed in the main text, the formula of $MoS_3$ used in this work does not signify that its ideal ratio of S/Mo should be 3. We use $MoS_3$ formula in this work for clarity purposes. In some works from the literature, more accurate representations such as $MoS_{3\,2/3}$ [20] or $[Mo_3S_{13}]^{2-}$ [21] are used, both of which indicate an S/Mo ratio greater than 3. This explains why the $MoS_3$@NF (*i.e.,* $MoS_x$@NF_250) exhibits a higher S/Mo ratio compared to the stoichiometric value suggested by the simplified $MoS_3$ formula.

**Table S2.** S/Mo atomic ratio of the $MoS_3$@NF catalyst before and after the 240 h HER stability test.

| Sample | ICP (catalyst) S/Mo, atm% | ICP (electrolyte) S/Mo, atm% | XPS (catalyst) S/Mo, atm% | *EDS[a] (catalyst) S/Mo, atm%* |
|---|---|---|---|---|
| $MoS_3$@NF_pristine | 3.76 | N.A. | 2.42 | *2.01* |
| $MoS_3$@NF_240h | 4.31 | 2.62 | 12.51 | *2.80* |

[a.] Less reliable due to the close dispersive energy of Mo L (2.293 eV) and S K (2.307 eV).

As summarized in **Table S2**, the ICP results acquired on electrolyte demonstrate that both S and Mo are partially lost from the $MoS_3$ catalyst after the 240 h stability test, with Mo being more easily leached compared to S. This result can be further supported by the increased S/Mo atomic ratio of catalyst after stability test, as observed using ICP, XPS and EDS techniques.

**Table S3.** Comparison of the HER activities for $MoS_x$-based catalysts in 1 M KOH/NaOH reported over the past few years.

| Sample | η10 (mV) | η20 (mV) | η50 (mV) | η100 (mV) | Tafel slope (mV/dec) | Reference |
|---|---|---|---|---|---|---|
| **$MoS_3$@NF** | **132** | **155** | **168** | **186** | **74** | **This work** |
| $Co_3S_4$-$MoS_2$ HNPls | 117 | / | ~175 | ~205 | 79 | Angew. Chem. Int. Ed., 2025, 64, e202414720 |
| Se-$CoS_2$@$MoS_2$/CNTs | 85.2 | ~110 | ~150 | ~200 | 91.59 | Small, 2025, 21, 2407049 |
| $MoS_2$@NiFe-LDH | 91 | / | ~320 | ~430 | 132 | J. Power Sources, 2025, 657, 238120 |
| a-$MoS_x$/$Ni_3S_2$/$Ni_3N$@NF-2 | 69 | ~100 | ~150 | ~180 | 63.7 | Chem. Eng. J., 2025, 510, 161077 |
| Fe-$MoS_2$/CoFe-MOF | 148 | / | ~280 | ~360 | 103.2 | J. Power Sources, 2025, 641, 236881 |
| $CoFe_2O_4$/$MoS_2$ | 154 | ~175 | ~200 | 229 | 42 | J. Colloid Interface Sci., 2025, 680, 541 |
| $MoS_2$/NGr-Co SACs | ~200 | ~230 | ~325 | ~375 | 87.8 | Chem. Eng. J., 2025, 521, 166979 |
| $MoS_2$-$Ni_xS_6$/NF | 136 | / | ~250 | ~330 | 92.2 | Int. J. Hydrogen Energy, 2025, 137, 95 |
| $CoNiP_x$/$NiS_x$/$MoS_2$/NF | 79 | ~110 | ~160 | / | 40 | J. Power Sources, 2025, 653, 237770 |
| $Ni_2P$/$MoS_2$-$CoMo_2S_4$@C | 73 | / | ~140 | ~175 | 94.2 | ACS Nano, 2024, 18, 6202 |
| $MoS_2$/$Ti_3C_2$ | 124 | ~160 | ~240 | / | 24.63 | Adv. Funct. Mater., 2024, 34, 2406188 |
| vr-1T $MoS_2$ | 184 | ~210 | ~255 | / | 92 | Appl. Catal. B: Environ., 2024, 352, 124037 |
| Re-$MoS_2$-Vs | 99 | / | ~200 | ~230 | 89 | Nano Res., 2024, 17, 9507 |

| Tm SAs-$MoS_2$ | 80 | ~110 | ~140 | ~170 | 82.6 | Adv. Energy Mater., 2024, 14, 2401716 |
|---|---|---|---|---|---|---|
| Co Nds/$MoS_{2-x}$-d | 151 | ~190 | ~280 | / | 99.8 | ACS Appl. Nano Mater., 2024, 7, 6596 |
| Cr–$Co_3S_4$/$NiMoS_4$ | 72 | / | ~160 | 209 | 104 | J. Power Sources, 2024, 614, 234969 |
| $Mo_2C$/$MoS_2$-CP | 191.4 | / | 220 | 258.4 | 64.5 | Chem. Commun., 2024, 60, 11112 |
| Core−Shell $MoS_2$/$TiO_2$ | 520 | / | 620 | 700 | 60 | ACS Appl. Energy Mater., 2019, 2, 2053 |
| $MoS_2$@Ni Core/Shell | 91 | 118 | 155 | 196 | 89 | ACS Appl. Mater. Interfaces, 2016, 8, 14521 |
| $MoS_2$:$MoNi_4$/$MoO_2$ | 155.6 | / | 220 | 320 | 67 | ACS Appl. Nano Mater., 2021, 4, 886 |
| Ni-$MoS_2$ | 98 | / | 150 | / | 60 | Energy Environ. Sci., 2016, 9, 2789 |
| $NiMoS_4$/Ti | 138 | / | 185 | / | 97 | J. Mater. Chem. A, 2017,5, 16585 |
| $MoS_2$-$NiS_2$/NGF | 172 | 200 | 250 | / | 70 | Appl. Catal. B: Environ., 2019, 254, 15 |
| meso-Fe-$MoS_2$/$CoMo_2S_4$ | 122 | 150 | 195 | / | 90 | ACS Nano 2020, 14, 4141 |
| $MoS_2$-$CoS_2$@PCMT | 200 | / | / | 300 | 93 | J. Power Sources, 2021, 514, 230580 |
| γ-FeOOH/$MoS_2$/CC | / | / | / | 226 | 106 | Appl. Catal. B: Environ., 2021, 297, 120456 |
| $MoS_2$/SS-1600 | 160 | / | / | 240 | 61 | Int. J. Hydrog. Energy, 2019, 44, 1555 |
| 1T–2H $MoS_2$ | 300 | 320 | / | / | 65 | Adv. Energy Mater., 2018, 8, 1801345 |
| $MoS_2$/NiCoS | 189 | 210 | 250 | / | 75 | J. Mater. Chem. A, 2019,7, 27594 |

**Table S4.** Comparison of the performance of the $MoS_3$@NF||SSMs AWE cell with previously reported metal sulfide-based electrolyzers.

| Electrolyzer type | Temperature (°C) | Cathode | Anode | @0.2 A/cm$^2$ (V) | @0.4 A/cm$^2$ (V) | @1 A/cm$^2$ (V) | Stability | Reference |
|---|---|---|---|---|---|---|---|---|
| **AWE** | **80** | **$MoS_3$@NF** | **5-stacked SS mesh** | **1.67** | **1.75** | **1.90** | 1A/cm$^2$ @ ~1.96V after 1000 h | **This work** |
| AWE | 80 | NF | 5-stacked SS mesh | 1.82 | 1.97 | 2.27 | / | |
| AWE | 80 | Raney Ni | NiFe-LDH/$Ni_3S_2$ | ~1.64 | ~1.74 | ~1.98 | 0.8A/cm$^2$ @ ~1.9V after 40 h | Angew. Chem. Int. Ed., 2025, 64, e202504972 |
| AWE | 80 | graphene $MoS_2$/FeCoNi$P_x$ | graphene/$MoS_2$/ FeCoNi$(OH)_x$ | 1.59 | ~1.75 | / | 0.1A/cm$^2$ @ ~1.64V after 100 h | Nat. Commun., 2021, 12, 1380 |
| AWE | 70 | Ni-Mo-steel | Low-carbon steel | / | / | ~2.18 | 1A/cm$^2$ @ ~2.13V after 100 h | J. Electrochem. Soc., 2020, 167, 114513 |
| AWE | 60 | $W_s/Mo_i$-$Ni_3S_2$@CoSe/NF | $W_s/Mo_i$-$Ni_3S_2$@CoSe/NF | / | ~1.74 | 1.88 | / | Appl. Catal. B: Environ., 2026, 380, 125800 |
| AWE | 80 | MW-ex-$ZrSe_2$:Ni | SS mesh | 1.85 | 1.92 | 2.03 | 0.4A/cm$^2$ @ ~1.95V after 120 h | Nano Select, 2022, 3, 1069 |

| PEMWE | 80 | MW-ex-$ZrSe_2$:Ni | $IrO_2$ | 1.80 | 1.88 | / | | |
|---|---|---|---|---|---|---|---|---|
| PEMWE | 80 | FeCo-$MoS_2$ | Ir black (Johnson–Matthey) | ~1.85 | ~2 | ~2.2 | n.d. | Adv. Funct. Mater., 2025, 10.1002/adfm.2025 03549 |
| PEMWE | 80 | Co-doped $MoS_2$ | $IrO_2$ | / | / | ~2.1 | n.d. | Mater. Today Adv., 2020, 6, 100020 |
| PEMWE | 80 | $MoS_x$/carbon black | Ir black | / | / | ~1.95 | n.d. | Catalysts, 2018, 8, 657 |
| PEMWE | 80 | $MoS_2$/C | $IrO_2$ | ~1.82 | / | / | ~0.35A/$cm^2$ @ 2V after 18h | Int. J. Hydrogen Energy, 2014, 39, 20837 |
| PEMWE | 80 | $MoS_2$ | $IrO_2$ | >2 | / | / | n.d. | |
| PEMWE | 80 | Pyrite $FeS_2$ | $IrO_2$ | / | / | 2.10 | n.d. | ACS Catal., 2016, 6, 2626 |
| PEMWE | 60 | 1T′-$MoTe_2$/CC | $RuO_2$/Ti-fiber felt | >2 | / | / | 0.1A/$cm^2$ @ ~1.9V after 20 h | ACS Sustainable Chem. Eng., 2024, 12, 1276 |
| AEMWE | n.d. | core–shell-structured $MoS_2$/CoS | core–shell-structured $MoS_2$/CoS | 1.63 | ~1.76 | / | 0.2A/$cm^2$ @ ~1.63V after 500 h | Inorg. Chem. Front., 2025,12, 5159 |

| AEMWE | 60 | 20 wt% Pt/C | NiFe/Ni-S sprayed Ni foam | / | ~1.64 | ~1.78 | 1A/cm$^2$ @ ~1.9V after 400 h | Small, 2024, 20, 2310737 |
|---|---|---|---|---|---|---|---|---|
| AEMWE | 70 | MD-Co/$NiS_2$ | $IrO_2$ | ~1.71 | ~1.81 | 1.97 | 0.5A/cm$^2$ @ ~1.88V after 50h | Surf. Interf., 2024, 46, 103987 |
| AEMWE | 60 | Pt–C core-shell@*h*-$MoS_2$/GNF | $NiCo_2O_4$ | ~1.82 | ~1.92 | 2.08 | ~0.1A/cm$^2$ @ 1.85V after 14 h | Carbon, 2024, 220, 118816 |
| AEMWE | n.d. | Co−1T−$MoS_2$-bpe-350 | NiFe-LDH | >2 | / | / | 0.1A/cm$^2$ @ 2V after 10 h | Angew. Chem. Int. Ed., 2023, 62, e202313845 |
| n.d. | 60 | $Ni_3S_2$/$NiMoO_4$ | $Ni_3S_2$/$NiMoO_4$ | ~1.7 | ~1.8 | 2.1 | 0.5A/cm$^2$ @ 1.87V after 100 h | Chem. Eng. J., 2025, 512, 162044 |
| n.d. | 60 | Ni | SS | 2.00 | / | / | n.d. | Angew. Chem. Int. Ed., 2015, 54, 11989 |
| n.d. | 60 | Ni@NiO/$Cr_2O_3$ | NiFe layered double hydroxide | 1.64 | / | / | 0.2A/cm$^2$ @ 1.64V after 500 h | |

## References

1. Tran, P.D., Tran, T. V., Orio, M., Torelli, S., Truong, Q.D., Nayuki, K., Sasaki, Y., Chiam, S.Y., Yi, R., Honma, I., et al. (2016). Coordination polymer structure and revisited hydrogen evolution catalytic mechanism for amorphous molybdenum sulfide. Nat. Mater. *15*, 640–646. https://doi.org/10.1038/nmat4588.

2. Kresse, G., and Joubert, D. (1999). From ultrasoft pseudopotentials to the projector augmented-wave method. Phys. Rev. B *59*, 1758. https://doi.org/10.1103/PhysRevB.59.1758.

3. Kresse, G., and Furthmüller, J. (1996). Efficiency of ab-initio total energy calculations for metals and semiconductors using a plane-wave basis set. Comput. Mater. Sci. *6*, 15–50. https://doi.org/10.1016/0927-0256(96)00008-0.

4. Kresse, G., and Furthmüller, J. (1996). Efficient iterative schemes for *ab initio* total-energy calculations using a plane-wave basis set. Phys. Rev. B *54*, 11169. https://doi.org/10.1103/PhysRevB.54.11169.

5. Blöchl, P.E. (1994). Projector augmented-wave method. Phys. Rev. B *50*, 17953. https://doi.org/10.1103/PhysRevB.50.17953.

6. Perdew, J.P., Burke, K., and Ernzerhof, M. (1996). Generalized Gradient Approximation Made Simple. Phys. Rev. Lett. *77*, 3865. https://doi.org/10.1103/PhysRevLett.77.3865.

7. Grimme, S. (2006). Semiempirical GGA-type density functional constructed with a long-range dispersion correction. J. Comput. Chem. *27*, 1787–1799. https://doi.org/10.1002/JCC.20495.

8. Furness, J.W., Kaplan, A.D., Ning, J., Perdew, J.P., and Sun, J. (2020). Accurate and Numerically Efficient r2SCAN Meta-Generalized Gradient Approximation. J. Phys. Chem. Lett. *11*, 8208–8215. https://doi.org/10.1021/ACS.JPCLETT.0C02405.

9. Ning, J., Kothakonda, M., Furness, J.W., Kaplan, A.D., Ehlert, S., Brandenburg, J.G., Perdew, J.P., and Sun, J. (2022). Workhorse minimally empirical dispersion-corrected density functional with tests for weakly bound systems: <math xmlns="http://www.w3.org/1998/Math/MathML"><mrow><msup><mrow><mi mathvariant="normal">r</mi></mrow><mn>2</mn></msup><mi>SCAN</mi><mo>+</mo><m…. Phys. Rev. B *106*, 075422. https://doi.org/10.1103/PhysRevB.106.075422.

10. Kothakonda, M., Kaplan, A.D., Isaacs, E.B., Bartel, C.J., Furness, J.W., Ning, J., Wolverton, C., Perdew, J.P., and Sun, J. (2022). Testing the r2SCAN Density Functional for the Thermodynamic Stability of Solids with and without a van der Waals Correction. ACS Mater. Au *3*, 102–111. https://doi.org/10.1021/ACSMATERIALSAU.2C00059.

11. Kaplan, A.D., Shahi, C., Sah, R.K., Bhetwal, P., Kanungo, B., Gavini, V., and Perdew, J.P. (2024). How does HF-DFT achieve chemical accuracy for water clusters? ChemRxiv. https://doi.org/10.26434/CHEMRXIV-2024-S1WHK.

12. Weber, T., Muijsers, J.C., and Niemantsverdriet, J.W. (1995). Structure of amorphous MoS3. J. Phys. Chem. *99*, 9194–9200. https://doi.org/10.1021/J100022A037.

13. Ting, L.R.L., Deng, Y., Ma, L., Zhang, Y.J., Peterson, A.A., and Yeo, B.S. (2016). Catalytic Activities of Sulfur Atoms in Amorphous Molybdenum Sulfide for the Electrochemical Hydrogen Evolution Reaction. ACS Catal. *6*, 861–867. https://doi.org/10.1021/ACSCATAL.5B02369.

14. Liang, K.S., Cramer, S.P., Johnston, D.C., Chang, C.H., Jacobson, A.J., deNeufville, J.P., and Chianelli, R.R. (1980). Amorphous MoS3 and WS3. J. Non-Cryst. Solids *42*, 345–356. https://doi.org/10.1016/0022-3093(80)90035-6.

15. Hibble, S.J., and Wood, G.B. (2004). Modeling the Structure of Amorphous MoS3: A Neutron Diffraction and Reverse Monte Carlo Study. J. Am. Chem. Soc. *126*, 959–965. https://doi.org/10.1021/JA037666O.

16. Hinnemann, B., Moses, P.G., Bonde, J., Jørgensen, K.P., Nielsen, J.H., Horch, S., Chorkendorff, I., and Nørskov, J.K. (2005). Biomimetic hydrogen evolution: MoS2 nanoparticles as catalyst for hydrogen evolution. J. Am. Chem. Soc. *127*, 5308–5309. https://doi.org/10.1021/JA0504690.

17. Kronberg, R., Hakala, M., Holmberg, N., and Laasonen, K. (2017). Hydrogen adsorption on MoS2-surfaces: a DFT study on preferential sites and the effect of sulfur and hydrogen coverage. Phys. Chem. Chem. Phys. *19*, 16231–16241. https://doi.org/10.1039/C7CP03068A.

18. Su, H.Y., Ma, X., and Sun, K. (2022). Single-atom metal tuned sulfur vacancy for efficient H2 activation and hydrogen evolution reaction on MoS2 basal plane. Appl. Surf. Sci. *597*, 153614. https://doi.org/10.1016/J.APSUSC.2022.153614.

19. Zuo, Y., Bellani, S., Saleh, G., Ferri, M., Shinde, D. V., Zappia, M.I., Buha, J., Brescia, R., Prato, M., Pascazio, R., et al. (2023). Ru-Cu Nanoheterostructures for Efficient Hydrogen Evolution Reaction in Alkaline Water Electrolyzers. J. Am. Chem. Soc. *145*, 21419–21431. https://doi.org/10.1021/JACS.3C06726.

20. Daeneke, T., Dahr, N., Atkin, P., Clark, R.M., Harrison, C.J., Brkljača, R., Pillai, N., Zhang, B.Y., Zavabeti, A., Ippolito, S.J., et al. (2017). Surface Water Dependent Properties of Sulfur-Rich Molybdenum Sulfides: Electrolyteless Gas Phase Water Splitting. ACS Nano *11*, 6782–6794. https://doi.org/10.1021/ACSNANO.7B01632.

21. Kibsgaard, J., Jaramillo, T.F., and Besenbacher, F. (2014). Building an appropriate active-site motif into a hydrogen-evolution catalyst with thiomolybdate [Mo3S13]2− clusters. Nat. Chem. *6*, 248–253. https://doi.org/10.1038/nchem.1853.